\PassOptionsToPackage{kerning}{microtype}

\documentclass[%
  format=manuscript,%
  anonymous=false,%
  review=false,%
  screen=true,%
]{acmart}
\acmJournal{TOSEM}
\setcopyright{acmlicensed}

\usepackage[main=UKenglish]{babel}
\usepackage{mathtools}
\usepackage{mismath}
\usepackage{xspace}
\usepackage[ruled,vlined,linesnumbered]{algorithm2e}

\usepackage[%
  add-integer-zero=false,%
  free-standing-units=true,%
  group-minimum-digits=4,%
  list-final-separator={, and },%
  mode=match,%
  round-mode=figures,%
  round-pad=false,%
  round-precision=3,%
  separate-uncertainty=true,%
  uncertainty-mode=separate,%
]{siunitx}

\usepackage[]{subcaption}

\usepackage{listings}
\lstdefinestyle{inlinelst}{%
  basicstyle=\normalsize\ttfamily,%
  keywordstyle=,%
  language=Python,%
}
\lstdefinestyle{numberedlst}{%
  language=Python,%
  basicstyle=\scriptsize\ttfamily,%
  numbers=left,%
  numberstyle=\tiny,%
  tabsize=2,%
  breaklines=true,%
  xleftmargin=1em,%
  xrightmargin=1em,%
}
\lstdefinestyle{nonnumberedlst}{%
  language=Python,%
  basicstyle=\scriptsize\ttfamily,%
  numbers=none,%
  tabsize=2,%
  breaklines=true,%
  xleftmargin=1em,%
  xrightmargin=1em,%
}
\lstdefinestyle{nonnumberedtxt}{%
  basicstyle=\scriptsize\ttfamily,%
  numbers=none,%
  tabsize=2,%
  breaklines=true,%
  xleftmargin=1em,%
  xrightmargin=1em,%
}
\NewDocumentCommand{\inlinelst}{m}{\lstinline[style=inlinelst]{#1}}
\usepackage{colortbl}
\usepackage{adjustbox}
\usepackage{multirow}
\usepackage{todonotes}
\usepackage{graphicx}
\usepackage{csquotes}
\usepackage{booktabs}
\usepackage{tikz}
\usepackage{pgf}

\usetikzlibrary{positioning,shapes.geometric,arrows.meta}

\SetExtraKerning{
  encoding = {T1}
}{
  \textemdash = {167,167},
  — = {167,167}
}

\let\origfootnote\footnote
\renewcommand{\footnote}[1]{\kern.06em\origfootnote{#1}}
\newcommand{\punctfootnote}[1]{\kern-.20em\origfootnote{#1}}

\usepackage[capitalise]{cleveref}

\newcommand{\TuneNumModules}{95}
\newcommand{\TuneNumProjects}{67}
\newcommand{\RQCoverageSigBetterFiveZero}{20}
\newcommand{\RQCoverageSigWorseFiveZero}{3}
\newcommand{\RQEffectSizeFiveZero}{0.5570434995534814}

\newcommand{\RQCoverageSigBetterFiveOneHundred}{16}
\newcommand{\RQCoverageSigWorseFiveOneHundred}{1}
\newcommand{\RQEffectSizeFiveOneHundred}{0.5378905021173623}
\newcommand{\AnnotatedNumModules}{630}
\newcommand{\AnnotatedNumProjects}{60}

\newcommand{\AnnotatedRemovedModules}{39}

\newcommand{\AnnotatedMeanAlgorithmIterationsTypeHints}{9633.1095}
\newcommand{\AnnotatedMeanCoverageTypeHints}{70.1921}
\newcommand{\AnnotatedMeanAlgorithmIterationsTypeHintsTypeTracing}{8599.0072}
\newcommand{\AnnotatedMeanCoverageTypeHintsTypeTracing}{71.0910}
\newcommand{\AnnotatedMeanAlgorithmIterationsRandomTypes}{9805.6590}
\newcommand{\AnnotatedMeanCoverageRandomTypes}{68.8128}
\newcommand{\AnnotatedMeanAlgorithmIterationsRandomTypesTypeTracing}{8727.5258}
\newcommand{\AnnotatedMeanCoverageRandomTypesTypeTracing}{70.5326}

\newcommand{\AnnotatedFullCoverageModulesRandomTypes}{237}

\newcommand{\RelCoverageTypeHintsTypeTracing}{87.8192}

\newcommand{\AnnotatedCoverageSigBetterRandomTypesTypeTracingRandomTypes}{151}

\newcommand{\AnnotatedCoverageBetterTypeHintsTypeTracingRandomTypes}{208}

\newcommand{\AnnotatedCoverageWorseTypeHintsTypeTracingRandomTypes}{27}

\newcommand{\AnnotatedCoverageBetterRandomTypesTypeTracingTypeHints}{128}

\newcommand{\AnnotatedCoverageWorseRandomTypesTypeTracingTypeHints}{95}

\newcommand{\AnnotatedCoverageBetterTypeHintsTypeTracingTypeHints}{153}
\newcommand{\AnnotatedCoverageSigBetterTypeHintsTypeTracingTypeHints}{100}
\newcommand{\AnnotatedCoverageWorseTypeHintsTypeTracingTypeHints}{60}

\newcommand{\AnnotatedCoverageBetterGPTJEDI}{93}

\newcommand{\AnnotatedCoverageWorseGPTJEDI}{100}

\newcommand{\AnnotatedCoverageMeanEffSizeGPTJEDI}{0.49666232134144017}
\newcommand{\AnnotatedCoverageBetterTypeHintsTypeTracingGPT}{201}
\newcommand{\AnnotatedCoverageSigBetterTypeHintsTypeTracingGPT}{156}
\newcommand{\AnnotatedCoverageWorseTypeHintsTypeTracingGPT}{36}
\newcommand{\AnnotatedCoverageSigWorseTypeHintsTypeTracingGPT}{12}

\newcommand{\AnnotatedCoverageMeanEffSizeTypeHintsTypeTracingGPT}{0.5855608917490804}
\newcommand{\AnnotatedCoverageBetterRandomTypesTypeTracingGPT}{195}
\newcommand{\AnnotatedCoverageSigBetterRandomTypesTypeTracingGPT}{134}
\newcommand{\AnnotatedCoverageWorseRandomTypesTypeTracingGPT}{42}
\newcommand{\AnnotatedCoverageSigWorseRandomTypesTypeTracingGPT}{15}

\newcommand{\AnnotatedCoverageMeanEffSizeRandomTypesTypeTracingGPT}{0.5716413641789935}
\newcommand{\AnnotatedCoverageBetterRandomTypesTypeTracingJEDI}{187}
\newcommand{\AnnotatedCoverageSigBetterRandomTypesTypeTracingJEDI}{140}
\newcommand{\AnnotatedCoverageWorseRandomTypesTypeTracingJEDI}{47}
\newcommand{\AnnotatedCoverageSigWorseRandomTypesTypeTracingJEDI}{15}

\newcommand{\AnnotatedCoverageMeanEffSizeRandomTypesTypeTracingJEDI}{0.5683420667220334}
\newcommand{\AnnotatedCoverageBetterTypeHintsTypeTracingJEDI}{197}
\newcommand{\AnnotatedCoverageSigBetterTypeHintsTypeTracingJEDI}{153}
\newcommand{\AnnotatedCoverageWorseTypeHintsTypeTracingJEDI}{40}
\newcommand{\AnnotatedCoverageSigWorseTypeHintsTypeTracingJEDI}{13}

\newcommand{\AnnotatedCoverageMeanEffSizeTypeHintsTypeTracingJEDI}{0.5827980789173425}

\newcommand{\CovEffectSizeRandomTypesTypeTracingRandomTypes}{0.5771611876286411}
\newcommand{\CovBetterRandomTypesTypeTracingRandomTypes}{151}
\newcommand{\CovWorseRandomTypesTypeTracingRandomTypes}{10}
\newcommand{\CovSameRandomTypesTypeTracingRandomTypes}{394}

\newcommand{\CovEffectSizeTypeHintsTypeTracingTypeHints}{0.5486795131811266}
\newcommand{\CovBetterTypeHintsTypeTracingTypeHints}{100}
\newcommand{\CovWorseTypeHintsTypeTracingTypeHints}{15}
\newcommand{\CovSameTypeHintsTypeTracingTypeHints}{417}

\newcommand{\ItEffectSizeRandomTypesTypeTracingRandomTypes}{0.33615271918581785}
\newcommand{\ItBetterRandomTypesTypeTracingRandomTypes}{52}
\newcommand{\ItWorseRandomTypesTypeTracingRandomTypes}{312}
\newcommand{\ItSameRandomTypesTypeTracingRandomTypes}{116}

\newcommand{\ItEffectSizeTypeHintsTypeTracingTypeHints}{0.3437758390346125}
\newcommand{\ItBetterTypeHintsTypeTracingTypeHints}{40}
\newcommand{\ItWorseTypeHintsTypeTracingTypeHints}{303}
\newcommand{\ItSameTypeHintsTypeTracingTypeHints}{119}

\newcommand{\AnnotatedMutationNumModules}{475}

\newcommand{\AnnotatedMutationScoreMeanTypeHints}{14.8498}

\newcommand{\AnnotatedMutationScoreMeanTypeHintsTypeTracing}{15.3118}

\newcommand{\AnnotatedMutationScoreMeanRandomTypes}{14.7975}

\newcommand{\AnnotatedMutationScoreMeanRandomTypesTypeTracing}{15.4797}

\newcommand{\AnnotatedMutationScoreMeanGPT}{14.8510}

\newcommand{\AnnotatedMutationScoreMeanJEDI}{15.0874}

\newcommand{\AnnotatedMutationBetterTypeHintsRandomTypes}{125}
\newcommand{\AnnotatedMutationSigBetterTypeHintsRandomTypes}{35}
\newcommand{\AnnotatedMutationWorseTypeHintsRandomTypes}{98}
\newcommand{\AnnotatedMutationSigWorseTypeHintsRandomTypes}{17}

\newcommand{\AnnotatedMutationMeanEffSizeRandomTypesTypeTracingTypeHints}{0.5007489158425823}

\newcommand{\AnnotatedMutationMeanEffSizeGPTJEDI}{0.4961942133897732}
\newcommand{\AnnotatedMutationBetterTypeHintsTypeTracingGPT}{146}
\newcommand{\AnnotatedMutationSigBetterTypeHintsTypeTracingGPT}{46}
\newcommand{\AnnotatedMutationWorseTypeHintsTypeTracingGPT}{86}
\newcommand{\AnnotatedMutationSigWorseTypeHintsTypeTracingGPT}{13}

\newcommand{\AnnotatedMutationMeanEffSizeTypeHintsTypeTracingGPT}{0.5237666695241951}
\newcommand{\AnnotatedMutationBetterRandomTypesTypeTracingGPT}{140}
\newcommand{\AnnotatedMutationSigBetterRandomTypesTypeTracingGPT}{34}
\newcommand{\AnnotatedMutationWorseRandomTypesTypeTracingGPT}{87}

\newcommand{\AnnotatedMutationMeanEffSizeRandomTypesTypeTracingGPT}{0.5195102075464316}
\newcommand{\AnnotatedMutationBetterRandomTypesTypeTracingJEDI}{129}
\newcommand{\AnnotatedMutationSigBetterRandomTypesTypeTracingJEDI}{35}
\newcommand{\AnnotatedMutationWorseRandomTypesTypeTracingJEDI}{95}

\newcommand{\AnnotatedMutationMeanEffSizeRandomTypesTypeTracingJEDI}{0.5162063526905306}
\newcommand{\AnnotatedMutationBetterTypeHintsTypeTracingJEDI}{144}
\newcommand{\AnnotatedMutationSigBetterTypeHintsTypeTracingJEDI}{40}
\newcommand{\AnnotatedMutationWorseTypeHintsTypeTracingJEDI}{91}
\newcommand{\AnnotatedMutationSigWorseTypeHintsTypeTracingJEDI}{13}

\newcommand{\AnnotatedMutationMeanEffSizeTypeHintsTypeTracingJEDI}{0.5217693513899941}

\newcommand{\RandomTypesTypeTracingMatchCoverage}{\qty{89.1}{\percent}}
\newcommand{\RandomTypesTypeTracingMatchCount}{552}

\newcommand{\RandomTypesTypeTracingMismatchCount}{413}

\newcommand{\TypeHintsMatchCoverage}{\qty{79.1}{\percent}}
\newcommand{\TypeHintsMatchCount}{1361}

\newcommand{\TypeHintsMismatchCount}{52}

\newcommand{\TypeHintsTypeTracingMatchCoverage}{\qty{79.6}{\percent}}
\newcommand{\TypeHintsTypeTracingMatchCount}{1354}

\newcommand{\TypeHintsTypeTracingMismatchCount}{121}

\newcommand{\AnnotatedCorrelationCoverageType}{0.5235}
\newcommand{\AnnotatedCorrelationCoverageTypeP}{4.476177462161305e-27}

\newcommand{\JediPrimitiveRetentionPercentage}{35.8595}
\newcommand{\TypeTracingPrimitiveRetentionPercentage}{56.1748}

\newcommand{\RandomTypesTypeTracingAbstractMatchRate}{\qty{34.3}{\percent}}
\newcommand{\RandomTypesTypeTracingCollectionMatchRate}{\qty{36.9}{\percent}}
\newcommand{\RandomTypesTypeTracingNoneNoReturnMatchRate}{\qty{32.2}{\percent}}
\newcommand{\RandomTypesTypeTracingPrimitiveMatchRate}{\qty{30.1}{\percent}}
\newcommand{\RandomTypesTypeTracingProjectSpecificMatchRate}{\qty{16.9}{\percent}}

\newcommand{\GptminiPrimitiveMatchRate}{\qty{90.5}{\percent}}
\newcommand{\GptminiProjectSpecificMatchRate}{\qty{76.6}{\percent}}

\newcommand{\JediPrimitiveMatchRate}{\qty{56.9}{\percent}}
\newcommand{\JediProjectSpecificMatchRate}{\qty{20.4}{\percent}}

\newcommand{\TTEffectHintsProjectSpecific}{\qty{0.6}{\percent}}
\newcommand{\TypeshedPynMoreAnnotations}{2493}
\newcommand{\TypeshedTotal}{5610}

\newcommand{\TypeshedOnlyPynguinCorrect}{499}

\newcommand{\TypeResolutionGptRateAfter}{78.0000}

\newcommand{\TypeResolutionJediRateAfter}{59.1000}

\newcommand{\TypeInfParamgptFiveFourminiFOneScore}{0.8713083397092578}

\newcommand{\TypeInfParamhityperTotalGT}{3424}
\newcommand{\TypeInfParamhityperTotalInferred}{372}
\newcommand{\TypeInfParamhityperMatches}{190}

\newcommand{\TypeInfParamhityperPrecision}{0.510752688172043}
\newcommand{\TypeInfParamhityperRecall}{0.05549065420560748}

\newcommand{\TypeInfParamjediPrecision}{0.5828009828009828}

\newcommand{\TypeInfParamjediFOneScore}{0.43451181535079686}

\newcommand{\TypeInfParampynguinPrecision}{0.20398970398970398}
\newcommand{\TypeInfParampynguinRecall}{0.18516355140186916}

\newcommand{\TypeInfReturngptFiveFourminiFOneScore}{0.9031808683538426}

\newcommand{\TypeInfReturnhityperTotalGT}{2186}
\newcommand{\TypeInfReturnhityperTotalInferred}{364}

\newcommand{\TypeInfReturnhityperRecall}{0.10933211344922233}

\newcommand{\TypeInfReturnjediPrecision}{0.16361021215390148}

\newcommand{\TypeInfReturnjediFOneScore}{0.2348993288590604}

\newcommand{\TypeInfReturnpynguinPrecision}{0.35717155484130175}
\newcommand{\TypeInfReturnpynguinRecall}{0.4066788655077768}

\newcommand{\TypeInfReturnpynguinFOneScore}{0.38032085561497325}

\usepackage{tcolorbox}
\usepackage{xspace}
\usepackage{todonotes}

\newcommand{\toolname}[1]{%
  \textsc{#1}\xspace%
}

\newcommand{\pynguin}{%
  \toolname{Pynguin}%
}
\newcommand{\Pynguin}{%
  \toolname{Pynguin}%
}
\newcommand{\mypy}{%
  \toolname{Mypy}%
}

\newcommand{\jedi}{%
  \toolname{Jedi}%
}

\newcommand{\dseval}{%
  \textsc{DS2-Evaluation}\xspace%
}

\newcommand{\dstuning}{%
  \textsc{DS1-Tuning}\xspace%
}

\newcommand{\typetracing}{%
  type tracing\xspace%
}

\newcommand{\TypeTracing}{%
  Type Tracing\xspace%
}

\newcommand{\Typetracing}{%
  Type tracing\xspace%
}

\newcommand{\RandomTypes}{%
  \emph{RandomTypes}\xspace%
}

\newcommand{\RandomTypesTypeTracing}{%
  \emph{RandomTypes-TypeTracing}\xspace%
}

\newcommand{\TypeHints}{%
  \emph{TypeHints}\xspace%
}

\newcommand{\TypeHintsTypeTracing}{%
  \emph{TypeHints-TypeTracing}\xspace%
}

\newcommand{\JediTypeHints}{%
  \emph{Jedi}\xspace%
}

\newcommand{\GptTypeHints}{%
  \emph{GPT}\xspace%
}

\newcommand{\codamosa}{%
  \toolname{CodaMosa}\xspace%
}

\newcommand{\match}{%
  MATCH\xspace%
}

\newcommand{\mismatch}{%
  MISMATCH\xspace%
}

\newcommand{\missing}{%
  MISSING\xspace%
}

\newcommand{\any}{%
  ANY\xspace%
}

\newcommand{\all}{%
  ALL\xspace%
}

\newcommand{\gpt}{%
  \toolname{gpt-5.4-mini}%
}
\newcommand{\Gpt}{%
  \toolname{Gpt-5.4-mini}%
}

\newcommand{\hityper}{%
  \toolname{HiTyper}%
}

\newcommand{\guesswhat}{%
  \toolname{Guess What}%
}

\newcommand{\quac}{%
  \toolname{QuAC}%
}

\newcommand{\Fone}{\ensuremath{\text{F}_1}\xspace}

\newcommand\effectsize{\ensuremath{\hat{A}_{12}}\xspace}

\newcommand*{\eg}{e.g.\@\xspace}
\newcommand*{\ie}{i.e.\@\xspace}

\newcommand{\exnum}[1]{\num[round-mode=none]{#1}}

\NewDocumentEnvironment{summary}{m}{%
  \begin{tcolorbox}[title={Summary~(#1)}]%
  }{%
  \end{tcolorbox}%
}

\begin{document}

\title{\TypeTracing: Using Runtime Information to Improve Automated
Unit-test Generation for Python}
\title{\TypeTracing}
\title{Tracing Types from Executions to\\ Augment Test Generation for Python}
\title{Augment Test Generation for Python\\ with Types from Execution Traces}
\title{Augment Test Generation for Python\\ with Traced and Predicted Types}
\title{Combined Type Inference and Test Generation for Python}
\title[%
  Combining Type Inference and Automated Unit Test Generation for Python%
]{Combining Type Inference and Automated Unit Test Generation for Python}

\author{Lukas Krodinger}
\email{lukas.krodinger@uni-passau.de}
\orcid{0009-0005-4571-4757}
\affiliation{%
  \institution{University of Passau}
  \city{Passau}
  \country{Germany}
}

\author{Stephan Lukasczyk}
\email{stephan@pynguin.eu}
\orcid{0000-0002-0092-3476}
\affiliation{%
  \institution{JetBrains Research}
  \city{Munich}
  \country{Germany}
}
\authornote{Research partially done while with the University of Passau}

\author{Gordon Fraser}
\email{gordon.fraser@uni-passau.de}
\orcid{0000-0002-4364-6595}
\affiliation{%
  \institution{University of Passau}
  \city{Passau}
  \country{Germany}
}



\begin{abstract}
  Automated unit test generation is an established research field
  that has so far
  focused on statically-typed programming languages.
  The lack of type information in dynamically-typed programming
  languages, such as
  Python, inhibits test generators, which heavily rely on information about
  parameter and return types of functions to select suitable arguments when
  constructing test cases.
  Since automated test generators inherently rely on frequent
  execution of candidate tests, we make use of these frequent
  executions to address this problem by introducing \emph{type
  tracing}, which extracts type-related information during execution
  and gradually refines the available type information.
  We implement \typetracing as an extension of the \pynguin
  test-generation framework for Python, allowing it (i) to infer parameter
  types by observing how parameters are used during runtime, (ii) to
  record the types of values that function calls return, and (iii) to
  use this type information to increase code coverage.
  The approach leads to up to
  \qty{\RelCoverageTypeHintsTypeTracing}{\percent} more branch coverage,
  improved mutation scores, and to type information of similar quality
  to that produced by other state-of-the-art type-inference tools.
\end{abstract}


\maketitle

\section{Introduction}\label{sec:introduction}

Testing is an important aspect of software development, yet it is often seen as
a tedious
chore~\cite{waychal_practitioners_2021,straubinger_engaging_2024,santos_would_2017,deak_challenges_2016,weyuker_clearing_2000}.
To reduce the burden on developers, automated unit-test
generation tools have been developed, e.g.,
\toolname{EvoSuite}~\cite{FA13} for Java or \pynguin~\cite{LF22} for Python.
These tools retrieve a \emph{subject under test}~(SUT), \eg, a Java class or
a Python module, and apply a meta-heuristic search algorithm, query
\emph{large language models}~(LLMs), or a combination of these approaches,
to generate test cases that invoke the SUT's routines, \ie, functions,
methods, and constructors. The goal is to maximise a given metric,
such as branch coverage.

Statically-typed programming languages, like Java, allow constructing inputs to
methods by inspecting the static type of each parameter. In dynamically-typed
programming languages such as Python, parameters do not have a static type but
only their values at runtime have a type. Not having to specify types allows
for fast prototyping~\cite{GPS+15}, but poses a significant challenge to
automated test generators, which no longer know what inputs to construct.

\begin{figure}[!t]
  \begin{subfigure}{0.95\linewidth}
\begin{lstlisting}[style=nonnumberedlst,frame=single]
def visit_ClassDef(self, n):
    if not n.bases:
        n.bases = [ast.Name(id='object')]
        self._tree_changed = True
    return self.generic_visit(n)
\end{lstlisting}%
    \Description{A Python method definition where function arguments
    and return types are defined without any type annotations.}
    \caption{\label{lst:intro-not-annotated}without type hints}
  \end{subfigure}
  \begin{subfigure}{0.95\linewidth}
\begin{lstlisting}[style=nonnumberedlst,frame=single]
def visit_ClassDef(self, n: ast.ClassDef) -> ast.ClassDef:
    if not n.bases:
        n.bases = [ast.Name(id='object')]
        self._tree_changed = True
    return self.generic_visit(n)
\end{lstlisting}%
    \Description{The same Python method definition as above, but now
      including a colon-separated type hint for the argument and an
    arrow notation for the return type.}
    \caption{\label{lst:intro-annotated}with type hints}
  \end{subfigure}
  \caption{\label{lst:intro-example}%
    Simplified excerpt from
    \inlinelst{py-backwards.transformers.class_without_bases}.
  }
\end{figure}

Recently, Python developers have started to alleviate the problem of
missing type information by using type hints~(also called \emph{type
annotations}). The example Python SUT shown in
\cref{lst:intro-example} consists of two versions of a simple
function, one with and one without type hints. Type hints are not only
beneficial for type checking (e.g., statically checking whether
functions are called with arguments of valid types) but also because
they provide test generators with crucial
information~\cite{LKF23}. The version with type
hints~(\cref{lst:intro-annotated}) allows a test generator for Python
to inspect the parameter \inlinelst{node} and retrieve the information
that an instance of \inlinelst{ast.ClassDef} is required to call the
function \inlinelst{visit_ClassDef}. Unfortunately, type
hints are rare in Python projects: A recent large-scale
study~\cite{GP22} of \num[round-precision=4]{9655} Python projects
revealed that only \qty{7}{\percent} of these projects use
type hints, and even when they do, only about \qty{8}{\percent} of the
parameter and return types are annotated. Furthermore, even when annotated,
the provided type hints may not be correct~\cite{RMM20}.

To automatically infer missing type annotations, various static
type inference systems have been proposed~\cite{LWQ22,VWL+23,wu_quac_2024}.
However, Python's dynamic nature makes it a particularly
difficult target for static analysis:
variables carry no declared types,
duck typing means that programs rely on structural attribute access
rather than declared interfaces,
and features such as late binding and metaprogramming
allow programs to alter their structure at runtime.
As a consequence, static tools are forced to make conservative assumptions,
leading to
over-generalisation
or imprecise predictions.
To side-step these limitations, approaches based on
machine-learning models~\cite{HBB+18,PGL+20,MLP+22}
and LLMs~\cite{PWW+23, yan_dlinfer_2023,
yang_llm-enhanced_2025, wei_typet5_2023}
have been proposed for type inference.
However, open questions remain regarding their
performance on data which is not in the training set, potential
overfitting due to data leakage~\cite{dong_unmasking_2026}, and the
financial costs involved in their application.
Dynamic recording tools like
\toolname{MonkeyType}\footnote{\url{https://monkeytype.readthedocs.io/en/latest/},
last accessed 2026-04-30.} or
\toolname{RightTyper}~\cite{pizzorno_2025_righttyper}
overcome the limitations of both%
---static and machine-learning approaches---%
by observing the SUT during runtime to collect type
information.
However, they rely on existing ways to execute the SUT
(\eg, manually written tests),
which is what an automated test generator is intended to
create in the first place.
Hybrid approaches that integrate type inference into test-case generation,
such as \guesswhat~\cite{SOP22} for JavaScript, address this dependency but
rely on an underlying static model (e.g., a probabilistic model built from
AST relations), and thus still suffer from the same limitations as static
type inference when applied to dynamic languages.
Overall, all existing approaches have limitations.

In this paper, we introduce \typetracing, a novel dynamic approach that uses
runtime information to infer missing type annotations \emph{during}
test generation.
We divide the problem of inferring missing type annotations into two
sub-problems:
finding missing
(1) parameter and (2) return types. One can approximate the latter by simply
recording the types of the values a routine returns. However, this requires that
the routine can be executed successfully to yield such a return
value. This reduces the problem of finding missing return types to the
problem of finding missing parameter types. To address this problem, our
approach traces the usage of parameters by wrapping their arguments in a
transparent proxy. The proxy records how the SUT's code interacts with the
argument object, \eg, which attributes or methods of the argument object get
accessed, thus gradually narrowing down possible types.

Consider again the function \inlinelst{visit_ClassDef} in
\cref{lst:intro-not-annotated}. Initially, there is no information
about the type of the parameter \inlinelst{n}. When a test generator
adds a statement calling this function to a test case, it can only
guess an arbitrary type and value for the formal parameter
\inlinelst{n}.  Assume it picks the value \inlinelst{42} of type
\inlinelst{int}. The test generator then executes the call
\inlinelst{visit_ClassDef(42)} to record the coverage achieved by the
test case containing this call. Furthermore, it wraps the argument
\inlinelst{42}---which is an object in Python---with a proxy. The proxy
records that there is an access to the attribute \inlinelst{bases} of
the parameter; the execution fails here because the \inlinelst{int}
type does not provide such an attribute. The failed execution prevents
us from recording which type the return value has because the
\inlinelst{return} statement is never executed.  However, when the
test generator constructs an input for \inlinelst{visit_ClassDef} the
next time it can use the information from the proxy, and pick a type
that contains an attribute \inlinelst{bases}.  In the scope of the
\inlinelst{py-backwards} project, from which the example is taken,
only the class \inlinelst{ClassDef} has this attribute and is thus
chosen and instantiated. This would lead to a successful invocation
of \mbox{\inlinelst{visit_ClassDef}},
i.e., no type errors or input validation checks fail,
which allows the execution to reach a return statement
and reveal the return type of the
function. Consequently, the approach has (1)~elicited precise type
information on the parameter and return types of \inlinelst{visit_ClassDef},
while at the same time (2)~guiding the test generator in achieving higher
code coverage on the SUT \inlinelst{visit_ClassDef}.

In detail, this paper makes the following contributions:
\begin{itemize}
  \item We introduce \typetracing, an approach that combines automated test
    generation with dynamic type inference.
  \item We implement \typetracing on top of \pynguin~\cite{LF22}, a
    well-established automated test generator for Python.
  \item To balance the gain of type information against the execution
    overhead caused by \typetracing, we make the use of \typetracing in
    \pynguin probabilistic, and tune the probability of doing a proxied
    execution.
  \item We evaluate our implementation on \num{\AnnotatedNumModules} modules
    taken from \num{\AnnotatedNumProjects} open-source Python projects.  We
    compare the types predicted by our \typetracing with several
    state-of-the-art tools on the \toolname{TypeEvalPy}~\cite{VSW+23} benchmark
    set.
  \item Lastly, we study the influence of \typetracing on the achieved branch
    coverage and mutation scores.
\end{itemize}
Our evaluation shows that our \typetracing approach allows \pynguin to generate
tests that yield up to
\qty{\RelCoverageTypeHintsTypeTracing}{\percent} more branch coverage.  It
has a beneficial influence on the achieved mutation scores and the types it
infers are of comparable quality to other state-of-the-art type-inference
approaches.

\section{Background}\label{sec:background}

\subsection{Typing in Python}\label{sec:background:typing}

In general, a type is a set of values with common operations,
\eg, the type \inlinelst{int} comprises the set of all possible integer
values\footnote{An \texttt{int} in Python is unbounded, contrary to
languages like Java.} along with their common operations, such as addition and
multiplication.%

The Python programming language makes use of several typing disciplines: first
and foremost, Python uses dynamic typing, \ie, variables do not have an
associated type but only their values at runtime do. Second, Python uses
strong typing. This means the language does not implicitly coerce values from
one type to another. Consider the expression \mbox{\inlinelst{'12' + 1};} it
raises a type error in Python because addition between strings and integers is
not defined. In contrast, a weakly-typed language, \eg, JavaScript, would
coerce \inlinelst{1} into a string and afterwards concatenate the two strings
to \mbox{\inlinelst{'121'}.} Third, Python also uses the concept of duck
typing~\cite{MGN17}, often summarised as \enquote{If it looks like a duck and
quacks like a duck, it must be a duck}. Instead of directly checking for the
type of an object, it rather tests for the existence of expected attributes or
methods on that object. Programmers often do this indirectly, by simply
accessing the expected member of the object, which either succeeds or raises
a type-related error.

While dynamic typing and duck typing allow for rapid prototyping~\cite{GPS+15},
the downside is that type-related errors happen only at runtime,
which may---in the worst case---be the production environment. The lack of
static type information can make maintenance of large codebases
harder~\cite{KHR12} because static type checking is not possible and the
support provided by an IDE might therefore be limited.

The designers of the Python language realised this problem and started to
introduce gradual typing in Python~3.5. In gradual typing, some variables can
have a static type assigned, while others may stay untyped~\cite{ST07}. This
allows developers to statically type check parts of a codebase and
can be seen as a
transition towards a more static type system in Python. The premise that some
variables can be left untyped is important because it allows developers to
gradually specify variable types, which makes it easier to adopt this practice
in existing codebases, \eg, during necessary code changes. Furthermore,
gradual typing enables seamless interactions between typed and untyped
codebases, \eg, code from third-party vendors, where the domain knowledge
to specify all types may be absent.

While gradual typing allows some variables to remain untyped, the
portions of the code that are annotated with types can be checked for
correctness using type checking. A type checker verifies type
correctness by determining whether a value of type~\(A\) can be
assigned to a variable of type~\(B\) using subtype relationships.
Python's type system has two ways to convey such a subtype relation to
a static type checker: first, \emph{nominal subtyping} uses the class
hierarchy. The class definitions \inlinelst{class B()} and
\inlinelst{class A(B)} define that class \inlinelst{A} is a subtype of
class \inlinelst{B}.
Second, \emph{structural subtyping} in Python is based on explicitly
declared \inlinelst{Protocol}s, which define a set of methods and
attributes that a class must implement to be considered a
subtype---regardless of the actual inheritance.  Consider a
\inlinelst{class Flyer(Protocol)} with a method \inlinelst{fly(self)}
and another \inlinelst{class Bird()} explicitly \textbf{not}
inheriting from \inlinelst{Flyer}. If \inlinelst{Bird} also defines a
method \inlinelst{fly(self)}, a static Python type checker accepts
\inlinelst{Bird} as an argument to the function
\inlinelst{lift_off(thing: Flyer)}.  This works because structural
subtyping in Python is \emph{implicit}: it relies solely on whether
the class provides the required methods and attributes defined by the
explicitly declared \inlinelst{Protocol}---not on inheritance.

\subsection{Type Checking and Type Inference}%
\label{sec:background:typeinference}

With the addition of type hints to the Python language, the
necessity for checking those type hints arose.
\toolname{MyPy},\punctfootnote{\url{https://mypy-lang.org/}, last accessed
2026-04-30.} was introduced and serves as the reference type
checker for the Python language developers. Further type-checkers include
among others,
\toolname{PyType},\punctfootnote{\url{https://google.github.io/pytype},
  last accessed
2026-04-30.}
\toolname{PyRight},\punctfootnote{\url{https://microsoft.github.io/pyright},
  last
accessed 2026-04-30.}
\toolname{Pyre},\punctfootnote{\url{https://pyre-check.org}, last accessed
2026-04-30.}
\toolname{Pyrefly},\punctfootnote{\url{https://pyrefly.org/}, last accessed
2026-04-30.} and
\toolname{Ty}.\punctfootnote{\url{https://github.com/astral-sh/ty},
  last accessed
2026-04-30.}
A type checker verifies that a program adheres to rules introduced by a
type system.  A \emph{type system} is a tractable syntactic method
which allows one to
prove the absence of certain program behaviours by classifying statements in a
programming language according to the kinds of values they
compute~\cite{Pie02}.  The
fundamental purpose of such a type system is to prevent the occurrence of
errors during program execution~\cite{Car04}, which avoids many common
bugs early in the development phase of a program.

While the addition of type checkers allows for static type checking of annotated
code, most of the existing Python codebases still do not provide type
annotations~\cite{GP22}. Adding those type annotations manually is
tedious and error-prone, thus automated tool support for type
inference is desirable.
To meet this need, several static tools, such as
\toolname{Jedi},\footnote{\url{https://jedi.readthedocs.io/en/latest},
last accessed 2026-04-30.} %
\toolname{Pyright},\footnote{\url{https://github.com/microsoft/pyright},
last accessed 2026-04-30.} %
\toolname{Scalpel}~\cite{LWQ22}, %
\toolname{HeaderGen}~\cite{VWL+23}, %
and \toolname{QuAC}~\cite{wu_quac_2024} %
have been developed.
These static tools work by parsing the program's abstract syntax
tree, performing control‐flow and data‐flow analyses to propagate
type information across definitions and uses, and then generating
type annotations based on inferred variable and function signatures.

However, static approaches face general and Python-specific limitations.
There are well-known general limitations of static analysis, such as the
undecidable aliasing problem~\cite{ramalingam1994undecidability,PY07}, which
forces static analysis tools to over-approximate; similarly, the path-explosion
problem~\cite{boonstoppel_rwset_2008} forces a trade-off between
analysis precision and computational scalability.
Python's dynamic nature adds further challenges: variables
carry no declared types, duck typing causes programs to rely on
structural attribute access rather than declared interfaces, and
features such as late binding and metaprogramming allow programs to
alter their structure at runtime.
As a consequence, static tools are forced to make conservative
assumptions, leading to over-generalisation or imprecise predictions.
To circumvent these limitations, dynamic type inference tools such as
\toolname{MonkeyType}\footnote{\url{https://monkeytype.readthedocs.io/en/latest/},
last accessed 2026-04-30.} or
\toolname{RightTyper}~\cite{pizzorno_2025_righttyper} have been developed.
They record execution traces of programs and extract type hints
from these executions. They use
tracing to observe the object types of parameters but
they rely on existing test cases to explore the program code.

To side-step the limitations of both static and dynamic
approaches, research has investigated predictive models for type
inference, leveraging machine-learning techniques. They range from
simple probabilistic models to predict identifier names and type
annotations for JavaScript~\cite{RVK15} to complex deep neural network
models~\cite{HBB+18,MPP19,PGL+20,PGL22,MLP+22}. Recently,
generative LLMs became available and are being
explored for their type-inference
capabilities~\cite{PWW+23, yan_dlinfer_2023, wei_typet5_2023}.

The earlier machine learning models are usually trained on large
corpora of projects hosted on GitHub. For JavaScript, it is possible
to generate labelled training data from TypeScript source code because
TypeScript is a superset of JavaScript with the possibility to add
type information; furthermore, the TypeScript transpiler checks the
correctness of the provided type information. For Python, in contrast,
the only source of labelled data is manually annotated code. However,
acquiring high-quality annotated code for training is challenging: A
recently conducted large empirical study~\cite{GP22} on
\num[round-precision=4]{9655} Python projects revealed that only about
\qty{7}{\percent} of these projects make use of type annotations.
Of the projects that do, only around \qty{8}{\percent} of
the parameter and return types are annotated. Consequently,
predictive approaches
may suffer from weak models in Python. Furthermore, predictive models
are by construction limited to predicting types contained in the training
data, and cannot predict project-specific types without further
fine-tuning.  Because the type annotations in the training data could
be incorrect in the first place~\cite{RMM20}, the prediction may also
be wrong.

Similar limitations extend to LLMs applied to type inference, where
their performance is often undermined by data leakage and a
limited generalisability to code they have not seen during
training~\cite{wei_typet5_2023, dong_unmasking_2026}.
LLMs' high accuracy on standard benchmarks is inflated because
benchmark datasets have been publicly available for years and are
most likely used during training of LLMs.
When evaluated on unseen code,
the performance of LLMs experiences decreases
by up to \qty{59}{\percent} in precision and up to
\qty{72}{\percent} in recall~\cite{dong_unmasking_2026}.
Furthermore, the accuracy of LLMs decreases when inferring less
commonly used types,
which limits their practical utility for project-specific or rare
types~\cite{wei_typet5_2023}.

\subsection{Unit Test Generation for Python}%
\label{sec:background:testgeneration}

Strong test suites are important, but creating tests manually can be
tedious~\cite{waychal_practitioners_2021,straubinger_engaging_2024,santos_would_2017,deak_challenges_2016,weyuker_clearing_2000}.
A semi-automated approach to generating tests is therefore
offered by property-based testing tools such as
\toolname{Hypothesis}~\cite{MH19,MD20}, which generates a wide range
of input values using composable strategies that define how to create
data of various types. These strategies are passed to a test function
via a \inlinelst{@given} decorator, which drives automatic input
generation. However, \toolname{Hypothesis} requires the user to manually
provide a basic test case that defines the expected behaviour of the
SUT~\cite{MH19,MD20}.

Common approaches to fully automate test generation include purely
random approaches, some of which incorporate feedback from the
test-case execution to guide the generation~\cite{PLE+07}; (dynamic)
symbolic execution, which systematically explores program paths to
generate test cases~\cite{SMC13, bucur_prototyping_2014,
ding_dynamic_2016, ryan_code_aware_2024}; and evolutionary algorithms,
which aim to find test cases that are considered best with respect to
a fitness function~\cite{Ton04}. Fitness functions can be based on,
\eg, coverage metrics such as branch coverage. Furthermore, previous
work explored the usage of generative models~\cite{DDS21} and the
usage of LLMs for Python test
generation~\cite{dakhel_effective_2024, yang_enhancing_2024,
  pizzorno_coverup_2024, lemieux_codamosa_2023, ryan_code_aware_2024,
yang_llm-enhanced_2025, xiao_optimizing_2024}.  We refer the
interested reader to the literature that discusses automated unit test
generation and various algorithms in great detail~\cite{CGA18}.

\Pynguin~\cite{LF22} is a well-established, actively developed test generator
for the Python programming language, implementing various
test-generation algorithms~\cite{LKF23}. Many novel LLM-based approaches,
such as \codamosa~\cite{lemieux_codamosa_2023}, extend
\Pynguin by incorporating LLM-guided test generation. These
state-of-the-art approaches have since been integrated into \Pynguin itself,
making it a competitive baseline that incorporates the latest advances.
Evaluating \typetracing in combination with LLM-based
approaches would introduce confounding factors, as LLMs
might ``guess'' correct types based on semantic information
(e.g., parameter names), which would disguise the
effectiveness of our dynamic type-inference technique. We therefore
focus our evaluation on the evolutionary search of \pynguin to
isolate the impact of \typetracing.

\Pynguin aims to generate test cases that yield high coverage
values. To achieve this, it analyses the SUT and collects all public
classes, functions, and methods and stores all their lines or
branches~(depending on the configuration) as targets for the test
generation. Upon creating test cases for a focal method that has at least
one parameter, \pynguin needs to generate suitable arguments to call
the method with.  In case type information is available, \pynguin uses
this information by attempting to generate an object of the type
required by the parameter's type annotation.  Frequently, however,
Python projects do not provide type hints~\cite{GP22}. In this case
\pynguin nevertheless has to choose a concrete type to generate a
value or object as input for a target, but it can only randomly choose
from the set of all available types in the project under test.  For
example, consider the method \inlinelst{visit_ClassDef} in
\cref{lst:intro-example}: Assume \pynguin picks the value
\inlinelst{42} of type \inlinelst{int} and executes the call
\inlinelst{visit_ClassDef(42)} without \typetracing. The execution
fails with an \inlinelst{AttributeError}, because \inlinelst{42} does not
provide the attribute \inlinelst{bases}.
As no further lines are executed, the achieved code coverage is low.
\Pynguin would next
select another type and value, say \inlinelst{'abc'} of type
\inlinelst{string} and execute \inlinelst{visit_ClassDef('abc')},
which also fails and does not increase coverage.
\Pynguin would continue to randomly select types
and values until it finds a value that increases coverage.
While doing so it can choose from all types that are built into the
language, but also from all types that are defined or (recursively)
imported in the module under test. This may lead to a high number of
types, which in turn may make the random selection inefficient and
costly.  In addition, previous work on \pynguin has also shown that
the availability of type information in Python modules can have a
significant positive influence on the coverage of the resulting test
suites~\cite{LKF20,LKF23}.  This, combined with the inefficiency
regarding the type handling of \pynguin, motivates our \typetracing
approach.

\section{\TypeTracing}\label{sec:approach}

The core concept of \typetracing is to leverage the iterative nature of
search-based test generation to dynamically discover and refine type
information. Instead of relying solely on static analysis or one-shot
predictions, \typetracing observes the actual runtime behaviour of a program
to infer the types required for its parameters and return values.
\Cref{fig:workflow} provides a high-level conceptual overview of this
process.
The steps are:

\begin{enumerate}
  \item \textbf{Static Analysis}: Before the search starts, static analysis
    identifies available types and their attributes and stores them
    (among other information, such as test targets) in the \emph{test
    cluster} (\cref{sec:approach:collect:classes}).
  \item \textbf{Test Generator}: The \pynguin test case generator
    creates test cases,
    using type information from the \emph{test cluster} to generate
    test input arguments of suitable types
    (\cref{sec:background:testgeneration}).
  \item \textbf{Execution}: \Pynguin executes the test cases.
    Regular and proxied executions are performed to gather type information
    on return types and on parameter usage (\cref{sec:approach:tracing}).
  \item \textbf{Tracing}: During proxied
    executions, \typetracing records parameter usage through \emph{proxies}
    (\cref{sec:approach:collect:proxy}) and \emph{shims}
    (\cref{sec:approach:collect:checks}).
    The system also records the types of
    returned values (\cref{sec:approach:tracing:return}).
  \item \textbf{Type Inference}: The usage traces and return types
    are used to infer potential types,
    which are then stored back in the test cluster to guide
    subsequent iterations
    (\cref{sec:approach:selection}).
\end{enumerate}

\begin{figure}[t]
  \centering
  \begin{tikzpicture}[
      node distance=1cm and 1.0cm,
      block/.style={rectangle, draw, fill=blue!10, text width=2.6cm,
      align=center, minimum height=0.8cm, rounded corners},
      subblock/.style={rectangle, draw, fill=white, text width=2.4cm,
      align=center, minimum height=0.6cm, rounded corners},
      arrow/.style={-Stealth, thick},
      database/.style={
        cylinder, draw, shape border rotate=90, aspect=0.25,
        fill=orange!10, text width=2.0cm, align=center, minimum
        height=1.1cm
      },
      input/.style={
        trapezium, draw, trapezium left angle=70, trapezium right angle=110,
        fill=gray!10, text width=2.0cm, align=center, minimum height=0.8cm
      }
    ]
    \node [block] (generator) {(2) Test Generator};

    \node [block, right=of generator, fill=blue!5, minimum
    height=2.2cm, minimum width=3.2cm, label=(3) Execution]
    (exec_env) {};
    \node [subblock, yshift=0.5cm] (reg_exec) at (exec_env.center)
    {Regular Execution};
    \node [subblock, yshift=-0.5cm] (prox_exec) at (exec_env.center)
    {Proxied Execution};

    \node [block, right=of exec_env, fill=green!5, minimum
    height=2.2cm, minimum width=3.2cm, label=(4) Tracing] (tracing) {};
    \node [subblock, yshift=0.5cm] (ret_trace) at (tracing.center)
    {Return Types};
    \node [subblock, yshift=-0.5cm] (usage_trace) at (tracing.center)
    {Input Usage};

    \node [block, below=of tracing] (inference) {(5) Type Inference};
    \node [database] (cluster) at (generator |- inference) {Test Cluster};
    \node [block, left=of cluster, fill=gray!10, text width=2.0cm]
    (analysis) {(1) Static Analysis};
    \node [input]
    (sut) at (analysis |- generator) {Subject under Test (SUT)};

    \coordinate (branch1) at ([xshift=0.5cm]generator.east);
    \draw [-] (generator.east) -- (branch1);
    \draw [arrow] (branch1) |- (reg_exec.west);
    \draw [arrow] (branch1) |- (prox_exec.west);

    \draw [arrow] (reg_exec) -- (ret_trace);
    \draw [arrow] (prox_exec) -- (usage_trace);

    \draw [arrow] (usage_trace.south) -- (inference.north);
    \draw [arrow] (ret_trace.east) -| ([xshift=0.4cm]tracing.east) |-
    (inference.east);

    \draw [arrow] (inference) -- node[above] {Update} (cluster);

    \draw [arrow] ([xshift=-2mm]generator.south) -- ([xshift=-2mm]cluster.north)
    node[midway, left] {Query Types};
    \draw [arrow] ([xshift=2mm]cluster.north) -- ([xshift=2mm]generator.south)
    node[midway, right] {Suitable Types};

    \draw [arrow, dashed] (analysis) -- (cluster);

    \draw [arrow, dashed] (sut) -- (analysis);

  \end{tikzpicture}
  \caption{Overview of \pynguin's test generation with \typetracing.}
  \label{fig:workflow}
\end{figure}

In the following, we describe the components of this workflow in detail,
starting with \pynguin's underlying type system.
We decided to implement \typetracing in the Python test generator
\pynguin, as we identified the lack of type handling as a
core limitation of the framework. We demonstrate the
effectiveness of \typetracing by augmenting the test generation
process with type information. However, \typetracing is neither
limited to the \pynguin framework nor to the Python
programming language; it can be incorporated into other dynamic
analysis tools, and it can be applied to other dynamically-typed programming
languages~(c.f. \cref{sec:approach:generalisation}).
We choose \pynguin because it is a widely used
automated unit test generation framework for Python. Its established
search-based algorithms provide a robust baseline for evaluating
the impact of improved type handling without the confounding
influence of LLM-based approaches, such as
\codamosa~\cite{lemieux_codamosa_2023}.
Due to the significant engineering effort required to adapt
\typetracing to other frameworks, we leave this as future work.

\subsection{Improving \pynguin's Type System}\label{sec:approach:pynguin}

During test-case construction, \pynguin must determine which values shall be
used to fill parameters of functions and methods (see
\cref{sec:background:testgeneration}) and therefore it must know
about types and their relationship.
Python provides capabilities to retrieve various type information from
an SUT's source code, \eg, one can programmatically determine nominal
subtyping relations of classes. For instance, \inlinelst{issubclass(B,
A)} is true when \inlinelst{B} inherits from \inlinelst{A} and
\inlinelst{isinstance(obj, A)} checks if \inlinelst{obj} is an
instance of class \inlinelst{A}.  However, this functionality is
limited to nominal subtyping and non-generic types.
It does not support structural subtyping, and even though, from a
type-theory perspective, \inlinelst{Sequence[bool]} is a subtype of
\inlinelst{Sequence[int]} since \inlinelst{bool} is a subtype of
\inlinelst{int}, Python has no built-in mechanism to check this.  Both
\inlinelst{issubclass(Sequence[bool], Sequence[int])} and
\inlinelst{isinstance([True, False], Sequence[int])} will raise an
exception.  \Pynguin needs to be able to reason about generic types,
such as \inlinelst{list}, as it is crucial for generating values of
the correct parameter type.  When \pynguin knows---either through
developer type hints or through \typetracing---that a parameter of
type \inlinelst{list[int]} is required, it should generate a value of
type \inlinelst{list} containing elements of type \inlinelst{int} or
\inlinelst{bool}.  This cannot be achieved by relying solely on
Python's built-in \inlinelst{issubclass} and \inlinelst{isinstance},
as they do not support generic types or structural subtyping.

Type-checking tools aim to overcome these
limitations. There exist several tools, \eg, \toolname{MyPy},
\toolname{PyRe}, \toolname{PyRight}, or \toolname{PyType}, all with
their own internal type systems which have subtle
differences~\cite{RMM20}. For our purposes, we require a type system
that is conceptually similar to those used in type checkers, but
tailored to the needs of dynamic test generation and type inference in
\pynguin. Therefore, we implemented our own type system based on
\toolname{MyPy} but focusing on the subset of features relevant
for our approach and omitting aspects unnecessary for test generation.
One unnecessary aspect is, e.g., local variable typing, as
\pynguin does not need to generate arguments for those but only for
parameters and return values of functions and methods.
While our system is conceptually identical to the type-theory
framework implemented by \toolname{MyPy}---adopting the same gradual typing
principles and consistency relationships---the implementation itself is a
technical adaptation.

A necessary and very central aspect of our type system is
supporting gradual typing (see \cref{sec:background:typing}), as
Python code is typically not fully annotated, requiring \pynguin to
handle a mixture of typed and untyped code. Instead of defining a
subtyping relationship, we define a so-called \emph{consistency}
relationship, which is a generalisation of the subtyping
relationship for gradual typing.
A consistency relationship not only exists between two types~\(S,
T\), where~\(S\) is a nominal subtype of~\(T\) but also if they fulfil a
structural subtype relation, \ie, they provide the same methods or attributes.
For two types~\(S\) and~\(T\), if \(S\) is consistent to \(T\),
\pynguin assumes that a
value of type~\(S\) can be used as a formal argument for a parameter of
type~\(T\). This allows us, \eg, to instantiate a concrete type, \ie,
a subtype of
a required abstract type, and use it as a parameter value.
A key property of the consistency relation is that it is \emph{not}
transitive, unlike the usual subtyping relation.
To illustrate this, consider the special type \(\mathtt{Any}\), which
is consistent with every type and vice versa: for any type~\(X\),
both \(X\) is consistent with \(\mathtt{Any}\) and \(\mathtt{Any}\)
is consistent with \(X\).
However, this does not mean that all types are consistent with each other.
For example, although \(\mathtt{int}\) is consistent with
\(\mathtt{Any}\) and \(\mathtt{Any}\) is consistent with
\(\mathtt{str}\), it does \emph{not} follow that \(\mathtt{int}\) is
consistent with \(\mathtt{str}\).

Having established a consistency relationship to handle gradual
typing, it is equally important to ensure that the type system treats
semantically equivalent types coherently.  Therefore, the type system
performs unification of types. For example, \inlinelst{list} and
\inlinelst{list[typing.Any]} semantically all represent the same type,
\ie, a list of arbitrary element types. Therefore, \pynguin unifies
these into a single type.  This is necessary to allow falling back for
unsupported types. In fact, the type system uses \inlinelst{Any} as a
fallback for all unsupported types. Additionally, it treats all
unspecified parameter or return types as \inlinelst{Any}, which is in
line with \toolname{MyPy}'s behaviour.

A current limitation of \pynguin is its limited support of generic
types~\cite{LKF23}, as it only supports the built-in collection types
\inlinelst{list}, \inlinelst{set}, \inlinelst{dict}, and
\inlinelst{tuple}.
This was done for reasons of technical complexity and practical
relevance: as a research prototype, \pynguin prioritises support for
the most frequently used types in real-world codebases. Supporting
arbitrary user-defined generic types would require significant
engineering while offering limited practical benefit, since such types
are relatively uncommon in typical Python projects. For example, the
official \mypy\footnote{
  \url{https://mypy.readthedocs.io/en/latest/generics.html}, last
accessed 2026-04-30.}  documentation describes user-defined generics
as an optional and advanced feature.
Consequently, \pynguin treats user-defined generic types as
non-generic, simplifying existing type annotations. For example,
\pynguin transforms \inlinelst{CustomGenericType[int]} into
\inlinelst{CustomGenericType}. This simplification makes \pynguin
assign values of arbitrary types when constructing instances of user
defined generic types, which may result in generic type instantiations
that would violate a subtype relation.
Because this is a fundamental limitation of
\pynguin itself, we decided to not add handling of generic types to
the type system, except for the aforementioned collection types.  For
\inlinelst{list}s and \inlinelst{set}s, all generated values will be
of the same type; the same holds for the values in \inlinelst{dict}s,
where the keys will be of type \inlinelst{str}.  For
\inlinelst{tuple}s the types of all values can be mutually different.

\subsection{Collecting Information on Parameter and Return Types}%
\label{sec:approach:collect}

To implement \typetracing we need to collect information on how routines
interact with their arguments, as well as which types the routines return.
Additionally, we have to know which classes exist and which attributes they
provide to be able to infer the types of parameters from that information.
Ultimately, this relates the available types into consistency relationships.

\subsubsection{Object Proxy and Usage Trace}\label{sec:approach:collect:proxy}

Consider the example function depicted in \cref{fig:motivating-the-proxy}. When
executing the function \inlinelst{test_me}, we would like to record that there
is an access to the attribute \inlinelst{bar} of the function's formal
parameter \inlinelst{foo}. Furthermore, we want to retrieve that there is a
comparison of the value of \inlinelst{foo.bar} and the integer literal
\inlinelst{42}. Note that, while a simple static analysis could potentially
resolve this particular example, the limitations of static analysis,
in particular in the case of Python
(see \cref{sec:background:typeinference}),
necessitate a dynamic approach for general real-world programs.

\begin{figure}[!t]
  \centering
\begin{lstlisting}[%
    style=nonnumberedlst,%
    frame=single,%
  ]
def test_me(foo):
    if foo.bar > 42:
        return 1
    return foo
\end{lstlisting}
  \Description{A Python code snippet defining a function
    \inlinelst{test_me} with
    one argument \inlinelst{foo}. The function accesses the attribute
    \inlinelst{bar} of \inlinelst{foo} and compares it to the integer
    literal \inlinelst{42}.
  }
  \caption{\label{fig:motivating-the-proxy}%
    A simple function to show the recording of argument interactions.%
  }
\end{figure}

To record the usage of arguments, we wrap each argument in a proxy. The proxy
acts as a thin wrapper around the actual argument and records all operations
that are performed on the proxy, \eg, accesses to attributes or invocations of
methods. The proxy forwards all operations to the wrapped object and passes the
return value or raised exceptions back to the caller. To capture
interactions with objects retrieved from the initial argument (\eg, elements
of a collection), the proxy recursively wraps any returned value in a new
proxy, thereby lazily extending the trace to nested objects. We implement the
proxy using a set of special methods in Python, the so-called
\emph{dunder} methods, \eg, \inlinelst{__getattr__}, \inlinelst{__setattr__},
or \inlinelst{__eq__}. One can use these special methods to implement certain
operators to arbitrary objects, \eg, \inlinelst{__add__} provides the
implementation for a \inlinelst{+} operator. Note that our proxy first records
an event and then forwards the call to the wrapped object. This ensures that
the operation is recorded even if the subsequent forwarding of the operation
fails, \eg, due to the attempted access to a non-existing attribute.
\Cref{fig:example-proxy} shows a pseudo-code excerpt of how our proxy
implementation works.
The \inlinelst{ObjectProxy} class takes any object to wrap as an
argument for its
constructor. It overwrites all dunder methods and within each dunder method
calls the \inlinelst{trace_call} method which records the event,
before forwarding the call to the wrapped object.

\begin{figure}[!t]
  \centering
\begin{lstlisting}[%
    style=nonnumberedlst,%
    frame=single,%
  ]
class ObjectProxy:
    def __init__(self, wrapped):
        self.__wrapped = wrapped
    def __add__(self, other):
        self.trace_call("__add__", other)
        return self.__wrapped + other
    def __eq__(self, other):
        self.trace_call("__eq__", other)
        return self.__wrapped == other
    ...
    def trace_call(self, operator, other):
        # store that <self.__wrapped> <operator> <other>
        # was executed.
\end{lstlisting}
  \Description{A Python code snippet defining a class \inlinelst{ObjectProxy}
    that overwrites the dunder methods \inlinelst{__init__},
    \inlinelst{__add__}, \inlinelst{__eq__}, and more. Each dunder
    method calls the \inlinelst{trace_call} method which records the
  event, before forwarding the call to the wrapped object.}
  \caption{\label{fig:example-proxy}%
    Pseudo code showing how our proxy implementation works.%
  }
\end{figure}

Besides being able to record all operations performed on it, the proxy should
be as transparent as possible, \ie, existing code should not be able to
distinguish between an arbitrary object and a proxy wrapping such an object.
Python's duck typing comes in handy here: routines normally do not perform
explicit type checks on their arguments but accept them if they provide all
required attributes. The proxy forwards all attribute accesses, thus it behaves
identically to the wrapped object; it is consistent to every wrapped object.
Special handling is necessary for type checks using the \inlinelst{isinstance}
function: it checks for the value of the \inlinelst{__class__} attribute of an
object and whether it fulfils the required subtype relation. Our proxy is able
to disguise its own type by returning the value of the wrapped object's
\inlinelst{__class__} attribute to the caller when its own
\inlinelst{__class__} attribute is accessed.

While this proxy-based approach allows us to record interactions with
arguments, one might wonder whether Python's trace
module\footnote{\url{https://docs.python.org/3/library/trace.html},
last accessed 2026-04-30.}  could be used instead.  We cannot use it
because, although it supports tracing function calls and returns
(e.g., via the \emph{call} and \emph{return} events using
\inlinelst{sys.settrace}), it does not offer sufficient granularity
for our purposes: To reason about the types of arguments, we need to
monitor not just which code paths are executed, but also which
attributes are accessed and which methods are invoked on parameter
objects from which call points. This level of detail requires a
proxy-based approach rather than using the trace module.

\subsubsection{Recording Type Checks}\label{sec:approach:collect:checks}

In addition to recording the method calls and attribute accesses on
the proxy, type checks provide a further source of type information
that helps successfully inferring the types of parameters and return
values.  In Python, one usually uses the \inlinelst{isinstance(obj,
type)} function for type checks. We replace the original function with
a shim, \ie, a function that replaces the original API
call. \Cref{fig:example-isinstance-shim} shows a simplified code
snippet how the shim works. In case of a type check on a proxy the
shim records the checked types; afterwards, and for non-proxy objects,
it forwards the type check to the original \inlinelst{isinstance}
function. Our shim can be disabled to allow the usage of
\inlinelst{isinstance} checks in \pynguin's code: we do not want to
record such checks performed within \pynguin's code itself but only
those performed by the SUT\@.

\begin{figure}[!t]
  \centering
\begin{lstlisting}[%
    style=nonnumberedlst,%
    frame=single,%
  ]
isinstance_orig = builtins.isinstance

def isinstance_shim(obj, type_):
    store_info_if_proxy(obj, type_)
    return isinstance_orig(obj, type_)

isinstance = isinstance_shim
\end{lstlisting}
  \Description{A python code snippet showing how the shim for
    \inlinelst{isinstance} works. First, the original
    \inlinelst{isinstance} function is saved. Then, the
    \inlinelst{isinstance_shim} function is defined, which records the
    checked types and forwards the type check to the original
    \inlinelst{isinstance} function. Finally, the
  \inlinelst{isinstance} function is replaced with the shim.}
  \caption{\label{fig:example-isinstance-shim}%
    A (simplified) code snippet of how the shim for \inlinelst{isinstance}
    works.
  }
\end{figure}

Another way to check the type of a given object is to call
\inlinelst{type(obj)}, which returns the runtime class of the object.
For this, we considered three possible approaches:
Approach~(1) is to perform a transformation on the
\emph{abstract syntax tree}~(AST) of the SUT, replacing
all invocations of \inlinelst{type(obj)} with \mbox{\inlinelst{obj.__class__}}.
We would then need to modify the proxy mechanism so that it wraps the
value returned by \inlinelst{class} in another proxy. This proxy
could record type checks of the form \mbox{\inlinelst{obj.__class__}}
by intercepting calls to \inlinelst{__eq__}. However, this
transformation has side effects: wrapping the result of
\inlinelst{class} can break \inlinelst{isinstance} checks, and checks
like \inlinelst{obj.class is SomeType} would remain unobservable
because the \inlinelst{is} operator compares memory addresses without
invoking any methods we could intercept.
Approach~(2) is to replace the \inlinelst{type} function
with a shim, similar to how \inlinelst{isinstance} can be wrapped.
However, this also introduces problems. The \inlinelst{type} function
is not only used to retrieve the runtime class of an object---it can
also be used to define new classes and serves as the superclass of
every class, including itself. Replacing it leads to unintended
side effects, such as breaking \inlinelst{isinstance(int, type)}.
Additionally, reliably resolving calls to \inlinelst{type} in the AST
is challenging, since \inlinelst{type} may be shadowed or rebound in
the analysed code.
Approach~(3) is instrumenting the bytecode to detect comparisons
between type objects. While this avoids issues like the type
being shadowed or breaking built-in functionality, it would interfere
with \pynguin's crucial instrumentation on measuring code coverage,
that also relies on bytecode instrumentation and is therefore
challenging.
Due to technical limitations caused by the mentioned side effects of
all three possible options we chose not to support recording
\inlinelst{type(obj)}.

Regardless of using \inlinelst{isinstance(obj, type)} or
\inlinelst{type(obj)}---type checks in code are often employed as
preconditions to ensure that arguments conform to expected types,
raising exceptions when this is not the case.  Upon encountering such
a type check, \typetracing records the type of the argument being
checked against. For example, if a function checks whether an argument
is of type \inlinelst{int} and raises an exception if not,
\typetracing records the type of the argument as a possible type for
the parameter. However, in a different case, the type check could be
used to check for an invalid type, \eg, if the function checks whether
an argument is of type \inlinelst{int} and raises an exception
\emph{if it is not}. In this case, \typetracing records the type check
nevertheless and \pynguin learns contradicting type information,
namely that the argument is of type \inlinelst{int} while the function
actually expects a type other than \inlinelst{int}.  However, most
of the time such guarding type checks are used to ensure that the
argument \emph{is} of a specific type and not vice versa.
Additionally, if \pynguin observes type checks of multiple (even
conflicting) types it keeps track of all observed
types (see \cref{sec:approach:selection}),
such that parameter types can be chosen among all possible
ones. Consequently, this is a negligible problem in practice.

\subsubsection{Class Analysis}\label{sec:approach:collect:classes}

Using proxies and the shim for \inlinelst{isinstance} allows us to
gather type information. To make use of this gathered information,
\Pynguin must know which classes exist in the SUT and what attributes
they provide. During its setup phase~(see~\cite{LF22}), \pynguin creates a test
cluster~\cite{WL05}, which contains information about all available classes,
their constructors and methods, as well as all available functions within the
SUT and its transitive dependencies. We enhance the class analysis such that
\pynguin creates an inheritance graph that relates all seen classes using their
subclass relationship. We furthermore extend the analysis to identify which
attributes are available on instances of each class. This is not straightforward
in Python, in contrast to other languages such as Java: one can change
the layout of a Python object almost arbitrarily during runtime, \eg,
by adding or removing attributes and methods dynamically~\cite{HH09}.
At best, we can construct an approximation of the attributes
available on instances of a class. To achieve this, \pynguin collects
all instance and static attributes of each class.
Instance attributes are attributes that are defined within the
\inlinelst{__init__} method of a class. \Pynguin uses
\toolname{astroid}\footnote{\url{https://github.com/pylint-dev/astroid},
last accessed 2026-04-30}
to parse the source code of the \inlinelst{__init__} method into
an AST and to then collect all
attribute assignments within the method body as instance attributes.
Static (class-level) attributes are attributes that are defined on the class
object itself and are collected by inspecting the class object at runtime
using the built-in \inlinelst{vars()} function.

\begin{figure}[!t]
  \centering
  \includegraphics[width=0.4\linewidth]{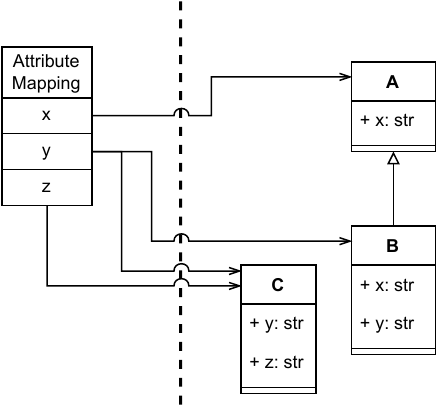}
  \Description{A UML class diagram showing how attributes
    \inlinelst{x}, \inlinelst{y}, and \inlinelst{z} are mapped to
    classes \inlinelst{A}, \inlinelst{B}, and \inlinelst{C}. Class
    \inlinelst{A} has the attribute \inlinelst{x: str}. Class
    \inlinelst{B} extends class \inlinelst{A} and has the attributes
    \inlinelst{x: str} and \inlinelst{y: int}. Class \inlinelst{C} has
    the attributes \inlinelst{y: str} and \inlinelst{z: str}. The diagram
    shows that the attribute \inlinelst{x} is mapped to class
    \inlinelst{A}, \inlinelst{y} is mapped to classes \inlinelst{B} and
  \inlinelst{C}, and \inlinelst{z} is mapped to class \inlinelst{C}.}
  \caption{\label{fig:attribute-mapping}%
    Example for mapping attribute names to classes which have that attribute.%
  }
\end{figure}

After approximating all instance and static attributes of each class,
\pynguin creates a mapping from each attribute name to a set of
classes providing that attribute.
\Cref{fig:attribute-mapping} depicts an example of such a mapping. Please note
that even though both class \inlinelst{A} and \inlinelst{B} define the
attribute \inlinelst{x}, the respective mapping only contains class
\inlinelst{A}; the mapping takes the inheritance between classes into account
and only includes the class defining an attribute which is highest in the
hierarchy. Therefore, inferring a parameter type always leads to the most
general class. When \pynguin tries to find a value for such a
parameter, it will consider all subclasses that adhere to the consistency
relationship, too (\ie, \inlinelst{B}).

\subsection{\TypeTracing During Test Case Execution}\label{sec:approach:tracing}

To implement \typetracing, we have to change the way how \pynguin
executes test cases. The primary goal of an execution without \typetracing
is to collect an
execution trace, \eg, to retrieve which branches and code objects a test
covers. \Typetracing adds a secondary goal, namely, how the subject uses
arguments and which types its routines return.
We capture argument usage via our proxies and a shim for
\inlinelst{isinstance}, and we record return types by logging the
types of returned values.

\begin{figure}[!t]
  \centering
\begin{lstlisting}[%
    style=nonnumberedlst,%
    frame=single,%
  ]
>>> def bar(arg):
...     return arg is None
...
>>> arg = None
>>> bar(arg)
True

>>> proxy = ObjectProxy(arg)
>>> bar(proxy)
False
\end{lstlisting}
  \Description{An interactive Python session where a proxied
    \inlinelst{None} is compared
    against \inlinelst{None} using \inlinelst{is} and the check returns
  \inlinelst{False}.}
  \caption{\label{fig:example-proxy-native-no-exception}%
    Example for a diverging control flow due to a native method.
  }
\end{figure}

\subsubsection{Proxied Executions on C
Code}\label{sec:approach:proxied-execution}

The proxy, as we described in \cref{sec:approach:collect:proxy}, aims to be
indistinguishable from the wrapped object. There is, however, an
exceptional case where this approach does not work,
\ie, for methods of objects that are implemented in C.
\Cref{fig:example-proxy-native-no-exception} depicts such a case,
where a function \inlinelst{bar} checks if its argument is \inlinelst{None}.
Calling this function with the argument \inlinelst{None} returns
\inlinelst{True} as expected, however, when wrapping the argument in
a proxy and calling
\inlinelst{bar} with that proxy, it returns \inlinelst{False}. The
reason for this is that
line 2 compares the memory addresses of the two operands.
Because \inlinelst{None} is a singleton object, the memory addresses
of two \inlinelst{None} objects are always the same,
but since the proxy is a different object,
the memory addresses are different
and the comparison returns \inlinelst{False}.

To mitigate the limitations of proxies in native method interactions,
we explored three strategies. However, as detailed below, each introduced
significant technical or performance-related drawbacks.
\begin{enumerate}
  \item \textbf{Patching}: Intercepting native method implementations
    to unwrap proxies before they reach the C code.
  \item \textbf{Instrumentation}: Instrumenting the Python bytecode
    to perform runtime unwrapping of proxies before passing them to
    native methods.
  \item \textbf{AST Transformation}: Transforming the AST of the
    Python code to inject unwrapping logic around call sites before
    passing them to native methods.
\end{enumerate}

\textbf{(1) Patching}.
While Python-level methods (such as \inlinelst{isinstance}) can be successfully
patched, C-implemented methods are generally immutable at runtime.
A workaround involves using \texttt{ctypes} to wrap the C implementation;
however, this approach introduces issues with exception handling.
Specifically, \texttt{ctypes} wrappers fail to propagate errors
from the underlying C implementation back to the Python interpreter.
For example, if we patch the \inlinelst{in} operator to unwrap proxies,
a valid \inlinelst{TypeError} raised by the original implementation
(e.g., \inlinelst{1 in "abc"}) is suppressed and cleared by the wrapper.
This suppression prevents the error from reaching the caller,
leading to divergent execution flows between \typetracing and regular execution.

\textbf{(2) Instrumentation}.
A second approach is to instrument Python bytecode to unwrap proxies before
passing them to native methods.
However, because proxies are passed transitively to any imported module,
this would require the instrumentation of the entire dependency graph,
including the Python standard library and all third-party dependencies.
Even though the performance overhead per call site is small,
adding unwrapping logic at every call site across all modules
introduces a prohibitive performance overhead,
undermining the efficiency required for search-based testing.

\textbf{(3) AST Transformation}.
Similarly, transforming the AST to inject unwrapping logic requires processing
not only the module under test but also its recursive dependencies.
Beyond the performance costs shared with the instrumentation approach,
AST transformation frequently conflicts with existing instrumentation
used for coverage measurement, leading to unstable or incorrect telemetry.

A common theme among these strategies is the requirement for intrusive
modifications to the entire execution environment. In contrast, \pynguin is
designed to be lightweight, instrumenting only the module under test. The
performance degradation caused by heavy modifications would negate the
benefits provided by \typetracing. Consequently, we opted not to implement
these mitigations. We instead accept that the execution trace during
\typetracing
may occasionally diverge from regular execution and perform a separate execution
of the test case whenever we want to record the usage of a parameter.
We call the
execution without \typetracing \emph{regular} execution and the one
with \typetracing \emph{proxied} execution.
Executing a test case twice may pose a substantial runtime overhead
and would slow down the search process. Unfortunately, a single
\emph{proxied} execution at the start of
the search is insufficient. As the search evolves, reaching higher
coverage or refining inferred types may alter the execution path
which may lead to new information about the usage of the
parameters, which in turn might benefit the search process from there
on.
Since an evolutionary search algorithm will likely execute very
similar test cases very frequently during its exploration
(as mutation and
  crossover produce only slightly modified individuals, which often
follow the same execution paths),
we use a simple probabilistic approach where an additional
\emph{proxied} execution
only follows a \emph{regular} execution with a certain
probability. This allows balancing the trade-off between the overhead
of executing a test case twice and the amount of type information we
can collect.

\subsubsection{Tracing Parameter Usage}\label{sec:approach:tracing:arguments}

Whenever a test case calls a routine during a proxied execution, we
wrap each parameter with a temporary proxy. We wrap only objects that
are created by the tests, not objects created internally in the SUT,
as parameters represent the only point where \Pynguin
needs to make choices based on types. We apply this transformation
on-the-fly before the execution. To be able to distinguish which
interactions were performed on which proxy, we need to create separate
proxies for each parameter. \Cref{fig:statement-transform} shows an
example of such a transformation: function \inlinelst{test} is the
original test case, function \inlinelst{test_transformed} shows the
code generated from the on-the-fly transformation.

\begin{figure}[!t]
  \centering
\begin{lstlisting}[
    style=nonnumberedlst,%
    frame=single,%
  ]
def foo(a, b): ...

def bar(d): ...

def test():
    int_0 = 42
    var_0 = bar(int_0)
    var_1 = foo(var_0, int_0)

def test_transformed():
    int_0 = 42
    tmp_0 = ObjectProxy(int_0)
    var_0 = bar(tmp_0)

    tmp_1 = ObjectProxy(var_0)
    tmp_2 = ObjectProxy(int_0)
    var_1 = foo(tmp_1, tmp_2)
\end{lstlisting}
  \caption{\label{fig:statement-transform}%
    A code example showing the original test case in function \inlinelst{test}
    and the transformed test case using proxies
    in function \inlinelst{test_transformed}.
  }
\end{figure}

We extract the usage trace from all generated proxies at the end of the
test-case execution, \ie, after the last statement of a test case was
executed or if a statement raised an error. The test cluster stores the usage
traces for each respective parameter. Storing a usage trace means that we merge
the new trace with the existing one in the test cluster.

\subsubsection{Recording Return Types}\label{sec:approach:tracing:return}

Besides information about parameter interactions, we are also interested in the
return types of routines. We trace these from the regular execution because
they might differ in the proxied execution. The tracing simply inspects the
return value of each successfully executed test case statement. If the return
type is a collection, we try to inspect the elements of the collection, if any,
to retrieve a more precise type. For \inlinelst{tuple}s we inspect all
elements, whereas for all other collection types, we only inspect the first
element. If no element is available for inspection we assume the parameter type
to be \inlinelst{Any}.

We update the return type of a routine in the test cluster whenever we record a
return type for it. If the return type previously was unknown, \ie,
\inlinelst{Any}, we replace it with the recorded type. Otherwise, we store the
union of the types. Note that we store both types in the union even if there
exists a subtyping relation between them to avoid traversing the
subtype relationships when searching for a routine that returns a specific
type. We restrict the size of these unions because \pynguin has very limited
support for generic types and our approach thus is unable to infer whether a
return type might be generic.

\subsection{Selecting Parameter Types}\label{sec:approach:selection}

\begin{algorithm}[thpb]
  \caption{Type Inference}\label{alg:infer_type}
  \KwIn{A usage trace node $N$}
  \KwOut{An inferred type $T$ or \texttt{Any}}
  $S \gets \emptyset$\;

  \If{$N.\text{type\_checks} \neq \emptyset$}{
    $S \gets S \cup \{\textbf{FromTypeCheck}\}$ \tcp*{Based on
    isinstance checks}
  }

  \If{$N.\text{invocations} \neq \emptyset$}{
    $S \gets S \cup \{\textbf{FromMethodInvocations}\}$ \tcp*{Based
    on recorded parameter types}
  }

  \If{$N.\text{attributes} \neq \emptyset$}{
    $S \gets S \cup \{\textbf{FromAttributeAccess}\}$ \tcp*{Based on
    required attributes}
  }

  \If{$S = \emptyset$}{
    \Return \texttt{Any}\;
  }

  $\mathit{strategy} \gets \text{SelectRandom}(S)$\;
  $T \gets \mathit{strategy}.\text{Execute}(N)$\;

  \If{$\text{IsCollection}(T)$}{
    $T \gets \text{\textbf{InferGenerics}}(T, N)$ \tcp*{Recursive application
    of strategies}
  }

  \Return $T$\;
\end{algorithm}

\Pynguin stores both, the developer-annotated types and a list of
recently recorded types, in the so-called test cluster. Whenever
\pynguin adds or modifies a call to a routine in a test case during
the generation it needs to create a variable to satisfy a parameter.
\Pynguin queries the test cluster which type to use for that
parameter. The generator chooses a type for the
parameter using a weighted random selection between either the
developer-annotated type~(if available, weight~$\omega_\text{dev}$),
the \inlinelst{None} type~(weight $\omega_\text{none}$), the
\inlinelst{Any} type~(weight $\omega_\text{any}$), or a union of the
previously inferred types for that parameter~(if available, weight
$\omega_\text{union}$).
Choosing \inlinelst{Any} allows \pynguin to use any value to satisfy
the parameter, whereas for the
\inlinelst{None} type it uses the \inlinelst{None} value.
In Python \inlinelst{None} is often used to convey that a parameter is
unused or not set, similar to \inlinelst{null} in Java, in which case
the type is irrelevant and which is why we handle \inlinelst{None} as
a separate case.

In case the union of the currently inferred types is selected, \pynguin
adds types to the union by randomly picking types with uniform probability
inferred by the strategies detailed in \cref{alg:infer_type}.
These strategies map the dynamic usage traces back to concrete types
available in the test cluster:
\begin{description}
  \item[FromTypeCheck (lines 2--3)] leverages explicit
    \inlinelst{isinstance} checks. If a routine checks
    \inlinelst{isinstance(param, int)}, \typetracing records
    \inlinelst{int} as a candidate. During inference, \pynguin
    randomly selects an observed type.
  \item[FromMethodInvocations (lines 4--5)] addresses the challenge
    of ambiguous method names, such as \inlinelst{__add__}. When an
    operation like \inlinelst{proxy + other} is recorded, the trace
    node for \inlinelst{proxy} stores both the method name
    (\inlinelst{__add__}) and the type of \inlinelst{other}. \Pynguin
    then uses the recorded argument type directly as the inferred
    type for \inlinelst{proxy}. This allows distinguishing between,
    e.g., \inlinelst{int.__add__(int)} and \inlinelst{str.__add__(str)}.
  \item[FromAttributeAccess (lines 6--7)] uses the mapping described
    in \cref{sec:approach:collect:classes}. If the trace shows access
    to an attribute \inlinelst{x}, \pynguin selects a type from the
    set of classes known to provide that attribute.
  \item[InferGenerics (lines 12--13)] provides support for collection
    types. When a collection's method is called (\eg,
    \inlinelst{list.append(item)} or \inlinelst{list.__getitem__(i)}),
    \typetracing creates a separate
    usage trace node for the elements. To maintain efficiency and avoid
    per-element tracking, all elements accessed via the same method
    (\eg, \inlinelst{append}) share a single trace
    node. \Pynguin then recursively applies the inference strategies to
    this element node. For example, if the trace of the element node
    contains an \inlinelst{isinstance(item, int)} check, \pynguin
    infers the element type as \inlinelst{int}, resulting in
    \inlinelst{list[int]}.
\end{description}
Finally, the test cluster answers the query for a type of a parameter
by probabilistically returning the developer-annotated type,
\inlinelst{None}, \inlinelst{Any},
or a union of the inferred types, depending on the $\omega$-weights.
\Pynguin uses this information to instantiate the
required type, e.g., by invoking a constructor of the type
itself or one of its subtypes.

\subsection{Generalisability}\label{sec:approach:generalisation}

While we decided to implement \typetracing for \pynguin in Python,
the two core mechanisms \typetracing relies on---transparent proxy
objects and type-check interception---are available in many
dynamically-typed languages: JavaScript provides a native
\texttt{Proxy}~API~(ES6)\footnote{
  \url{https://developer.mozilla.org/en-US/docs/Web/JavaScript/Reference/Global_Objects/Proxy},
  last
accessed 2026-04-30.}
that can intercept property accesses and method invocations and
Ruby enables similar behaviour via \texttt{method\_missing} and
\texttt{BasicObject}.\punctfootnote{
  \url{https://docs.ruby-lang.org/en/2.1.0/BasicObject.html}, last
accessed 2026-04-30.}
In Ruby, subclassing \texttt{BasicObject}---which exposes almost no
predefined methods---ensures that all method calls fall
through to the \texttt{method\_missing} hook, where they can be
recorded and forwarded to the wrapped object.
Likewise, Python-specific type-check functions such as
\texttt{isinstance} have direct analogues in other languages~(\eg,
JavaScript's \texttt{instanceof}, Ruby's \texttt{is\_a?}) that can
be shimmed in the same way. The class-level introspection used for
attribute mapping also has equivalents in other runtimes~(\eg,
  \texttt{Object.keys} in JavaScript, \texttt{instance\_variables} in
Ruby). While the specific Python implementation details---such as
CPython's dunder methods or the \texttt{isinstance} shim---would need
to be re-engineered for each target language and runtime, the approach
does not rely on any Python-unique semantic that is absent from other
dynamically-typed languages.

\section{Evaluation}\label{sec:eval}

Given the proposed approach of \typetracing, this section now aims to
empirically evaluate its effects.

As described in \cref{sec:approach:tracing}, \typetracing may slow
down the search process of \pynguin through additional \emph{proxied}
execution. Before running further experiments, we therefore need to
determine an optimal probability for adding \emph{proxied} executions,
such that the positive effects of type information are maximised while
the negative impact of the overhead is minimised. We thus ask:

\begin{itemize}
  \item[\textbf{RQ1:}] What is the optimal probability of
    \emph{proxied} executions for \typetracing in \pynguin?
\end{itemize}

Once we know how \emph{proxied} executions should be configured in
\pynguin, we want to know whether the type information inferred by
\typetracing is helpful for generating tests. Thus, we ask:

\begin{itemize}
  \item[\textbf{RQ2:}] What effect does \typetracing have on the achieved
    branch coverage?
\end{itemize}

\begin{figure}[!t]
  \centering
\begin{lstlisting}[%
    style=nonnumberedlst,%
    frame=single,%
  ]
def less_equal(x: int, y: int) -> bool:
    return x <= y  # Correct implementation

def less_equal_mutated(x: int, y: int) -> bool:
    return x < y  # Mutated version

def test_less_equal():
    assert less_equal(1, 1) == True  # Pass

def test_less_equal_mutated():
    assert less_equal_mutated(1, 1) == True  # Fail (mutation detected)

def test_less_equal_wrong_type():
    assert less_equal("1", 1) == True  # TypeError

def test_less_equal_mutated_wrong_type():
    assert less_equal_mutated("1", 1) == True  # TypeError (mutation not detected)
\end{lstlisting}
  \caption{\label{fig:types-affect-mutant-discovery}%
    An example of how types can affect the discovery of mutants regardless
    of the code coverage.
  }
\end{figure}

A major quality aspect for tests is their fault-finding capability.
Using incorrect types can have a negative impact on the fault-finding
capabilities, as \cref{fig:types-affect-mutant-discovery} exemplifies:
The function \inlinelst{less_equal} checks if the first parameter is
less than or equal to the second parameter.  In the mutated version
\inlinelst{less_equal_mutated} the operator is changed to
\inlinelst{<} which leads to a different behaviour.  The test case
\inlinelst{test_less_equal} calls the function with two integers as
intended and achieves full coverage.
\inlinelst{test_less_equal_mutated} fails and detects the mutant
successfully.  When the function is called with wrong types, e.g., a
string and an integer as in \inlinelst{test_less_equal_wrong_types},
the test case achieves full coverage\footnote{
  A line is marked as covered once it has been executed, independent
  of whether the
  execution was successful or raised an exception.
} but fails for both the original
and the mutated version of the function.  Thus, even though full
coverage is achieved in both cases, the fault is only detected when
the correct types are used.
To shed light on the effects of \typetracing on the generated tests'
fault-finding
capabilities, we thus ask:

\begin{itemize}
  \item[\textbf{RQ3:}] What is the effect of \typetracing on the
    fault-finding capabilities of the generated tests?
\end{itemize}

Reliable type inference from \typetracing could serve as a
lightweight alternative to existing (static) inference tools---similar
to \toolname{MonkeyType} but without the need for human-written tests.
To determine the quality of the types inferred by \typetracing, we
compare our \typetracing approach with existing type-inference tools based on
the \textsc{TypeEvalPy}~\cite{VSW+23} benchmark.  We therefore ask:

\begin{itemize}
  \item[\textbf{RQ4:}] What is the quality of the parameter and return
    types inferred by \typetracing?
\end{itemize}

\subsection{Experimental Setup}\label{sec:eval:setup}

We conducted an empirical evaluation as follows to answer our research
questions.

\subsubsection{Experiment Subjects}\label{sec:eval:setup:subjects}

We use two datasets: \dstuning to evaluate the hyperparameter tuning
of RQ1 and \dseval to answer RQ2, RQ3, and RQ4. We chose to evaluate
RQ1 on a separate dataset to avoid biasing the results of the other RQs.

\dstuning.
The goal of this dataset is to provide a representative sample of
real-world Python modules, where \pynguin
successfully generates tests and the modules are neither trivial to cover
nor problematic edge cases, overall enabling reliable hyperparameter tuning.
To construct the dataset, we first fetched all available projects
from \toolname{PyPI}\footnote{\url{https://pypi.org/}, last accessed
2026-04-30.} and then randomly sampled \exnum{10000} projects.
For these we fetched all metadata from \toolname{PyPI}, including the
GitHub URL\@.
Using the \toolname{GitHub API}\footnote{%
  \href{https://docs.github.com/en/rest}{docs.github.com/en/rest},
  last accessed 2026-04-30.%
} we obtained all Git tags of the projects. The rationale for this is that
most developers use Git's tags to mark releases and that the tag's label is
similar to the version on \toolname{PyPI}.  By matching the Git tag
included in the \toolname{PyPI} metadata
with the Git tags from the \toolname{GitHub API} we determined which
commit was used for
the release. We removed all cases where no GitHub URL was found or
where no tag could be matched.
In some cases multiple \toolname{PyPI} projects point to the same
GitHub repository,
e.g., when a single GitHub project publishes multiple related
packages (e.g., \emph{tool-core} and \emph{tool-extension1}).
We dropped such duplicates based on identical GitHub URLs.
This resulted in
\exnum{5959} projects for further processing.
Using the GitHub URLs we cloned the repositories and checked out the
matched tags using \toolname{Git}.\punctfootnote{%
  \href{https://git-scm.com/}{git-scm.com/},
  last accessed 2026-04-30.%
}

We executed \pynguin \exnum{30} times on each of the modules of these
projects without \typetracing
and removed all trivially covered modules, i.e., modules that
reach \qty{100}{\percent}
coverage within \qty{10}{\second} or less. Furthermore, we removed all modules
that could not be executed successfully due to \pynguin terminating
with an error message. The reasons for this are:
\begin{itemize}
  \item \num{121} modules where the SUT dependencies could not be
    installed, due to missing or incompatible packages.
  \item \num{230} modules where the SUT contained nothing one can
    test, for example only private methods.
  \item \num{330} modules where the SUT was broken, e.g., due to
    errors when importing the SUT\@.
  \item \num{63} modules where the module used code that \pynguin
    cannot handle, e.g., due to the use of coroutines.
\end{itemize}
Finally, we sampled \num{\TuneNumModules} modules randomly from the remaining
\exnum{2403} modules resulting in a dataset of \num{\TuneNumModules} non-trivial
modules from \num{\TuneNumProjects} projects.

\dseval.
The goal of this dataset is to provide a sample of real-world Python modules
with type annotations, which can be used to evaluate both the improvement
in test generation using existing and/or inferred types as well as
the quality of the inferred types.
We started with a dataset used by previous work on \pynguin~\cite{LKF23},
which had already excluded projects without any type hints or those
with dependencies on native code. We refined this initial set by
removing projects that no longer work with Python 3.10, violate
invariants of the Python runtime, lead to crashes, or contain only trivial code.
To further enhance the
dataset's generalisability for evaluating type inference and test generation,
we extended it with modules from \toolname{typeshed}\footnote{%
  \href{https://github.com/python/typeshed}{github.com/python/typeshed},
  last accessed 2026-04-30.%
} and thereby incorporate more type annotations from various modules across a
broader range of widely used projects.
\toolname{typeshed} provides manually crafted type annotations for both
the Python standard library and popular third-party libraries.
We considered only projects that are marked as \enquote{fully annotated}
to produce ground truth data for evaluation that is as complete as possible.
Because \pynguin expects type annotations to be part of the subject's
source code, we automatically processed these selected annotations
and merged them with the respective library's source files.
Our merging script uses heuristics to determine the
most likely library version from the metadata provided by
\toolname{typeshed} and
uses \toolname{libcst}\footnote{%
  \href{https://github.com/Instagram/LibCST}{github.com/Instagram/LibCST},
  last accessed 2026-04-30.%
} for the actual merging of annotations and implementation code.
In total, the dataset comprises \num{669} modules from \num{63}
projects where developers provided type annotations.

During our initial experiments, we observed that \pynguin did not
successfully complete all execution runs for every module. To ensure
the robustness of \pynguin itself, we investigated these
failing runs. This process involved identifying and fixing several
bugs within \pynguin. However, even after these improvements, a
number of modules continued to exhibit frequent failures.
The predominant remaining failure category was attributed to intrinsic defects
within the SUT, such as importation errors,
syntax errors, or null bytes within its source code. A secondary, less
frequent, source of error stemmed from resource limitations (e.g.,
memory, runtime). Given that current resource allocations were
sufficient for the vast majority of executions, and that this error
category was comparatively minor, an increase in resource
provisioning was not pursued to maintain efficiency.
As these remaining failures were largely due to SUT-specific
issues or fundamental technical limitations, we decided to focus our
evaluation on modules where \pynguin could operate with reasonable
consistency.
Therefore, we adopted a heuristic for the results reported in our
evaluation: we only considered those modules that yielded at least
\num{25} out of \num{30} successful repetitions in each of the studied \pynguin
configurations.
Applying this heuristic removed \num{\AnnotatedRemovedModules} modules
from the evaluation, and we ended up with
\num{\AnnotatedNumModules} modules from \num{\AnnotatedNumProjects} projects,
which we considered in our evaluation.

\begin{figure}[!t]
  \centering
\begin{lstlisting}[
    style=nonnumberedtxt,%
    frame=single,%
  ]
## Task Description

**Objective**: Examine and identify the data types of various elements such as function parameters, local variables, and function return types in the given Python code.

**Instructions**:
1. For each question below, provide a concise, one-word answer indicating the data type.
2. For arguments and variables inside a function, list every data type they take within the current program context as a comma separated list.
3. Do not include additional explanations or commentary in your answers.

**Python Code Provided**:

{code}


**Questions**:
{questions}

**Format for Answers**:
- Provide your answer next to each question number, using only one word.
- Example:
    1. int
    2. float
    3. str

**Your Answers**:
{answers}
\end{lstlisting}
  \caption{\label{fig:gpt-54-mini-prompt}%
    Prompt used for the \gpt model.
  }
\end{figure}

\subsubsection{Experiment Settings}\label{sec:eval:setup:settings}

We implemented \typetracing by extending \pynguin and executed this
extended version on each of the constituent subjects.
We use a \toolname{Docker} container to
isolate the executions from their environment based on Python~3.10.16. All tool
runs were executed on dedicated compute servers equipped with an AMD EPYC 7443P
CPU and \qty{256}{\giga\byte} RAM\@. We assigned a single CPU core and
\qty{4}{\giga\byte} of RAM, as well as a hard time limit of \qty{3600}{\second}
for each run.

We used \pynguin's implementation of the \toolname{DynaMOSA}~\cite{PKT18b}
algorithm
for our executions,
which was effective for test generation in \pynguin,
along with the same parameter settings used in previous
work~\cite{LKF23}.
We chose the weights for the weighted random selection of which parameter type
to use (as described in \cref{sec:approach:selection}) as follows:
type annotated by the developers~$\omega_{\text{ann}} = 10$,
\inlinelst{None} type~$\omega_{\text{none}} = 1$,
\inlinelst{Any} type~$\omega_{\text{any}} = 5$, and union of
the currently inferred types for that parameter~
$\omega_{\text{union}} = 10$.  These values reflect that the developer-annotated
types shall be considered with high probability; the same holds for unions,
which are created when type tracing discovers a new type that also leads to a
successful execution of the code under test.  To ensure diversity and let
\pynguin occasionally try new types on the code under test, the
value~\(\omega_{\text{any}}\) is used; in rare cases, \pynguin should attempt to
call the code under test with a \inlinelst{None} value, which is often used as
a default by developers, even if it may not reflect a realistic input.
We did not perform any parameter tuning for \pynguin~\cite{LF25} beyond the
\typetracing probability, because the chosen values provided
reasonable results. Similar to
previous work~\cite{LKF23}, we limit \pynguin's
search budget to \qty{600}{\second}. To minimise the influence of randomness we
execute \Pynguin on each subject \num{30} times in each
configuration.  When we refer to
configurations, we use \RandomTypes for a random type selection and
\TypeHints if \pynguin uses the developer-provided type annotations; the
suffix \emph{TypeTracing} denotes activated \typetracing.
To directly compare \typetracing against other type-inference approaches,
we additionally introduce configurations in which the types
inferred by the best static analysis tool and the LLM evaluated in RQ4---%
\jedi and \gpt---are fed into \pynguin as type hints in the
same way as developer annotations, yielding the configurations
\JediTypeHints and \GptTypeHints.
Neither \jedi nor \gpt have access to the developer type
annotations at any point, as both tools are run on the
source code without type annotations used throughout RQ4~(see
\cref{sec:eval:setup:procedure}), and only their own inferred types
are used by \pynguin as hints.
Type information in the \textsc{TypeEvalPy} format is often not fully
qualified, so \pynguin must resolve an unqualified type name emitted
by \gpt or \jedi to a concrete class before it can use the prediction
as a type hint. We implement this resolution as a sequence of steps.
The resolver first checks builtins and \inlinelst{typing} primitives,
then the subject module's own namespace, i.e., classes defined in or
imported into the module under test. Where both checks fail, it
performs an \textsc{ast}-based import lookup combined with a dynamic
dotted-path import for standard-library and third-party modules.
Finally, it performs a project-wide unique-name lookup across modules
that have already been imported during the run.

\subsubsection{Evaluation Metrics}\label{sec:eval:setup:metrics}

For each execution of a module, we keep track of the branch coverage over time
as well as the overall branch coverage at the end of the search. Additionally,
we compute the relative coverage~\cite{AF13} for each module as a percentage.
Given the coverage of a subject~\(s\) in an execution~\(e\), denoted
by~\(\cov(s, e)\), we refer to the best coverage on~\(s\) in any execution
by~\(\max \left( \cov(s) \right)\) and to the worst coverage on~\(s\) in any
execution by~\(\min \left( \cov(s) \right) \).  We define the relative
coverage~\( \cov_r(s, e) \) as
\[
  \cov_r(s, e) =
  \frac{%
    \cov(s, e) - \min \left( \cov(s) \right)%
  }{%
    \max \left( \cov(s) \right) - \min \left( \cov(s) \right)%
  }.\label{eq:background-relcov}
\]
In cases where \( \min \left( \cov(s) \right) = \max \left( \cov(s) \right) \),
i.e., minimum and maximum coverage for a subject~\(s\) are equal, we define
the relative coverage~\( \cov_r(s, e) = \qty{100}{\percent}\)~\cite{CGA18}.
Additionally, we compute the mutation score~\cite{JH11} as a proxy
measure for fault-finding~\cite{ABL05,JJI14,PSY+18,OGK+23}.  We use
a custom version of the
\toolname{cosmic-ray}\footnote{%
  \href{https://cosmic-ray.readthedocs.io/en/latest/}{cosmic-ray.readthedocs.io},
  last accessed 2026-04-30.
} mutation testing tool for Python for this, which is in
line with previous research on test generation for Python~\cite{TMG22,EMB+24}.
The custom version allows a more fine-grained
classification of mutants into
mutants killed because an assertion failed,
\emph{mutants killed (assertion)},
and mutants killed because an exception was thrown
during test execution,
\emph{mutants killed (exception)}.
We use \emph{mutants killed (assertion)} for calculating
the mutation score,
as it ensures
that the score reflects meaningful functional verification
through assertions
rather than accidental kills from runtime exceptions,
which can occur when using invalid input types.
For instance,
a mutation from \inlinelst{is False} to \inlinelst{<= False}
could be killed by a test case passing \inlinelst{None},
which triggers a \inlinelst{TypeError}
in the mutant but not in the original code.
Such mutants are unlikely to be representative of an actual logical
bug in the program.
The mutation score due to assertion failure $M_{\text{assertion}}$ is
computed as
$ M_{\text{assertion}} = \frac{\text{mutants killed
(assertion)}}{\text{total mutants}} $.

To statistically evaluate if one configuration performs better
than another configuration in terms of overall coverage or mutation score, we
use the Mann-Whitney U-test~\cite{MW47} with $\alpha =  \num{0.05}$.
Additionally, we compute the Vargha and Delaney effect size
\effectsize~\cite{VD00} to investigate the effect sizes of the
difference in the achieved overall
coverage or mutation score between two configurations.  Lastly, we use Pearson's
correlation coefficient~\(r\)~\cite{Pea95} to study correlations.

To evaluate the quality of the \typetracing retrieved type
information, we use precision, recall, and the \Fone score
based on partial matches.
A partial match is given if the expected type is part of the inferred type.
For example, if the ground truth is the type \inlinelst{int}
and \pynguin yields a type \inlinelst{Union[int, float]}, we consider this
a partial match.
We use partial over exact matching as \pynguin's \typetracing can
provide multiple types without preference for each parameter and return type.
In principle, a Python method can be called with any type, but values
of most types will lead to a type error or other runtime exception. We consider
these types invalid. Although type annotations can be used to specify expected
types for parameters and return values, this might just be a subset of the
valid types.
For example, consider a function \inlinelst{add} that adds two
numbers using the \inlinelst{+} operator. Its intended use might be to
add two integers; floats or even strings (which would be concatenated) are also
valid parameter types. However, calling \inlinelst{add} with a
dictionary results
in a runtime exception and thus \inlinelst{dict} is an invalid type.
Our goal is to generate tests that cover as much code as possible for
which using
any valid type is beneficial. Thus, we consider partial matching as most
suitable for test generation.

When comparing the inferred types against
a ground-truth type (\eg \inlinelst{int}) the result is either a partial match
(\match, \eg for \inlinelst{Union[int, float]}) or no inferred types matches
(\mismatch, \eg for \inlinelst{Union[bool, float]}).
To gain more detailed insights we treat the cases
in which no type was inferred~(\missing) and those in which only the
\inlinelst{Any} type
was inferred (\any) separately from \mismatch when referring to
\match, \mismatch, \missing, and \any. \all refers to the combination
of all four cases.
We further use precision, recall, and \Fone score to evaluate type inference
correctness.
The precision~\(P\) measures the fraction of inferred types that are
correct (i.e., those that constitute a partial match with the ground truth). It
is calculated as the number of matches divided by the total number of
types our tool inferred: $ P = \frac{\text{matches}}{\text{total
inferred types}} $.  Recall~\(R\) measures the fraction of all
ground-truth types that were correctly identified. It is calculated as
the number of matches divided by the total number of types in the
ground truth: $ R = \frac{\text{matches}}{\text{total ground-truth
types}} $.
For example, for the parameters of the benchmark subjects,
\hityper matched
\num{\TypeInfParamhityperMatches} ground-truth types
out of a total of
\num{\TypeInfParamhityperTotalInferred} inferred types,
resulting in a precision of
$ P \approx
\frac{\num{\TypeInfParamhityperMatches}}{\num{\TypeInfParamhityperTotalInferred}}
\approx \num{0.511} $.
The \Fone score is used to provide a single performance
measure that balances both. It is calculated as the harmonic mean of
precision and recall: \( \Fone = 2 \cdot \frac{P \cdot R}{P + R} \).

We report all measured numbers rounded to three significant digits except for
numbers that can be counted exactly, e.g., number of modules or types.

\subsubsection{Experiment Procedure}\label{sec:eval:setup:procedure}

To address RQ1 and RQ2, we monitor the branch coverage achieved by
\pynguin over time. Specifically, after each iteration of the genetic
algorithm employed by \pynguin, we record the current branch
coverage, storing the most recent value for each elapsed second.
This includes measuring the final branch coverage at the end of the search.
The data is stored as CSV, along with relevant metadata such as the
configuration used, and the project and module under test.
After executing all configurations across all modules, we merge
the individual CSV records into a single file, which serves as
the basis of our evaluation.
To answer RQ1, we executed \pynguin using configurations that vary
the percentage of performing a proxied execution from
\qtyrange{0}{100}{\percent} in \qty{5}{\percent} increments on \dstuning. For
all configurations, \typetracing is enabled while the use of existing
type hints is disabled, \ie, we used \RandomTypesTypeTracing.
For RQ2, we executed \pynguin using the configurations
\TypeHints, \TypeHintsTypeTracing, \RandomTypes,
\RandomTypesTypeTracing, \JediTypeHints, and \GptTypeHints
(see \cref{sec:eval:setup:settings})
with a probability of \qty{5}{\percent} (the optimal probability according to
RQ1) for proxied executions on \dseval.
For \JediTypeHints and \GptTypeHints the inferred types are read from the
\toolname{TypeEvalPy} output produced during the RQ4 runs and
injected into \pynguin's type hinting mechanism prior to test generation.
Besides measuring coverage we extract the final \pynguin-inferred types
at the end of the search. We therefore used the
\textsc{TypeEvalPy}~\cite{VSW+23} JSON format.
We further extracted existing type hints from the \dseval benchmark
and stored them in the same \toolname{TypeEvalPy} JSON format as ground
truth.

To address RQ3, we measured the mutation score obtained by
\toolname{cosmic-ray} for each configuration across all modules. For
each run, we initialised a \toolname{cosmic-ray} session using the
\inlinelst{init} command with a configuration file specifying the
module under test path, the test command
\inlinelst{pytest --exitfirst} followed by the generated test suite,
a timeout of \num{5} minutes, and the distributor set to
\inlinelst{name = "local"}.
Before conducting mutation testing, we confirmed that the generated
test suite passed on the non-mutated code using the
\inlinelst{baseline} command. We then executed mutation testing via
\inlinelst{exec}.
Overall, the setup adheres to the procedure outlined in the official
\toolname{cosmic-ray} documentation\footnote{%
  \href{https://cosmic-ray.readthedocs.io/en/latest/tutorials/intro/index.html}{cosmic-ray.readthedocs.io/intro},
  last accessed 2026-04-30.
} and was similarly used in previous work~\cite{EMB+24}.
After completion, we extracted
the number of successful cosmic-ray runs,
the number of mutants killed through a failing assertion,
the number of mutants killed through an exception,
and the total number of generated mutants
from the mutation-testing database.
In total,
we extracted mutation scores for \exnum{\AnnotatedMutationNumModules} modules.
We computed the mutation score as
the number of mutants killed through an assertion
divided by the total number of generated mutants.
These metrics were subsequently added to the respective
CSV row produced by \pynguin.
To provide a granular analysis of the test suite effectiveness, we
extended the categorization of survived and killed mutants beyond
the standard classification into two subcategories of killed mutants:
\textit{killed by assertion} and \textit{killed by exception}.
A mutant is categorised as \textit{killed by assertion} if a test failure
is triggered by an explicit assertion violation, suggesting that the test
case correctly identified a change in the program's logic. In contrast,
\textit{killed by exception} identifies mutants that caused the code under
test to raise an unhandled runtime error (e.g., a \texttt{TypeError}
resulting from an invalid operator mutation). This distinction is
particularly relevant for automatically generated tests, as it allows us
to differentiate between tests that verify functional properties and
those that merely trigger crashes through unexpected input types.
The computed mutation scores serve as the basis for evaluating RQ3.

To address RQ4,
we used the \textsc{TypeEvalPy}~\cite{VSW+23} infrastructure.
\textsc{TypeEvalPy} currently incorporates six non-LLM type-inference tools
based on different inference techniques, ranging from static analysis
to deep learning:
\toolname{HeaderGen}~\cite{VWL+23},
\toolname{Jedi},\punctfootnote{%
  \href{https://jedi.readthedocs.io/en/latest}{jedi.readthedocs.io},
  last accessed 2026-04-30.%
}
\toolname{Pyright},\punctfootnote{%
  \href{https://github.com/microsoft/pyright}{github.com/microsoft/pyright},
  last accessed 2026-04-30.%
}
\toolname{HiTyper}~\cite{PGL22}, and
\toolname{Scalpel}~\cite{LWQ22}.
While \toolname{Type4Py}~\cite{MLP+22} is also supported, the tool did not work
with our datasets due to bugs in the \toolname{TypeEvalPy}
infrastructure regarding
this tool. Thus, we excluded it from our evaluation.
Additionally, we added \toolname{QuAC}~\cite{wu_quac_2024}
to \toolname{TypeEvalPy} for our evaluation.
\toolname{TypeEvalPy} also supports the usage of LLMs. We decided to add
\gpt to the evaluation, as its predecessor,
\toolname{GPT-3.5},
had already shown itself to be effective in type-inference
tasks~\cite{PWW+23}. We used the prompt that performed best in the
evaluation of \toolname{TypeEvalPy} (see
\cref{fig:gpt-54-mini-prompt}). Reusing the \toolname{TypeEvalPy}
benchmark setup means evaluating the model without codebase-wide
retrieval (RAG) or cross-module signature provision. This choice
ensures direct comparability with the other static type inference
tools in \toolname{TypeEvalPy}, which operate under the same
module-level conditions.

Regarding the evaluation data itself, the default dataset provided by
\textsc{TypeEvalPy} has significant limitations for our purposes.
Although it contains \num{154} code snippets with \num{845} type
annotations, these mostly concern local variables rather than the
parameter or return types we focus on. Furthermore, its artificial
construction does not reflect real-world usage of type hints. To
overcome these limitations, we used the \dseval dataset, which we
adapted for the \textsc{TypeEvalPy} infrastructure
and compared the types inferred by the
\textsc{TypeEvalPy} tools
against \pynguin with \typetracing. As ground truth, we
used the extracted existing type hints from the \dseval benchmark.
We then removed all hints from the
benchmark's modules and ran each tool once while collecting their inferred
types as JSONs using the \toolname{TypeEvalPy} infrastructure.
Similarly, we executed \Pynguin with \typetracing
(\RandomTypesTypeTracing) with a probability of
\qty{5}{\percent} for the proxied execution~(best according to RQ1).
To ensure fairness toward the other tools, we executed \pynguin only once when
answering this RQ, although this approach ignores the potential influence of
randomness on the results.
To measure performance, we computed the number of partial matches
for each tool's output, compared it to the ground truth, and calculated
precision, recall, and \Fone score as described in
\cref{sec:eval:setup:metrics}.

\subsection{Threats to Validity}\label{sec:eval-threats}

As any study of this type, our empirical evaluation is subject to
several threats to validity.

\subsubsection{Internal Validity}

\Typetracing might influence the observed branch coverage during test-case
execution because the SUT might behave differently when the given arguments are
wrapped in proxies. To mitigate the problem we measured the coverage in a
separate test-case execution, such that the reported coverage cannot be
influenced by the proxy.
However, the runtime overhead of the second, proxied execution reduces
the number of generations explored within the fixed time budget.
This is a threat to internal validity: if \typetracing explores fewer
generations than the baseline, any observed difference in coverage or
mutation score could be partly attributable to this disparity rather
than to the technique itself.
We mitigate this by employing a probabilistic approach and by measuring
the overhead during evaluation.
It may happen that a proxy is stored as some form of
global state, which might influence the resulting coverage; however, similar to
the detection of flaky behaviour, this is a current limitation of
\pynguin~\cite{GRP+24}. We
introduce a new abstraction for \pynguin's internal type handling. While we
cannot guarantee the absence of bugs, we provide sufficient unit tests to
ensure that the new abstraction is at least as capable as the previously used
one.
The proxy, together with the new abstraction, allows recording of type
checks using \inlinelst{isinstance()}, but type checks using
\inlinelst{type()} are currently not supported, which could affect
completeness of the collected type information.
We explored several implementation strategies
but we concluded that all viable options introduced significant
technical challenges and unacceptable side effects~(see
\cref{sec:approach:collect:checks}).
Consequently, if the SUT uses \inlinelst{type()} for its runtime checks,
\pynguin's \typetracing will not observe these checks, potentially reducing its
effectiveness.
However, the impact of this limitation is contained, as using
\inlinelst{isinstance()} is generally considered a better alternative.
A search of open-source Python code on GitHub suggests that
\inlinelst{isinstance()}
is also preferred over \inlinelst{type()} in practice:
Querying for conditional checks using \inlinelst{isinstance()} yielded
approximately \num{700000} results, whereas similar checks using
\inlinelst{type()} yielded only around \num{90000}.
A further threat to internal validity in RQ4 arises
from the methodological asymmetry between \pynguin and the static
type inference baselines. While static tools are limited to one-shot
predictions, \pynguin's type tracing is
inherently iterative, allowing it to refine type information through
dynamic feedback during test generation. To mitigate this advantage
and ensure a more meaningful comparison, we also evaluate the static
tools by using their generated type hints as inputs for \pynguin.
Additionally, we evaluate \gpt by reusing the established
\toolname{TypeEvalPy} benchmark setup, which operates on individual
modules without codebase-wide retrieval/RAG or cross-module signature
injection. Reusing this standard setup ensures direct comparability
with the static type inference tools evaluated under the exact same
module-level conditions.
Besides those technical limitations, there are newly-added parameters
that we did not fine-tune:
the weights for the weighted random selection of which parameter type
to use~(as described in \cref{sec:approach:selection}), we chose values
that provided reasonable results for a manual small-scale evaluation during
development.
Future work might include fine-tuning these parameters as well.

\subsubsection{External Validity}

For \dstuning we only use a sample of \num{100} modules from \num{71} projects
from \toolname{PyPI}.
While this is a reasonable sample size, it is not as large as the
number of modules from \dseval. However, we only use this dataset to
tune the hyperparameter of \typetracing in RQ1. By using a separate
dataset, we avoid biasing the results of the other RQs.  For \dseval
we use a subset of the evaluation subjects from a previous evaluation
of \Pynguin~\cite{LKF23}, which excluded projects
without type hints or with dependencies on native code.  We
furthermore removed projects that no longer work with Python~3.10,
Similarly, we merged the annotations from \toolname{typeshed} with the
respective modules' source code.  We considered only those third-party
libraries from \toolname{typeshed} that are marked to be fully
annotated, which is a decision by the \toolname{typeshed} contributors
and might not reflect the truth.  Additionally, we only used those
projects where we were able to find a suitable version of a project
from the metadata provided in \toolname{typeshed}.
Due to this curation and the limited dataset size
the results
may differ for modules that are not included in our dataset.

\subsubsection{Construct Validity}

We argued in \cref{sec:eval:setup:metrics} that using partial matching is more
suitable for test generation and for finding valid parameter types than exact
matches. However, this metric might not be the best choice for type inference
in general. The problem of what constitutes the correct type suggests
two answers:
(1) the developer-intended type and (2) the type that is valid~(no exception
during test case execution) for the SUT\@.
The developer-intended type might be given by a type annotation, but
might not be
valid or---in case of generics---imprecise. On the other hand the valid type
for the SUT might be a large set of types, while the SUT is never called with
most of them.
We argue that whether using partial matching or exact matching depends on the
context and goal, and partial matching is the better choice for test
generation.
Nevertheless, compared to exact matching, we might overestimate the performance
of the evaluated tools.
Finally, our refined mutation score counts only mutants killed by assertions,
excluding those killed by runtime exceptions, which we consider
incidental kills that do not reflect genuine functional
verification~(cf.~\cref{sec:eval:setup:metrics}).
This makes our reported mutation scores not directly comparable to
studies that use the standard mutation score, which includes
exception-based kills and therefore tends to yield higher values.

\subsection{RQ1: \TypeTracing Probability}\label{sec:rq1_results}

\begin{figure}[!t]
  \input{generated/img/prob-type-tracing-scatter-combined.pgf}
  \caption{\label{fig:rq1-coverage-mean-auc}%
    Mean branch coverage after \qty{600}{\second}
    on the \dstuning dataset for different probabilities of proxied execution.
  }
\end{figure}

\begin{figure}[!t]
  \centering
  \input{generated/img/prob-type-tracing-time-scatter-combined.pgf}
  \caption{\label{fig:rq1-time-mean-auc}%
    Time spent on type tracing's proxied executions
    out of a total
    of \qty{600}{\second} (lower is better)
    and number of algorithm iterations (higher is better)
    for different probabilities of proxied execution.
  }
\end{figure}

We executed \pynguin on \dstuning as outlined in
\cref{sec:eval:setup:procedure}, varying the probability of proxied
execution (see \cref{sec:approach:tracing}) from \qtyrange{0}{100}{\percent}
in \qty{5}{\percent} increments.
\Cref{fig:rq1-coverage-mean-auc} presents the mean branch coverage
achieved for each of these probabilities.
The results show that when the probability of proxied execution is
\qty{0}{\percent}, i.e., \typetracing is disabled, and when it is set to
\qty{100}{\percent}, the mean branch coverage is lowest.
The branch coverage is optimal for a probability of \qty{5}{\percent}
for the absolute final branch coverage after \qty{600}{\second},
thus providing the most effective balance between the benefits of dynamic
type inference and its associated costs.
When the probability is further increased, the mean coverage decreases.

Statistical analysis confirms that the \qty{5}{\percent} probability
setting is significantly better than the \qty{0}{\percent} setting~(no tracing)
on \exnum{\RQCoverageSigBetterFiveZero} modules, while only
\exnum{\RQCoverageSigWorseFiveZero} modules show a
significant decrease in
coverage~(mean \effectsize of \num{\RQEffectSizeFiveZero}). It is also
significantly better than
the \qty{100}{\percent} setting on
\exnum{\RQCoverageSigBetterFiveOneHundred} modules, while only
\exnum{\RQCoverageSigWorseFiveOneHundred} module shows a
significant decrease in coverage~(mean \effectsize of
\num{\RQEffectSizeFiveOneHundred}).  Compared
to the other probabilities, the performance at the \qty{5}{\percent}
setting was comparable to that observed for the
\qtyrange{10}{30}{\percent} range~(mean \effectsize of \num{0.51}).
At higher probabilities~(i.e., \qtyrange{35}{95}{\percent}), the mean
\effectsize increased up to \num{0.54}. This indicates more modules with
decreased coverage and fewer modules with increased coverage compared to the
\qty{5}{\percent} setting.
Consequently, for all subsequent RQs in this paper we utilise this
empirically-determined optimal probability of \qty{5}{\percent} for proxied
execution.

\begin{summary}{RQ1: \TypeTracing Probability}
  In our experiments, the best probability for adding proxied
  execution within the
  \typetracing strategy is \qty{5}{\percent}.
\end{summary}

\paragraph*{Discussion}
The results for RQ1 illustrate the trade-off between the information
gained from \typetracing's proxied executions and the computational
overhead these executions introduce.
Proxied executions furnish type information that aids test generation,
hence not using \typetracing (i.e., \qty{0}{\percent}) performs worse. However,
the proxied executions also introduce computational overhead due to the
duplicated executions~(cf.\ \cref{sec:approach:tracing}). As this overhead
consumes portions of the search budget, it potentially hinders overall
exploration; thus always using \typetracing (i.e., \qty{100}{\percent})
shows diminished coverage.
The computational cost associated with \typetracing is further
visualised in \cref{fig:rq1-time-mean-auc}, which depicts the mean
cumulative time spent on proxied executions and the resulting number
of algorithm iterations.
The overhead increases with higher tracing probability, which
directly curtails the available search budget and leads to a
corresponding decrease in the total number of generations explored.
At the upper bound of \qty{100}{\percent}, about a third of the
\qty{600}{\second} search budget (\qty{200}{\second}) is spent on
type tracing; this expenditure of resources markedly reduces the
number of iterations within the same time budget, thereby negatively
affecting coverage.
The optimal trade-off is achieved at a \qty{5}{\percent} probability,
as shown in \cref{fig:rq1-coverage-mean-auc}.
This suggests that a relatively small amount of dynamic \typetracing is
sufficient to provide benefits to \pynguin's search process,
while minimising the computational overhead
(shown in \cref{fig:rq1-time-mean-auc}).

Although the performance is overall optimal at \qty{5}{\percent},
there are
cases where \qty{0}{\percent} performed better are modules
where the overhead of even minimal tracing outweighed any marginal
type-discovery benefits, possibly because types were not the primary
challenge.
The performance of the \qty{5}{\percent} probability is remarkably
similar to probabilities in the \qtyrange{10}{30}{\percent}
range.  This suggests a plateau where increasing
the tracing probability in this lower range does not yield
substantially more useful type information relative to the added cost,
while the evolutionary search has sufficient time to converge despite
the additional overhead.
Beyond \qty{30}{\percent}, the detrimental effect of the execution
overhead becomes increasingly apparent, confirming that excessive
proxied execution consumes valuable search time without a proportional
gain in type-related insights, leading to a net loss in coverage.
The search budget is finite, and each proxied execution reduces
the number of possible test executions for the global search.

It is also noteworthy that for a majority of modules~(around \numrange{50}{60}
across comparisons), the coverage achieved was ``equal'' regardless of
the tracing probability.  As we are only using non-trivial modules
without type annotations~(see \cref{sec:eval:setup:subjects}), the
reason can neither be that the modules are trivial and thus easy to
cover, nor that the types are already given.  This leaves us with the
following possibilities: (1) the number of inferred possible types is
not substantially smaller than the general type search space and thus
randomly trying types is as good as using \typetracing or (2) the
bottleneck for coverage in these modules is not type-related but due
to other complexities like specific value generation or intricate
object states that \typetracing does not directly address.

While we conjecture that computational overhead is the likely main
reason why smaller probabilities of proxied executions yield better
results, this is not the only possible explanation for the observed
trend. An alternative explanation involves the diversity of generated
types: The more often the proxied execution is executed, the more type
information can be gathered by \typetracing. Thus, it might be the
case that having more type information available reduces the ratio of
randomly chosen types. This, in turn might have a negative impact on
the search, \eg, because randomly-chosen types might explore more
precondition checks, thus increasing overall coverage.
However, we conjecture that this is not the main reason, since the weights for
selecting types~(in particular $\omega_{\text{any}}$ for selecting any
  type and $\omega_{\text{union}}$ for selecting from the types inferred
by \typetracing) are not affected by the \typetracing
probability, always leaving the possibility of \Pynguin choosing a random type.

In essence, a \qty{5}{\percent} probability for proxied execution
appears to be optimal:
it is frequent enough to provide \pynguin with type information
to reduce the search space, but sparse enough to avoid impeding
the overall search exploration due to execution overhead.

\subsection{RQ2: Effects on Coverage}\label{sec:rq2_results}

\begin{figure}[!t]
  \input{generated/img/annotated-coverage-over-time.pgf}
  \caption{\label{fig:coverage-over-time}%
    Development of the branch coverage of the configurations over the
    algorithm runtime of \qty{600}{\second}.%
  }
\end{figure}

We executed \pynguin on \dseval as outlined in
\cref{sec:eval:setup:procedure}, with a probability of \qty{5}{\percent}
for proxied execution~(as determined in RQ1) and six different configurations:
\RandomTypes, \TypeHints, \RandomTypesTypeTracing,
\TypeHintsTypeTracing, \JediTypeHints, and \GptTypeHints.
To determine the performance gain in code coverage due to \typetracing,
we analysed its effect on the achieved branch
coverage compared to random type selection.
We therefore visualise the development of the mean
coverage per configuration over the algorithm runtime of
\qty{600}{\second} in \Cref{fig:coverage-over-time}.
The \RandomTypes approach reaches
\qty{\AnnotatedMeanCoverageRandomTypes}{\percent}, while respecting
the existing type hints~(\TypeHints) or adding \typetracing to
the random typing approach~(\RandomTypesTypeTracing)
yields \qty{\AnnotatedMeanCoverageTypeHints}{\percent} and
\qty{\AnnotatedMeanCoverageRandomTypesTypeTracing}{\percent}
branch coverage,
respectively.  Best performing is the combination of the existing type
annotations with our \typetracing technique~(\TypeHintsTypeTracing); it results
in \qty{\AnnotatedMeanCoverageTypeHintsTypeTracing}{\percent} branch
coverage.
Using either type hints or \typetracing improves the coverage achieved
by \pynguin compared to random type selection. Combining both leads to the best
results. The configurations based on \jedi and \gpt also show
improvements over \RandomTypes, yet they remain below the performance of
\TypeHints.

\begin{table}[!t]
  \caption{\label{tab:coverage-summary}%
    Comparison of the tested configurations (Treatment) against their
    baselines (Control) based on final branch coverage.
    The table reports the number of modules where each configuration
    performed (significantly) better, equal or (significantly) worse
    and the Vargha and Delaney \effectsize~effect size.
  }
  \begin{tabular}{@{}llrrrr@{}}
  \toprule
  \textbf{Treatment} & \textbf{Control} & \textbf{Better (sig)} &
  \textbf{Equal} & \textbf{Worse (sig)} & \textbf{Average \effectsize} \\
  \midrule
  TypeHints & RandomTypes & 168 (104) & 410 & 52 (13) & 0.552 \\
  RandomTypes-TypeTracing & RandomTypes & 204 (151) & 394 & 32 (10) & 0.577 \\
  TypeHints-TypeTracing & RandomTypes & 208 (175) & 395 & 27 (12) & 0.591 \\
  RandomTypes-TypeTracing & TypeHints & 128 (88) & 407 & 95 (46) & 0.526 \\
  TypeHints-TypeTracing & TypeHints & 153 (100) & 417 & 60 (15) & 0.549 \\
  GPT & Jedi & 93 (18) & 437 & 100 (25) & 0.497 \\
  TypeHints-TypeTracing & GPT & 201 (156) & 393 & 36 (12) & 0.586 \\
  RandomTypes-TypeTracing & GPT & 195 (134) & 393 & 42 (15) & 0.572 \\
  RandomTypes-TypeTracing & Jedi & 187 (140) & 396 & 47 (15) & 0.568 \\
  TypeHints-TypeTracing & Jedi & 197 (153) & 393 & 40 (13) & 0.583 \\
  \bottomrule
\end{tabular}

\end{table}

While the mean coverage scores show an improvement when using
type hints or \typetracing, this might not hold for all modules.
To differentiate the impact of \typetracing on coverage per module,
\cref{tab:coverage-summary} shows the number of (significantly)
better, equal, and
(significantly) worse modules, alongside the mean Vargha and Delaney
effect size~\effectsize, for the \TypeHints, \RandomTypesTypeTracing,
and \TypeHintsTypeTracing configurations against the \RandomTypes and
\TypeHints baselines.
The first three rows of the table demonstrate that when
compared to the \RandomTypes configuration, applying existing type
hints (\TypeHints), inferring them with \typetracing
(\RandomTypesTypeTracing), or using both in combination
(\TypeHintsTypeTracing) results in a (significantly) positive effect
on coverage for more modules than a negative one.
For instance, \TypeHintsTypeTracing improves coverage
in \exnum{\AnnotatedCoverageBetterTypeHintsTypeTracingRandomTypes} modules
while worsening it in only
\exnum{\AnnotatedCoverageWorseTypeHintsTypeTracingRandomTypes}. This is
supported by the Vargha and Delaney effect size~\effectsize
values, which are all above \num{0.5}, indicating a consistent
positive effect of using type information on coverage.

Given that both existing and inferred type hints improve coverage
over having no hints, the question arises whether
\typetracing is still needed when type hints are already present, and
vice versa.
The fourth row of \cref{tab:coverage-summary}, comparing
\RandomTypesTypeTracing against the \TypeHints configuration, shows
that inferring types with \typetracing has a (significantly) positive
effect on coverage in more modules
(\num{\AnnotatedCoverageBetterRandomTypesTypeTracingTypeHints}) than a
negative one
(\num{\AnnotatedCoverageWorseRandomTypesTypeTracingTypeHints}).
Furthermore, the final row shows that combining existing hints with
\typetracing (\TypeHintsTypeTracing) yields an even greater benefit
compared to using existing hints alone, with
\num{\AnnotatedCoverageBetterTypeHintsTypeTracingTypeHints} modules
improving versus only
\exnum{\AnnotatedCoverageWorseTypeHintsTypeTracingTypeHints} declining. The
effect sizes~\effectsize, being greater than \num{0.5} for both
comparisons, support these findings. Therefore,
we conclude that for improving coverage, \typetracing is not only
beneficial but superior to relying solely on existing type hints, and
the combination of both approaches is the most effective strategy.

Beyond comparing \typetracing to manual annotations,
we want to understand how our \typetracing approach compares to
other type inference approaches.
Thus, we evaluate the performance of \typetracing against other
state-of-the-art type inference methods, specifically \jedi (as a
representative of static analysis) and \gpt (as a representative of
LLM-based inference).
The last two rows of \cref{tab:coverage-summary} present these comparisons.
Comparing \GptTypeHints against \JediTypeHints yields an
average effect size of
\num{\AnnotatedCoverageMeanEffSizeGPTJEDI}, with a similar
number of modules showing improved
(\num{\AnnotatedCoverageBetterGPTJEDI}) versus declined
(\num{\AnnotatedCoverageWorseGPTJEDI}) coverage. This indicates that,
when used as type hints for test generation, both techniques provide
comparable utility despite their fundamentally different underlying
mechanisms and performance in terms of type quality (c.f.
\cref{sec:rq4_results}). However, when our combined approach
(\TypeHintsTypeTracing) is compared directly against \GptTypeHints, we
observe a substantial improvement. Our approach outperforms \gpt-based
hints on \num{\AnnotatedCoverageBetterTypeHintsTypeTracingGPT} modules,
with \num{\AnnotatedCoverageSigBetterTypeHintsTypeTracingGPT} of these
improvements being statistically significant. In contrast, it performs
worse on only \num{\AnnotatedCoverageWorseTypeHintsTypeTracingGPT}
modules (\num{\AnnotatedCoverageSigWorseTypeHintsTypeTracingGPT} significant).
The resulting average effect size of
\num{\AnnotatedCoverageMeanEffSizeTypeHintsTypeTracingGPT}
demonstrates that \typetracing
within \pynguin is more effective for
coverage exploration than the predictions provided by the LLM.
The comparison above compares \TypeHintsTypeTracing against
\GptTypeHints even though the former also carries developer-provided
type hints, which by themselves already account for the bulk of the
improvement over \RandomTypes. To control for this confound, we repeat
the comparison with \RandomTypesTypeTracing, which relies exclusively
on \typetracing without any developer-provided hints.
\RandomTypesTypeTracing still outperforms \GptTypeHints on
\num{\AnnotatedCoverageBetterRandomTypesTypeTracingGPT} modules
(\num{\AnnotatedCoverageSigBetterRandomTypesTypeTracingGPT} significant)
against \num{\AnnotatedCoverageWorseRandomTypesTypeTracingGPT} modules
where it performs worse
(\num{\AnnotatedCoverageSigWorseRandomTypesTypeTracingGPT} significant),
for an average effect size of
\num{\AnnotatedCoverageMeanEffSizeRandomTypesTypeTracingGPT}, and the
comparison against \JediTypeHints yields a near-identical pattern
(\num{\AnnotatedCoverageBetterRandomTypesTypeTracingJEDI} better,
  \num{\AnnotatedCoverageSigBetterRandomTypesTypeTracingJEDI}
  significant, versus
  \num{\AnnotatedCoverageWorseRandomTypesTypeTracingJEDI} worse,
  \num{\AnnotatedCoverageSigWorseRandomTypesTypeTracingJEDI} significant,
  average effect size
\num{\AnnotatedCoverageMeanEffSizeRandomTypesTypeTracingJEDI}). These
results confirm that the advantage of \typetracing over \jedi and
\gpt is not an artefact of the developer-provided hints. For completeness,
\cref{tab:coverage-summary} also reports \TypeHintsTypeTracing against
\JediTypeHints, where the same pattern holds
(\num{\AnnotatedCoverageBetterTypeHintsTypeTracingJEDI} better,
  \num{\AnnotatedCoverageSigBetterTypeHintsTypeTracingJEDI} significant,
  versus \num{\AnnotatedCoverageWorseTypeHintsTypeTracingJEDI} worse,
  \num{\AnnotatedCoverageSigWorseTypeHintsTypeTracingJEDI} significant,
  average effect size
\num{\AnnotatedCoverageMeanEffSizeTypeHintsTypeTracingJEDI}), indicating
that \jedi and \gpt are similarly outperformed by \typetracing
regardless of which baseline configuration is used.

\begin{table}[!t]
  \caption{\label{tab:coverage-overview}%
    Mean relative coverage values and algorithm iterations per
    configuration, along with the \effectsize effect size compared to
    the respective base configuration.
    The effect size columns also show the number of modules with a
    significant improvement~(\(\uparrow\)), significant
    decrease~(\(\downarrow\)), or no significant effect~(\(\approx\)).
    The baseline for
    \TypeHintsTypeTracing is \TypeHints, and for
    \RandomTypesTypeTracing it is \RandomTypes.
  }
  \begin{tabular}{@{}lrrrr@{}}
  \toprule
  \textbf{Configuration} & \textbf{Rel.~Cov.~(\%)} &
  \textbf{Rel.~Cov.~Eff.~Size} & \textbf{Mean~Alg.~It.} &
  \textbf{Mean~Alg.~It.~Eff.~Size} \\
  \midrule
  TypeHints & \num{82.86540663210825} &  & \num{9633.109472008491} &  \\
  TypeHints-TypeTracing & \num{87.75547977879515} &
  \num{0.5486795131811266}~(\(\uparrow 100 \downarrow 15 \approx
  515\)) & \num{8599.007165605095} &
  \num{0.3437758390346125}~(\(\uparrow 40 \downarrow 303 \approx 287\)) \\
  RandomTypes & \num{76.99989120782477} &  & \num{9805.65903307888} &  \\
  RandomTypes-TypeTracing & \num{85.5935664902601} &
  \num{0.5771611876286411}~(\(\uparrow 151 \downarrow 10 \approx
  469\)) & \num{8727.525777023444} &
  \num{0.33615271918581785}~(\(\uparrow 52 \downarrow 312 \approx 266\)) \\
  \bottomrule
\end{tabular}

\end{table}

\begin{table}[!t]
  \caption{\label{tab:configuration-match-coverage}%
    Mean relative coverage values for the four configurations against
    whether the inferred types result in \match, \mismatch, \missing, or
    \any. Overall includes all four groups.
    Additionally, the number of how many cases there are is given
    in braces. A case here refers to a single \match (or any other)
    of an inferred parameter type with a ground truth type.
  }
  \begin{tabular}{@{}llllll@{}}
  \toprule
  \textbf{Analysis Result} & \textbf{MATCH} & \textbf{MISMATCH} &
  \textbf{MISSING} & \textbf{ANY} & \textbf{Overall} \\
  \textbf{Configuration} &  &  &  &  &  \\
  \midrule
  RandomType & NaN (0) & NaN (0) & \qty{23.8}{\percent} (555) &
  \qty{73.3}{\percent} (1673) & \qty{61.0}{\percent} (2228) \\
  RandomType-TypeTracing & \qty{89.1}{\percent} (552) &
  \qty{81.1}{\percent} (413) & \qty{25.3}{\percent} (526) &
  \qty{62.9}{\percent} (654) & \qty{63.9}{\percent} (2145) \\
  TypeHints & \qty{79.1}{\percent} (1361) & \qty{61.4}{\percent} (52)
  & \qty{23.3}{\percent} (591) & \qty{68.1}{\percent} (149) &
  \qty{62.6}{\percent} (2153) \\
  TypeHints-TypeTracing & \qty{79.6}{\percent} (1354) &
  \qty{75.3}{\percent} (121) & \qty{25.9}{\percent} (602) &
  \qty{46.1}{\percent} (67) & \qty{63.2}{\percent} (2144) \\
  \bottomrule
\end{tabular}

\end{table}

Since Pynguin covers many branches always, regardless of the used
configuration, we additionally consider relative coverage, which removes the
influence of the non-changing branches~(either always covered or never
covered) and makes differences easier to observe.
\Cref{tab:coverage-overview} reports the mean relative coverage values
and number of performed iterations for the different configurations,
as well as the Vargha and Delaney effect size \effectsize between the
techniques.  The table compares the influence of \typetracing to the
respective baseline techniques: For \TypeHintsTypeTracing the baseline
is \TypeHints and for \RandomTypesTypeTracing the baseline is
\RandomTypes.
On \exnum{\CovBetterTypeHintsTypeTracingTypeHints} modules \TypeHintsTypeTracing
significantly improved compared to \TypeHints, while on
\exnum{\CovBetterRandomTypesTypeTracingRandomTypes} modules
\RandomTypesTypeTracing
significantly improved compared to \RandomTypes.
At the same time, the relative coverage
decreased significantly for only a few
modules~(\exnum{\CovWorseTypeHintsTypeTracingTypeHints}
and \exnum{\CovWorseRandomTypesTypeTracingRandomTypes}).
On most modules~(\exnum{\CovSameTypeHintsTypeTracingTypeHints} and
\exnum{\CovSameRandomTypesTypeTracingRandomTypes}) there is no
significant effect.
Altogether, there are more modules with a coverage increase than with
a coverage decrease.
The mean effect sizes~\effectsize of
\num{\CovEffectSizeTypeHintsTypeTracingTypeHints} and
\num{\CovEffectSizeRandomTypesTypeTracingRandomTypes} reflect this similarly.
Removing non-changing branches~(either always covered or never covered)
shifts the focus towards the remaining hard-to-cover branches.
The use of \typetracing therefore allows covering more of these
hard-to-cover branches across most modules.

This coverage improvement is a consequence of the
interdependence between type information and coverage: better type
information enables the test generator to construct more suitable
inputs and reach more branches; conversely, reaching more branches
exposes more runtime behaviour from which \typetracing can infer
better types.
To examine to what extent the observed improvement is caused by this
interdependence, we stratify the achieved coverage by the correctness of the
inferred types, using existing type annotations as ground truth.
We therefore differentiate between partial matches~(\match),
mismatches~(\mismatch), no type inferred~(\missing) or the \inlinelst{any}
type inferred~(\any).
After \pynguin finishes, we obtain both: the achieved coverage for each method
and the types that \pynguin inferred and used during this test generation
process for each method. As we know the ground truth for many types
in the \dseval
dataset based on existing code hints, we compare \pynguin's inferred
types against this ground truth, check for partial matching and assign each
resulting case to the coverage achieved for the method the type belongs to.
\Cref{tab:configuration-match-coverage} shows the mean coverage values per
configuration and type match
along with the number of cases in each group.
We observe that using type hints~(\TypeHints and
\TypeHintsTypeTracing)
achieves more \match cases~(\exnum{\TypeHintsMatchCount} and
\exnum{\TypeHintsTypeTracingMatchCount}) than not using type hints but solely
relying on \typetracing~(\exnum{\RandomTypesTypeTracingMatchCount}).
However, the achieved relative coverage for \match when using
\typetracing is higher~(\RandomTypesTypeTracingMatchCoverage and
\TypeHintsTypeTracingMatchCoverage)
than without it~(\TypeHintsMatchCoverage).
The distribution for the \mismatch case is similar to the \match case
in terms of achieved relative branch coverage,
with both \RandomTypesTypeTracing and \TypeHintsTypeTracing
outperforming \TypeHints.
In terms of \mismatch cases,
\RandomTypesTypeTracing~(\RandomTypesTypeTracingMismatchCount) comes
first, followed by
\TypeHintsTypeTracing~(\TypeHintsTypeTracingMismatchCount), and then
\TypeHints~(\TypeHintsMismatchCount).
This ordering aligns with expectations, as relying solely on
\typetracing is less likely to achieve a partial match compared to
using existing type hints.
Furthermore, inferring a wrong type~(\mismatch) leads to higher coverage than
not being able to infer a type at all~(\missing or \any). We observe this
for all but the \RandomTypes configuration, which is a special case
we will discuss later.
Overall, we gain two insights from this analysis. Firstly,
type hints inferred by \typetracing are more valuable for reaching
higher coverage than existing type hints even though they are less likely
to match manually annotated types.
Secondly, even types that are wrong according to partial matching
against manual type hints have a positive impact on the achieved coverage.

While the quality of the inferred types clearly influences the
resulting test coverage, another important dimension of the
evaluation is how much overhead \typetracing introduces,
as it requires executing the code. We therefore
analyse the effect of \typetracing on the search process by considering the
number of test generation iterations performed within the fixed time
budget, as reported in \Cref{tab:coverage-overview}.
In both cases, \TypeHintsTypeTracing compared to \TypeHints and
\RandomTypesTypeTracing compared to \RandomTypes, the number of mean
iterations that were performed within the search time of \qty{600}{\second}
are less (\num{\AnnotatedMeanAlgorithmIterationsTypeHintsTypeTracing} vs.
  \num{\AnnotatedMeanAlgorithmIterationsTypeHints} and
  \num{\AnnotatedMeanAlgorithmIterationsRandomTypesTypeTracing} vs.
\num{\AnnotatedMeanAlgorithmIterationsRandomTypes}).
Similarly, the number of significantly worse modules
(\exnum{\ItWorseTypeHintsTypeTracingTypeHints} and
\exnum{\ItWorseRandomTypesTypeTracingRandomTypes}), meaning more
iterations were needed, are more than the number of significantly
better modules (\exnum{\ItBetterTypeHintsTypeTracingTypeHints} and
\exnum{\ItBetterRandomTypesTypeTracingRandomTypes}). On most modules,
however, (\exnum{\ItSameTypeHintsTypeTracingTypeHints} and
\exnum{\ItSameRandomTypesTypeTracingRandomTypes}), there is no
significant effect. This is also reflected in the mean \effectsize of
\num{\ItEffectSizeTypeHintsTypeTracingTypeHints} and
\num{\ItEffectSizeRandomTypesTypeTracingRandomTypes} respectively.
While configurations with
\typetracing achieve fewer iterations in the same time due to duplicated
executions~(see \cref{sec:approach:tracing}), they still achieve higher
coverage than the configurations without \typetracing.
This shows that the benefits of incorporating type information
outweigh the drawback of requiring some duplicated executions.

\begin{summary}{RQ2: Effects on Coverage}
  \Typetracing has a positive and significant influence on the resulting branch
  coverage and can improve upon developer-provided type
  hints, yielding additional gains in branch coverage.
\end{summary}

\paragraph*{Discussion}

Augmenting \RandomTypes with \typetracing yields significantly
higher coverage values.  However, this is limited to a few modules: For
\RandomTypes,
we see a significant change in the coverage by adding \typetracing
for \num{\AnnotatedCoverageSigBetterRandomTypesTypeTracingRandomTypes} of the
\num{\AnnotatedNumModules} modules; using the developer-provided type
annotations as a
basis, \typetracing changes coverage significantly for
\num{\AnnotatedCoverageSigBetterTypeHintsTypeTracingTypeHints} of the
\num{\AnnotatedNumModules}
modules (see \cref{tab:coverage-summary}).
We have two conjectures why coverage does not significantly improve
on the other modules: either they are trivial to cover without
\typetracing~(\ie, even without type information \qty{100}{\percent} coverage
  is achieved, which is the case for
  \num{\AnnotatedFullCoverageModulesRandomTypes}
of \num{\AnnotatedNumModules} modules),
or they require a specific object state to cover a branch, which
\pynguin cannot generate, \eg, because it does not instantiate the required
object or object state.
\Cref{fig:example-known-type-fail} shows a code snippet where the
expected parameter type is known through the type annotation.
Without type annotations, \typetracing can successfully narrow the
search space to four possible types, including
\inlinelst{ast.Import}.  However, even though \pynguin then tries to
instantiate \inlinelst{ast.Import}, it fails to do so in such a way that
all required nested object states are set properly.  As a result,
the execution of \inlinelst{visit_Import} causes exceptions, thus
neither achieving full coverage nor producing information about the
return types.

Another possible explanation for the limited improvement in coverage
is that, even without \typetracing, \pynguin has a reasonable chance
of selecting the correct types randomly. Before \pynguin starts the search,
it analyses the SUT statically to collect all available types. This includes
Python built-in types, types from other imported modules, and types/classes
declared in the module under test. Thus, the number
of types that are considered without \typetracing is a finite set of types.
During test generation \pynguin needs to pass values for the parameters of
methods and functions when invoking them. For this, \pynguin needs
to create primitive values and instantiate objects of some type.
Without \typetracing, \pynguin has to choose this type at random,
which may lead to a low probability of selecting a suitable type. While
this might be sufficient for primitive types over many iterations, it is
far less likely for complex user-defined types and classes. For these
types, \typetracing provides a benefit by prioritising
relevant types, making their selection more efficient and increasing the
likelihood of reaching higher coverage faster.

\begin{figure}[!t]
  \centering
\begin{lstlisting}[
    style=nonnumberedlst,%
    frame=single,%
  ]
def visit_Import(
    self, node: ast.Import
) -> Union[ast.Import, ast.Try]:
    rewrite = self._get_matched_rewrite(node.names[0].name)
    if rewrite:
        return self._replace_import(node, *rewrite)
    return self.generic_visit(node)
\end{lstlisting}
  \caption{\label{fig:example-known-type-fail}%
    Simplified excerpt from
    \texttt{py-backwards.transformers.base} where the
    expected parameter type is known but \pynguin fails to instantiate the
    required object.
  }
\end{figure}

One way in which \typetracing provides benefits in addition to the
annotated type information~(\TypeHintsTypeTracing) is by suggesting
types that are not annotated but referenced in a routine's body.
\Cref{fig:example-wrong-type} provides a simple example: the checking
branch will only be covered with a very low probability when not using
\typetracing, because it is unlikely that \pynguin uses a
\inlinelst{str} as an input here.  \Typetracing records the explicit
type check and allows \pynguin to explicitly instantiate a
\inlinelst{str} because it knows from \typetracing that there is a
check for that type.  \Pynguin will not eliminate the type information
provided by the annotation in the code, \ie, the expected type
\inlinelst{int}; thus, it effectively records the union of both types
here.

\begin{figure}[!t]
  \centering
\begin{lstlisting}[
    style=nonnumberedlst,%
    frame=single,%
  ]
def foo(a: int):
    if isinstance(a, str):
        raise ValueError()
    ...
\end{lstlisting}
  \caption{\label{fig:example-wrong-type}%
    A code example showing a function that expects an \inlinelst{int} value
    and has an explicit guard check for the erroneous type \inlinelst{str}.
  }
\end{figure}

We attribute the superior performance of \typetracing compared to \gpt
and \jedi to the disadvantage of the static and one-shot
inference tools, which only analyse the module under test, whereas
\typetracing dynamically observes types across the entire project
during the search process, providing a more comprehensive view of
valid inputs.
Type information in the \textsc{TypeEvalPy} format is furthermore
often not fully qualified, which requires \pynguin to resolve an
unqualified type name emitted by \gpt or \jedi to a concrete class
before it can guide test generation. We describe the resolver
responsible for this step in \cref{sec:eval:setup:settings}. With this
resolver in place, \qty{\TypeResolutionGptRateAfter}{\percent} of
\gpt's and \qty{\TypeResolutionJediRateAfter}{\percent} of \jedi's
type predictions resolve successfully.

As reported in \cref{sec:rq2_results}, the achieved relative coverage
for \match when using
\typetracing is higher than without it.
One reason for this might be that for the goal of achieving coverage
the type hints inferred by \typetracing are more valuable than existing ones.
Another reason could be that the \match
cases for \RandomTypesTypeTracing with high coverage might be easier to achieve.
However, due to the \match cases for
\TypeHints~(\exnum{\TypeHintsMatchCount}) being
in a similar range to the ones for
\TypeHintsTypeTracing~(\exnum{\TypeHintsTypeTracingMatchCount}) this
is likely not
the only reason.
A combination of both is therefore the most plausible explanation.
In terms of \mismatch cases, the ordering aligns with expectations,
as relying solely on
\typetracing is less likely to achieve a partial match compared to
using existing type hints.

In \cref{tab:configuration-match-coverage}, the \inlinelst{NaN (0)}
values for the \RandomTypes
configuration under the \match/\mismatch categories are expected.
When \pynguin neither utilises existing type hints nor infers any types,
it cannot generate meaningful type information. In such cases,
it either assigns the generic \any type or provides no type at all.
For the \TypeHints configuration, \pynguin relies solely on existing
type hints without performing type inference. This suggests that we should
typically observe either a \match, when a type hint is present, or a
\missing/\any, when no type hint is available. However, the table
also includes \mismatch values, which may seem unexpected at first.
The reason for these \mismatch values is Python's support for
type aliases.
\Cref{fig:example-type-alias} provides a code example demonstrating
how a type alias
can lead to a mismatch in our analysis. Essentially, \pynguin
reports a type based
on its internal processing, which may differ from the alias used in
our dataset.
One might argue that both the alias and the actual type are
valid types for the parameter. However, when constructing our dataset,
we chose to treat the alias as the ground truth, because resolving
type aliases to their underlying types would cause distinct aliases to
collapse to the same type.
As a consequence, a tool that predicts the
underlying type would receive credit for matching the alias, even when
it has no knowledge of it. This would inflate the precision.

\begin{figure}[!t]
  \centering
\begin{lstlisting}[
    style=nonnumberedlst,%
    frame=single,%
  ]
Name = str

def greet(name: Name):
    return f"Hello, {name}!"
\end{lstlisting}
  \caption{\label{fig:example-type-alias}
    A code example showing a type alias along with a function that
    uses the alias.
    \Pynguin resolves the alias to the actual type \emph{str} and reports it
    while the ground truth is the alias \emph{Name}.
    This leads to a \emph{\mismatch} in the type inference.
  }
\end{figure}

\subsection{RQ3: Effects on Fault Finding}\label{sec:rq3_results}

\begin{table}[!t]
  \caption{\label{tab:mutation-scores}%
    Mean mutation scores for all six evaluated configurations:
    \RandomTypes, \TypeHints, \RandomTypesTypeTracing,
    \TypeHintsTypeTracing, \JediTypeHints, and \GptTypeHints.
  }
  \begin{tabular}{@{}lr@{}}
  \toprule
  \textbf{Configuration} & \textbf{Mean Mutation Score} \\
  \midrule
  GPT & \qty{14.8522}{\percent} \\
  Jedi & \qty{15.0832}{\percent} \\
  RandomTypes & \qty{14.7975}{\percent} \\
  RandomTypes-TypeTracing & \qty{15.4797}{\percent} \\
  TypeHints & \qty{14.8498}{\percent} \\
  TypeHints-TypeTracing & \qty{15.3118}{\percent} \\
  \bottomrule
\end{tabular}

\end{table}
We use the test suites generated as part of the experiment for RQ2 to answer
this question about their fault-finding capabilities.
The baseline configuration \RandomTypes achieves a mean
mutation score of
\qty{\AnnotatedMutationScoreMeanRandomTypes}{\percent}~(see
\cref{tab:mutation-scores}).
Incorporating the developer-provided type hints~(\TypeHints)
raises the mean mutation score only marginally, to
\qty{\AnnotatedMutationScoreMeanTypeHints}{\percent}.
Adding \typetracing~(\RandomTypesTypeTracing) to the
\RandomTypes configuration yields a mean mutation score of
\qty{\AnnotatedMutationScoreMeanRandomTypesTypeTracing}{\percent},
outperforming all other evaluated configurations, including
\TypeHints, \JediTypeHints~(\qty{\AnnotatedMutationScoreMeanJEDI}{\percent}),
and \GptTypeHints~(\qty{\AnnotatedMutationScoreMeanGPT}{\percent}).
This suggests that the dynamic information
gathered by \typetracing is effective for generating tests
that can detect faults.
Finally, using~\typetracing on top of the type-hint respecting
configuration (\TypeHintsTypeTracing) results in a mean mutation score
of \qty{\AnnotatedMutationScoreMeanTypeHintsTypeTracing}{\percent}.
While the mean score is slightly lower than that of \RandomTypesTypeTracing,
it still represents an improvement over the \TypeHints baseline.

\begin{table}[!t]
  \caption{\label{tab:mutation-summary}%
    Comparison of the tested configurations (Treatment) against their
    baselines (Control) based on mutation scores.
    The table reports the number of modules where each configuration
    performed (significantly) better, equal or (significantly) worse
    and the Vargha and Delaney \effectsize~effect size.
  }
  \begin{tabular}{@{}llrrrr@{}}
  \toprule
  \textbf{Treatment} & \textbf{Control} & \textbf{Better (sig)} &
  \textbf{Equal} & \textbf{Worse (sig)} & \textbf{Average \effectsize} \\
  \midrule
  TypeHints & RandomTypes & 125 (35) & 227 & 98 (17) & 0.520 \\
  RandomTypes-TypeTracing & RandomTypes & 145 (35) & 226 & 78 (10) & 0.522 \\
  TypeHints-TypeTracing & RandomTypes & 149 (51) & 215 & 88 (18) & 0.528 \\
  RandomTypes-TypeTracing & TypeHints & 123 (25) & 207 & 115 (24) & 0.501 \\
  TypeHints-TypeTracing & TypeHints & 128 (20) & 228 & 94 (8) & 0.513 \\
  GPT & Jedi & 84 (8) & 277 & 99 (12) & 0.496 \\
  TypeHints-TypeTracing & GPT & 146 (46) & 221 & 86 (13) & 0.524 \\
  RandomTypes-TypeTracing & GPT & 140 (34) & 224 & 87 (12) & 0.520 \\
  RandomTypes-TypeTracing & Jedi & 129 (35) & 225 & 95 (12) & 0.516 \\
  TypeHints-TypeTracing & Jedi & 144 (40) & 217 & 91 (13) & 0.522 \\
  \bottomrule
\end{tabular}

\end{table}

However, the differences in the mean mutation scores are rather small,
and may partly reflect data noise rather than genuine effects.
To distinguish between the two, we consider
\cref{tab:mutation-summary}, which reports the
number of~(significantly) better, equal, and~(significantly) worse
modules and the mean Vargha and Delaney effect size~\effectsize~\cite{VD00}
for the \TypeHints, \RandomTypesTypeTracing, and
\TypeHintsTypeTracing configurations against the
\RandomTypes and \TypeHints baselines.
The first three rows of the table show that compared to the
\RandomTypes configuration, adding \typetracing~(\RandomTypesTypeTracing)
or combining it with existing type hints~(\TypeHintsTypeTracing) has a
positive effect on the mutation score of more modules than a negative
one. For the \TypeHints configuration slightly more modules decline than
improve~(\num{\AnnotatedMutationWorseTypeHintsRandomTypes} against
\num{\AnnotatedMutationBetterTypeHintsRandomTypes}), yet more than twice
as many improve significantly as decline
significantly~(\num{\AnnotatedMutationSigBetterTypeHintsRandomTypes}
against \num{\AnnotatedMutationSigWorseTypeHintsRandomTypes}).
The Vargha and Delaney effect size~\effectsize values of
slightly above \num{0.5} indicate a small positive effect
of using type hints and/or \typetracing on the mutation score.
While both using existing type hints and inferring them with
\typetracing are superior to using no type hints at all, the question
remains which of the two approaches is more effective.  The fourth
row of \cref{tab:mutation-summary} with an average \effectsize~value of
\num{\AnnotatedMutationMeanEffSizeRandomTypesTypeTracingTypeHints}
shows that the \TypeHints configuration with
\typetracing~(\RandomTypesTypeTracing)
and without \typetracing~(\TypeHints)
perform almost equally well.

\begin{figure}[!t]
  \centering
\begin{lstlisting}[
    style=nonnumberedlst,%
    frame=single,%
  ]
class cached_property:
    def __init__(self, func):
        self.__doc__ = getattr(func, "__doc__")
        self.func = func

    def __get__(self, obj: Any, cls):
        if obj is None:
            return self

-       if asyncio.iscoroutinefunction(self.func):
+       if not asyncio.iscoroutinefunction(self.func):
            return self._wrap_in_coroutine(obj)

        value = obj.__dict__[self.func.__name__] = self.func(obj)
        return value
\end{lstlisting}
  \caption{\label{fig:cached-property-get}%
    Simplified excerpt from \inlinelst{flutils.decorators}
    with a mutated version of
    \inlinelst{cached_property.}\allowbreak\inlinelst{\__get__}.
  }
\end{figure}

When using
\typetracing in combination with existing type
hints~(\TypeHintsTypeTracing), there are more modules with
a~(significantly) positive effect on the mutation score than with a
negative one, and the average effect size~\effectsize is also slightly
above \num{0.5}.
We conclude that type hints are as effective as \typetracing in
improving mutation scores, and the combination of both
approaches is the most effective strategy.

Finally, we compare our \typetracing approach against other sources of
type information, specifically type hints inferred by the static
analysis tool \jedi and the LLM \gpt. The last two rows of
\cref{tab:mutation-summary} show these comparisons. When comparing the
two external tools, \GptTypeHints and \JediTypeHints perform
similarly, with an average effect size of
\num{\AnnotatedMutationMeanEffSizeGPTJEDI}. However, our
proposed approach of combining developer-provided type hints with
\typetracing~(\TypeHintsTypeTracing) outperforms the
configuration using \gpt-inferred type hints~(\GptTypeHints).
\TypeHintsTypeTracing is better on
\num{\AnnotatedMutationBetterTypeHintsTypeTracingGPT} modules
(with \num{\AnnotatedMutationSigBetterTypeHintsTypeTracingGPT}
being significant) and worse on only
\num{\AnnotatedMutationWorseTypeHintsTypeTracingGPT} modules
(with \num{\AnnotatedMutationSigWorseTypeHintsTypeTracingGPT} being
significant),
achieving an
average effect size of
\num{\AnnotatedMutationMeanEffSizeTypeHintsTypeTracingGPT}.
This suggests that the dynamic
information captured during the search process is more effective for
guiding test generation toward fault-detecting inputs than the
predictions from existing inference tools.
As with coverage, the \TypeHintsTypeTracing vs.\ \GptTypeHints
comparison above conflates the effect of \typetracing with that of
the developer-provided hints. Repeating the comparison with
\RandomTypesTypeTracing, which excludes those hints, still favours
\typetracing, though by a narrower margin than for coverage,
consistent with the smaller mutation-score effects observed
throughout this section. \RandomTypesTypeTracing improves the
mutation score on \num{\AnnotatedMutationBetterRandomTypesTypeTracingGPT}
modules against \GptTypeHints
(\num{\AnnotatedMutationSigBetterRandomTypesTypeTracingGPT}
significant) and
\num{\AnnotatedMutationBetterRandomTypesTypeTracingJEDI} against
\JediTypeHints
(\num{\AnnotatedMutationSigBetterRandomTypesTypeTracingJEDI}
significant), against
\num{\AnnotatedMutationWorseRandomTypesTypeTracingGPT} and
\num{\AnnotatedMutationWorseRandomTypesTypeTracingJEDI} modules where
it performs worse, respectively, with average effect sizes of
\num{\AnnotatedMutationMeanEffSizeRandomTypesTypeTracingGPT} and
\num{\AnnotatedMutationMeanEffSizeRandomTypesTypeTracingJEDI}. The
additional comparison of \TypeHintsTypeTracing against
\JediTypeHints shows the same pattern as against \GptTypeHints, with
\num{\AnnotatedMutationBetterTypeHintsTypeTracingJEDI} modules
improving (\num{\AnnotatedMutationSigBetterTypeHintsTypeTracingJEDI}
significant) against
\num{\AnnotatedMutationWorseTypeHintsTypeTracingJEDI} declining
(\num{\AnnotatedMutationSigWorseTypeHintsTypeTracingJEDI} significant)
and an average effect size of
\num{\AnnotatedMutationMeanEffSizeTypeHintsTypeTracingJEDI}. Overall,
\typetracing retains its advantage over both external tools on
mutation score even without the benefit of developer-provided hints.

\begin{summary}{RQ3: Effects on Fault Finding}
  The types used during test generation barely influence the mutation
  score and thereby the fault-finding capabilities of generated
  tests. Using as much type information as possible---including
  existing type hints and type information gathered by our
  \typetracing approach---yields the best results nevertheless.
\end{summary}

\paragraph*{Discussion}

One possible reason for \typetracing not often improving the mutation score
and on certain modules decreasing the mutation score significantly
is that \typetracing biases test generation toward
types observed during actual executions,
which represent
non-exceptional usage of the module under test.
Mutations that require inputs of a type not observed by \typetracing
during real executions---%
such as a non-callable value for a parameter that \typetracing
only ever saw initialised with actual functions---%
may therefore be
harder to detect with \typetracing
than without it.
\Cref{fig:cached-property-get} illustrates this with an excerpt from
\texttt{flutils.decorators}.
The method
\inlinelst{cached_property.}\allowbreak\inlinelst{\__get__} checks whether
\inlinelst{self.func} is a coroutine function and, if so, delegates
to \inlinelst{_wrap_in_coroutine}.
A mutation operator negates this condition, changing
\inlinelst{if asyncio.iscoroutinefunction(self.func)} to
\inlinelst{if not asyncio.iscoroutinefunction(self.func)}, so that
the method now wraps \emph{non}-coroutine functions and falls
through to direct invocation for actual coroutines.
To detect this mutant, a test must supply a non-callable
\inlinelst{func}---such as \inlinelst{None}, a string, or a
boolean---to the constructor
and call \inlinelst{__get__} with a non-\texttt{None}
\inlinelst{obj}: on the original code this triggers a
\inlinelst{TypeError} when attempting to call the non-callable,
which the test catches with \inlinelst{pytest.raises(TypeError)};
the mutant, however, routes through \inlinelst{_wrap_in_coroutine}
and does not raise the error, causing the assertion to fail.
Without \typetracing, \pynguin freely passes non-callable values as
the \inlinelst{func} argument to \inlinelst{cached_property},
reliably exercising this divergence and killing the mutant in
\exnum{30} out of \exnum{30} runs.
\TypeHintsTypeTracing, however, observes that \inlinelst{cached_property}
is initialised with actual methods during real executions---%
because \inlinelst{__init__} accesses \inlinelst{func.__doc__} via
\inlinelst{getattr}, \typetracing records method objects as valid
argument types for \inlinelst{func}---and therefore
focuses on callable \inlinelst{func} arguments;
tests using a callable \inlinelst{func} do not trigger the
\inlinelst{TypeError} divergence,
and the mutant is killed in only \exnum{10} out of \exnum{30}~runs.

We have two conjectures for why \typetracing increases the mutation
score on many modules: higher coverage and fewer exception-throwing
tests.
Higher coverage allows us to cover more mutants and thus enables their killing.
The additional type information from \typetracing allows us to generate better
inputs, which may lead to fewer exception-throwing tests due to
wrong input types.
Consider for example \cref{fig:is_com_url}. The method \inlinelst{is_com_url}
requires an argument \inlinelst{url}. \Pynguin without \typetracing
needs to try
out all built-in types including various generic types. For most types
an \inlinelst{AttributeError} is raised during test execution, which
allows covering neither the True nor the False branch. At some point
\Pynguin will by chance
choose a string as input for the method and it will cover either of
the two branches.
With \typetracing \Pynguin requires exactly two iterations: In the
first one it
will choose an arbitrary type. From this it will learn that the
method requires
an object with an \inlinelst{endswith} attribute. In the second
iteration it will
choose a string, which satisfies the requirement of having an
\inlinelst{endswith}
attribute and the test will cover one of the branches.
Finally, \typetracing allows for a better usage of the resources
for the test generation.

Overall, the changes of the mutation score by adding \typetracing to the
test-generation process are small but still beneficial. The mean
mutation scores are quite low for all configurations~(see
\cref{tab:mutation-scores}).  These small numbers are in line with
previous research using a similar tool chain~(\pynguin and
\textsc{cosmic-ray})~\cite{TMG22,EMB+24}.  Thus, we argue that the
low mutation scores are an inherent problem of \pynguin's approach
to assertion generation and independent of type information.  While
improving the assertion generation is not part of the scope of this
work, it imposes an interesting challenge for future research.

Despite these limitations in assertion generation,
our evaluation demonstrates that dynamic type information provides an
advantage over static and LLM-based alternatives.
As for branch coverage, we attribute the superior performance of
\typetracing compared to \gpt and \jedi to the same reason identified
in \cref{sec:rq2_results}, namely the broader scope of dynamic type
observation across the entire project, compared to the module-only
scope of static and one-shot inference. We resolve \gpt's and
\jedi's type predictions for this analysis using the same type-name
resolver described in \cref{sec:eval:setup:settings}.

\begin{figure}[!t]
  \centering
\begin{lstlisting}[
    style=nonnumberedlst,%
    frame=single,%
  ]
def is_com_url(url):
    if url.endswith(".com"):
        return True
    else:
        return False
\end{lstlisting}
  \caption{\label{fig:is_com_url}
    A method requiring an object with an \inlinelst{endswith}
    attribute, such as a \inlinelst{string}, as input to not raise
    an \inlinelst{AttributeError}.
  }
\end{figure}

\subsection{RQ4: Quality of Inferred Types}\label{sec:rq4_results}

\begin{table}[!t]
  \caption{\label{tab:type-matches-params}%
    Type inference results for the parameters of the tools from
    \toolname{TypeEvalPy}
    and \typetracing. The tools are sorted alphabetically; we highlighted the
    best-performing tool (\gpt) and our own (\pynguin).%
  }
  \begin{tabular}{l S S S S}
  \toprule
  \textbf{Tool Name} & {\textbf{Matches}} & {\textbf{Precision}} &
  {\textbf{Recall}} & {\textbf{\Fone score}} \\
  \midrule
  \rowcolor{lightgray}
  \gpt & 2847 & 0.915140 & 0.831484 & 0.871308 \\
  \toolname{HeaderGen} & 12 & 0.107143 & 0.003505 & 0.006787 \\
  \toolname{HiTyper} & 190 & 0.510753 & 0.055491 & 0.100105 \\
  \toolname{HiTyper-DL} & 190 & 0.510753 & 0.055491 & 0.100105 \\
  \toolname{Jedi} & 1186 & 0.582801 & 0.346379 & 0.434512 \\
  \rowcolor{lightgray}
  \pynguin & 634 & 0.203990 & 0.185164 & 0.194121 \\
  \quac & 320 & 0.501567 & 0.093458 & 0.157558 \\
  \toolname{RightTyper} & 11 & 0.379310 & 0.003213 & 0.006371 \\
  \toolname{Scalpel} & 275 & 0.058140 & 0.080315 & 0.067452 \\
  \bottomrule
\end{tabular}

\end{table}

\begin{table}[!t]
  \caption{\label{tab:type-matches-returns}%
    Type inference results for the return values of the tools from
    \toolname{TypeEvalPy}
    and \typetracing. The tools are sorted alphabetically; we highlighted the
    best-performing tool (\gpt) and our own (\pynguin).%
  }
  \begin{tabular}{l S S S S}
  \toprule
  \textbf{Tool Name} & {\textbf{Matches}} & {\textbf{Precision}} &
  {\textbf{Recall}} & {\textbf{\Fone score}} \\
  \midrule
  \rowcolor{lightgray}
  \gpt & 1945 & 0.917020 & 0.889753 & 0.903181 \\
  \toolname{HeaderGen} & 11 & 0.051887 & 0.005032 & 0.009174 \\
  \toolname{HiTyper} & 239 & 0.656593 & 0.109332 & 0.187451 \\
  \toolname{HiTyper-DL} & 239 & 0.656593 & 0.109332 & 0.187451 \\
  \toolname{Jedi} & 910 & 0.163610 & 0.416285 & 0.234899 \\
  \rowcolor{lightgray}
  \pynguin & 889 & 0.357172 & 0.406679 & 0.380321 \\
  \quac & 233 & 0.581047 & 0.106587 & 0.180131 \\
  \toolname{RightTyper} & 7 & 0.250000 & 0.003202 & 0.006323 \\
  \toolname{Scalpel} & 680 & 0.088553 & 0.311070 & 0.137861 \\
  \bottomrule
\end{tabular}

\end{table}

To evaluate the quality of the inferred types, we executed \pynguin on \dseval
without type hints. Additionally, as outlined in
\cref{sec:eval:setup:procedure}, we executed six further type-inference tools
that are part of \toolname{TypeEvalPy}~\cite{VSW+23}
and \toolname{QuAC}~\cite{wu_quac_2024},
which we added to \toolname{TypeEvalPy}.
\Cref{tab:type-matches-params,tab:type-matches-returns} show the type
inference results for the tools from \toolname{TypeEvalPy} and
\typetracing in terms of the number of partial matches, precision,
recall, and \Fone score, for parameters and return types
respectively.
Among the non-LLM tools,
\jedi and \pynguin emerge as the most effective.
Specifically, \jedi achieves the best \Fone score for
parameter types (\num{\TypeInfParamjediFOneScore}),
where it outperforms other tools including \pynguin.
However,
for return types,
\pynguin (\num{\TypeInfReturnpynguinFOneScore}) is the most effective
non-LLM tool,
also outperforming \jedi (\num{\TypeInfReturnjediFOneScore}).
These results reflect different trade-offs in inference strategies.
For instance, \pynguin generally achieves a higher recall than \hityper
for both parameter
(\num{\TypeInfParampynguinRecall} vs.
\num{\TypeInfParamhityperRecall}) and return
types (\num{\TypeInfReturnpynguinRecall} vs.
\num{\TypeInfReturnhityperRecall}). However, this comes at the cost of
lower precision for parameter types, where \pynguin's precision
(\num{\TypeInfParampynguinPrecision}) is lower than that of both
\hityper (\num{\TypeInfParamhityperPrecision}) and \jedi
(\num{\TypeInfParamjediPrecision}). Interestingly, while \jedi is
highly precise for parameters, its precision for return types
is lower (\num{\TypeInfReturnjediPrecision}), whereas \pynguin
has a higher precision for return types (\num{\TypeInfReturnpynguinPrecision}).
The discrepancy between \hityper's comparatively high precision and very
low recall is because the tool makes predictions for only a small
fraction of all parameter and return types.
While the ground truth covers all manually annotated elements
(\num{\TypeInfParamhityperTotalGT}~parameters,
\num{\TypeInfReturnhityperTotalGT}~return types),
\hityper produces type predictions for far fewer
(\num{\TypeInfParamhityperTotalInferred}~and
\num{\TypeInfReturnhityperTotalInferred}, respectively).
Of those predictions, roughly half are correct---yielding moderate to
high precision---but the vast majority of ground-truth types receive no
prediction at all, resulting in very low recall.
This may lead to an overestimation of the tool's practical usefulness
for annotating unannotated codebases.

Compared to \gpt
(parameter: \num{\TypeInfParamgptFiveFourminiFOneScore}, return:
\num{\TypeInfReturngptFiveFourminiFOneScore}), \pynguin and all other tools
perform worse in terms of \Fone score, precision, and recall.
There are two possible reasons that mainly
influence the high performance of \gpt:
there is a reasonable chance of data leakage, i.e., the code in our dataset
has been used during the training of the LLM\@. We only removed the type
annotations for our experiments; thus, they could have been part of the LLM's
training data, which would allow the LLM to basically re-add the annotations.
Another reason could be that \gpt is able to
predict the types of a focal method based on the variable names, code
structure, or other information that the model retrieves from our prompt.
In both cases a test generator profits from having the type information
available~\cite{VPK04,CGO15}. Having a tool that can predict types with high
accuracy suggests that integrating \gpt into \pynguin
is a promising direction for future work, provided the LLM-predicted types
can be reliably resolved and matched to their fully qualified counterparts.

\begin{summary}{RQ4: Quality of the Types}
  The quality of the types inferred by \typetracing is comparable to
  and often better than other non-LLM state-of-the-art tools of the
  \toolname{TypeEvalPy} benchmark. \gpt outperforms all tools.
\end{summary}

\paragraph*{Discussion}
\begin{figure}[!t]
  \centering
  \input{generated/img/ds1_coverage_vs_type_quality.pgf}
  \caption{\label{fig:coverage_vs_type_quality}%
    Type Score (\match over \all) against average code coverage. The
    correlation line is shown in red.
    The correlation is significant with a Pearson's correlation
    coefficient~\( r = \num{\AnnotatedCorrelationCoverageType} \)~(\( p =
    \num{\AnnotatedCorrelationCoverageTypeP} \)).
  }
\end{figure}
\begin{figure}[!t]
  \centering
\begin{lstlisting}[
    style=nonnumberedlst,%
    frame=single,%
  ]
    def execute(self, lexer:Lexer):
        lexer.skip()
\end{lstlisting}
  \caption{\label{fig:full_coverage_no_type}%
    A method that is fully covered without requiring any type information.
  }
\end{figure}
The quality of type information inferred by \pynguin is directly
influenced by the test generation's ability to achieve code
coverage. For return types, the \typetracing
approach requires a successful, exception-free invocation of a
routine to collect a return type.
For parameter types, \typetracing allows \pynguin to infer more and
better parameter types as more code is covered. Deeply nested
branches may contain statements that provide additional constraints
on parameter types, thereby limiting the number of possible types.

To investigate the correlation we plot the type score (\match over
\all) of both parameter and return inference results, against the
final average code coverage for our best-performing configuration in
\cref{fig:coverage_vs_type_quality}.  The analysis reveals a positive
and significant Pearson's correlation coefficient~\( r =
\num{\AnnotatedCorrelationCoverageType} \)~(\( p =
\num{\AnnotatedCorrelationCoverageTypeP} \)). This supports the hypothesis
that achieving higher code coverage has a direct, positive impact on
the quality of inferred types. Despite this correlation, edge cases
exist; for instance, \cref{fig:full_coverage_no_type} shows a trivial
method \inlinelst{execute(self, lexer:Lexer)} with a single line of
functional code in its method body. It is fully covered without
needing specific types, so \pynguin's type inference process is not
triggered. While such edge cases exist, the overall correlation
confirms that coverage drives type quality, which is best understood
through a qualitative analysis of examples for successes and failures.

\begin{figure}[!t]
  \centering
\begin{lstlisting}[
    style=nonnumberedlst,%
    frame=single,%
  ]
def parse(self, filename=None):
	"""Load an applications.menu file."""
	# ... resolve and validate the filename ...
	tree = etree.parse(filename)
	menu = self.parse_menu(tree.getroot(), filename)
	# ... handle moves, post-parse, generate, sort ...
	return menu
\end{lstlisting}
  \caption{\label{fig:typeshed-only-pynguin-match}
    \Pynguin is the only tool that infers the correct return type
    \inlinelst{Menu} of the method \inlinelst{parse}.
  }
\end{figure}

\begin{figure}[!t]
  \centering
\begin{lstlisting}[
    style=nonnumberedlst,%
    frame=single,%
  ]
    def visit_Import(self, node: ast.Import) -> Union[ast.Import, ast.Try]:
        rewrite = self._get_matched_rewrite(node.names[0].name)
        if rewrite:
            return self._replace_import(node, *rewrite)

        return self.generic_visit(node)
\end{lstlisting}
  \caption{\label{fig:typeshed-pyn-match-gpt-not}%
    \gpt is not able to infer the correct type for the
    \textit{node} parameter of the \textit{visit\_import}
    method, while \pynguin infers the correct type \inlinelst{ast.Import}
    among other possible types.
  }
\end{figure}

\begin{figure}[!t]
  \centering
\begin{lstlisting}[
    style=nonnumberedlst,%
    frame=single,%
  ]
def loguniform(a, b):
    return _loguniform(10**a, 10 ** (a + b))
\end{lstlisting}
  \caption{\label{fig:flapytype-late-binding}%
    \Pynguin is the only tool that infers the correct return type
    \inlinelst{rv_continuous_frozen} of \inlinelst{loguniform}, which
    is determined by \toolname{SciPy} only at runtime.
  }
\end{figure}

The successes of \typetracing correspond directly to the
scenarios that we identified in \cref{sec:introduction} as
drawbacks of static type inference. In each of these
scenarios the relevant type information becomes available only at
runtime, so \typetracing recovers it by observing the objects that
occur during covering executions, whereas a tool that reasons over the
source alone cannot.

The first scenario is that a routine carries no declared types, so its
types are fixed only by the objects that reach it at runtime.
\Cref{fig:typeshed-only-pynguin-match} shows such a case. The return type
of \inlinelst{parse} is determined solely by the object that
\inlinelst{parse_menu} constructs, so a tool that does not execute the
code has to guess it. \Pynguin is the only tool that reports the
correct type~(\inlinelst{Menu}), whereas \gpt infers
\inlinelst{XmlMenuBuilder} and all other tools infer nothing. This
ability to outperform other tools is not an isolated incident. In
\num{\TypeshedPynMoreAnnotations} of \num{\TypeshedTotal} cases,
\pynguin infers a wider set of possible types than the other
\toolname{TypeEvalPy} tools, and in \num{\TypeshedOnlyPynguinCorrect}
cases only \pynguin infers the correct type, often because the other
tools report no type at all or cannot resolve project-specific classes.

The second scenario is duck typing, where a routine constrains its
parameters only structurally rather than through a declared interface.
In \cref{fig:typeshed-pyn-match-gpt-not}, the body of
\inlinelst{visit_Import} constrains its parameter \inlinelst{node} only
through the access \inlinelst{node.names[0].name}. Over multiple
iterations, \typetracing infers from this access that \inlinelst{node}
must expose a \inlinelst{names} attribute, that the \inlinelst{names}
value must be an \inlinelst{iterable}, and that its elements must in
turn expose a \inlinelst{name} attribute. This structural constraint
alone does not identify a single class.
\inlinelst{ast.ImportFrom} and \inlinelst{ast.Import} both satisfy it.
\Gpt predcits \inlinelst{ast.ImportFrom}. \Pynguin searches the project's
loaded modules for a class that satisfies the observed constraints,
selects \inlinelst{ast.Import}, and uses it to generate a test that
achieves higher coverage and yields the correct type. \Typetracing
resolves the ambiguity because it observes which object actually
reaches the parameter during execution, rather than reasoning over the
module text alone, whereas \gpt lacks the execution context to resolve
such project-specific dependencies even if the code was part of its
training data. The same example also illustrates why resolving type
\emph{names} is hard. The ground truth records the name under which the
enclosing module refers to the class~(\inlinelst{ast.Import}), whereas
\typetracing reports the module in which the class is actually
defined~(\inlinelst{typed_ast._ast3.Import}).

The third scenario is late binding, where the routine that a call
ultimately reaches is determined only at runtime.
\Cref{fig:flapytype-late-binding} shows a module in which
\inlinelst{loguniform} forwards to \inlinelst{_loguniform}, imported
from \toolname{SciPy} as \inlinelst{scipy.stats.loguniform}.
Syntactically this is an ordinary function call, yet
\inlinelst{scipy.stats.loguniform} is not a function, but an instance
of a class that \toolname{SciPy} creates at
import time. Calling it therefore dispatches to
\inlinelst{rv_generic.__call__}, which forwards to \inlinelst{freeze},
which selects its return type through a runtime \inlinelst{isinstance}
check on \inlinelst{self} and returns an
\inlinelst{rv_continuous_frozen}. None of these three steps is visible
at the call site, and a static tool would have to model instance
creation, \inlinelst{__call__} dispatch, and a value-dependent branch
inside a third-party library to recover the return type. \Typetracing
simply observes the returned object and is the only evaluated tool that
infers this type.

\begin{figure}[!t]
  \centering
\begin{lstlisting}[
    style=nonnumberedlst,%
    frame=single,%
  ]
def convert_path(pathname: str) -> str:
    """Return 'pathname' as a name that will work on the native filesystem,
    i.e. split it on '/' and put it back together again using the current
    directory separator.  Needed because filenames in the setup script are
    always supplied in Unix style, and have to be converted to the local
    convention before we can actually use them in the filesystem.  Raises
    ValueError on non-Unix-ish systems if 'pathname' either starts or
    ends with a slash.
    """
    if os.sep == '/':
        return pathname
    if not pathname:
        return pathname
    if pathname[0] == '/':
        raise ValueError("path '%s' cannot be absolute" % pathname)
    if pathname[-1] == '/':
        raise ValueError("path '%s' cannot end with '/'" % pathname)

    paths = pathname.split('/')
    while '.' in paths:
        paths.remove('.')
    if not paths:
        return os.curdir
    return os.path.join(*paths)
\end{lstlisting}
  \caption{
    \Pynguin fails to overcome the input verification of
    \inlinelst{convert_path}
    \inlinelst{if os.sep == '/'} which is not dependent on the
    input type or value.
  }
  \label{fig:typeshed-no-type-low-coverage}
\end{figure}

\begin{figure}[!t]
  \centering
\begin{lstlisting}[
    style=nonnumberedlst,%
    frame=single,%
  ]
    def optimizeConfigs(self, interpreter:ATNSimulator):
        if self.readonly:
            raise IllegalStateException("This set is readonly")
        if len(self.configs)==0:
            return
        for config in self.configs:
            config.context = interpreter.getCachedContext(config.context)
\end{lstlisting}
  \caption{
    \Pynguin infers \inlinelst{None} as the return type of the
    \textit{optimizeConfigs} method, while the ground truth is
    \inlinelst{ASTNSimulator}.
  }
  \label{fig:typeshed-pynguin-mismatch}
\end{figure}

However, reliance on coverage also introduces drawbacks. When
\pynguin fails to achieve sufficient coverage, type inference
quality degrades. Failures typically occur when a method under test
is inside a class that \pynguin cannot instantiate, or when it
contains logic that is not dependent on test inputs, such as
environment checks. For instance, in
\cref{fig:typeshed-no-type-low-coverage}, the guard \inlinelst{if
os.sep == '/'} is dependent on the operating system rather than
the test inputs, making the branch unreachable and preventing \pynguin
from exploring the method's core logic.
Another concrete example of such a failure is shown in
\cref{fig:typeshed-pynguin-mismatch}. Here, the core logic of the
\inlinelst{optimizeConfigs} method is protected by checks that depend on how
its class is initialised. \Pynguin is unable to synthesise a
suitable class instance to satisfy these checks, and thus \inlinelst{None}
as input behaves like an object of type \inlinelst{ASTNSimulator}.
Thus, \Pynguin infers \inlinelst{None} as type for the parameter
\inlinelst{interpreter} instead of the ground truth \inlinelst{ASTNSimulator},
directly linking the failure in test generation to the failure in
type inference.

These successes and failures raise a more fundamental question about
what constitutes a \enquote{correct} type in a dynamic context. We
designed \typetracing to find any parameter type that allows for
successful execution, which may not always align with the single type
a developer intended.  For example, while a developer might write a
function expecting integers, Python may still successfully execute if
floats are passed instead: \inlinelst{add(1.0, 2.0)} works even if the
original intent was \inlinelst{add(1, 2)}.  We argue this is often
beneficial.  Discovering that a function correctly handles unexpected
but valid input types can reveal guard-checking branches (as in
\cref{fig:example-known-type-fail}) or suggest to a programmer that a
more abstract parameter type could be used, thereby improving the
code's design.
This also implies that our RQ4 metrics are pessimistic about
\typetracing: cases where \typetracing infers a valid but
developer-unintended type are counted as mismatches against the ground
truth, even though the inferred type may be equally correct or more
general.

\begin{figure}[!t]
  \centering
  \input{generated/img/ds1_type_counts.pgf}
  \caption{\label{fig:type_counts}%
    Type counts of the most frequent types.
  }
\end{figure}

Besides the question of what is the \enquote{correct} type,
the evaluation based on \toolname{TypeEvalPy} is also limited
regarding generic types: \toolname{TypeEvalPy}'s ground truth discards
element types from generic containers, so \inlinelst{list[int]} is
represented simply as \inlinelst{list}, and similarly for other
parameterised types such as \inlinelst{dict} or \inlinelst{set}.
Thus, the evaluation cannot distinguish between a tool
that correctly infers \inlinelst{list[int]} and one that infers
\inlinelst{list}. These two outcomes are treated equally.
While this affects all tools, \typetracing is particularly affected
by this limitation, because it records not just the container
type but also the element type observed during successful executions
and thus produces more fine-grained information than the ground
truth can represent.
As a consequence, the scores for RQ4 are a lower bound for the
inference quality.

Another reason for missing certain matches is that the type system
behind \typetracing performs type unification, merging, \eg, a type and
its super type to only the super type.
While this is sound for test generation (as the test generator selects
inputs from the super type), this can cause the inferred type to
diverge from the
ground-truth annotation in \toolname{TypeEvalPy}, which affects the
score negatively even though the inferred type is arguably correct.
Furthermore, we do not study the correctly determined types
on local variables because our \typetracing approach currently does not take
them into account---a variable local to a function cannot usually be
set from the
outside of the function during execution and is thus not beneficial for our
use case of \typetracing: test generation.

When interpreting the results of this RQ, we must also consider the
constitution of types in our dataset, because some types are
easier or harder to infer, which affects the calculated metrics.
\dseval is dominated by primitive types, as
shown in \cref{fig:type_counts}, where \inlinelst{str} is the
most frequent type, followed by \inlinelst{bool} and \inlinelst{int}.
While built-in generic types (\inlinelst{list}, \inlinelst{dict},
\inlinelst{iterable}) are also quite common and there is even a custom
type \inlinelst{money}, the distribution is heavily in favour of
primitive types. This may bias the evaluation and the apparent
performance of the inference approaches, since correctly inferring
simple and common types might be generally easier than inferring
generic or user-defined types. In turn, this might limit the
generalisability of the results to more diverse or domain-specific
codebases. However, as our dataset is sampled from real-world
projects we argue that the distribution of types is at least somewhat
close to the real-world distribution of types.
Overall, a large proportion of real-world project type annotations are
primitive ones, which our dataset reflects.

\begin{figure}[!t]
  \includegraphics{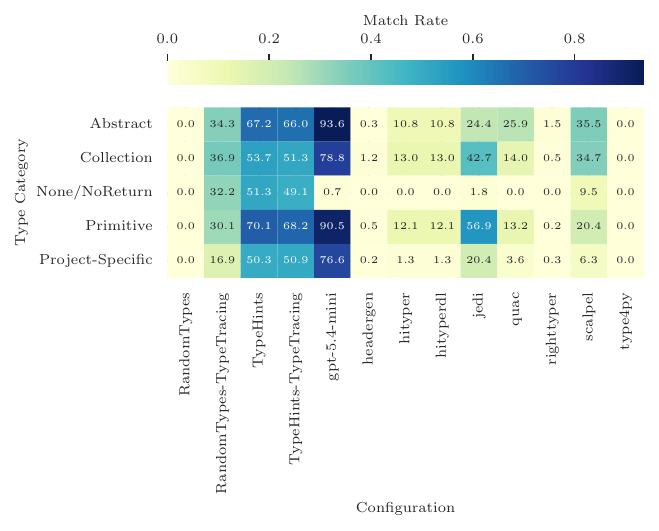}
  \caption{\label{fig:rq4-match-rate-heatmap}%
    Type match rates per type category for each evaluated tool and
    \typetracing configuration.
  }
\end{figure}

\begin{figure}[!t]
  \includegraphics{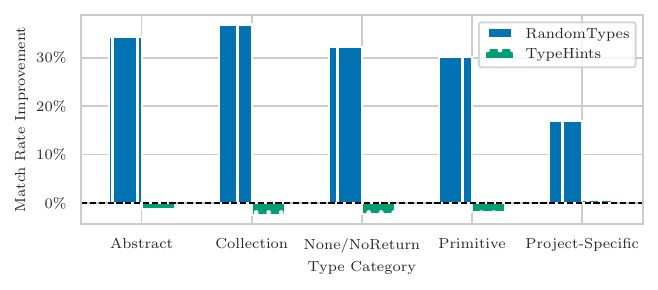}
  \caption{\label{fig:rq4-delta-by-category}%
    Effect of \typetracing on the type match rate per category,
    measured as the difference in match rate when adding \typetracing
    to the \RandomTypes and \TypeHints configurations.
  }
\end{figure}

Given this uneven distribution, aggregate match rates can mask
substantial variation in how well each tool handles different type
categories.
\Cref{fig:rq4-match-rate-heatmap} provides a per-category breakdown
across all evaluated tools, showing where each tool is most and
least effective.
Without type hints, \typetracing achieves match rates of
\RandomTypesTypeTracingCollectionMatchRate{} for collections,
\RandomTypesTypeTracingAbstractMatchRate{} for abstract types,
\RandomTypesTypeTracingNoneNoReturnMatchRate{} for
\texttt{None}/\texttt{NoReturn},
and \RandomTypesTypeTracingPrimitiveMatchRate{} for primitives---all
broadly similar.
The \texttt{None}/\texttt{NoReturn} category is mostly unique to
\pynguin in the heatmap: static and LLM-based tools rarely
produce predictions for this category, because functions that
return \texttt{None} implicitly often receive no prediction.
The match rate for project-specific types is markedly lower at
\RandomTypesTypeTracingProjectSpecificMatchRate{}, which aligns with
the qualitative failure cases discussed above: project-specific types
occupy a much larger search space, making them
harder to observe during the short proxied executions that
\typetracing relies on.
The heatmap further reveals that the pattern of strong performance on
primitive types relative to project-specific types is not unique to
\typetracing.
Static tools such as \jedi often also achieve their highest
match rates on primitives while struggling with project-specific and
abstract types.
We have two possible explanations for why primitives are
easier to infer across all approaches.
First, primitive types may be inherently easier to infer,
because their usage is constrained by arithmetic or
string operations that
reveal the type without requiring deep contextual reasoning.
Second, existing tools may be specifically optimised for
the common case of primitive types.
We believe a mixture of both explanations is most likely.

When comparing the performance on project-specific types
to that on primitive types, we can see that the gap for \pynguin is
smaller than the one for the static tools.
\Typetracing achieves
\RandomTypesTypeTracingProjectSpecificMatchRate{} on project-specific
types against \RandomTypesTypeTracingPrimitiveMatchRate{} on primitives,
retaining \qty{\TypeTracingPrimitiveRetentionPercentage}{\percent} of
its primitive-type performance,
whereas \jedi achieves \JediProjectSpecificMatchRate{} on project-specific
types against \JediPrimitiveMatchRate{} on primitives, retaining only
\qty{\JediPrimitiveRetentionPercentage}{\percent}.
By contrast, \gpt retains \qty{84.64}{\percent} of its primitive-type
performance on project-specific types
(\GptminiProjectSpecificMatchRate{} vs.\ \GptminiPrimitiveMatchRate{}),
even though LLMs often cannot know about project-specific types
from code signatures alone if the types are declared in a different module.
The high relative performance of \gpt on project-specific types may
be because they appeared in the model's training data.
Overall, \pynguin retains a larger share of its primitive-type
performance on project-specific types than most static tools do,
although LLM-based tools such as \gpt close this gap even further.

The analysis so far has characterised the per-category match rates in
the unannotated setting.
To isolate the contribution of \typetracing itself, and to see how it
behaves when annotations are already present,
\cref{fig:rq4-delta-by-category} reports the change in match rate per
category from adding \typetracing to each configuration.
When type hints are already present, adding \typetracing yields only
a small change across all categories---slightly negative for
most~(around \qty{2}{\percent}), and near zero for project-specific
types~(\TTEffectHintsProjectSpecific).
This confirms that the benefit of \typetracing is concentrated in the
unannotated setting, where it provides type information that would
otherwise be entirely absent.
In general, as long as there are non-primitive types, the quality
of inferred types can be improved by increasing the coverage of
generated tests.  Therefore, more research on the generation of input
objects is required~\cite{VPK04,CGO15}.

\section{Related Work}\label{sec:related}

Most existing work on type inference uses static analysis of source
code. Among these, the attribute-centric type
inference~\cite{wu_quac_2024} is maybe closest to our work. This
approach consists of collecting and querying sets of attribute
accesses to predict types, including for container parameters. Unlike
our approach, this is done statically, with the aim to predict types
for existing code, whereas our aim is to generate new test cases.
Probabilistic type inference~\cite{SOP22} is also closely related
since it uses its type model to guide test generation.  The approach,
\guesswhat,
implemented for JavaScript, constructs relations between the elements
of the SUT's AST.\@ Based on these relations, it creates a
probabilistic type model using a semantic analysis of the AST, which
assigns probabilities to each element being of a certain type; \eg,
the expression \inlinelst{x = y + z} can be applied to both strings
and numbers, which is why each type would be assigned a probability
of \qty{50}{\percent}.  The approach uses the probabilities from this
model during test-case construction to choose the types for
parameters.
While \guesswhat relies on a probabilistic model
derived from AST relations, it is not entirely a static approach as it
is executed within a dynamic environment where types are evaluated and
only the most suitable types end up in the final result.
However, the creation of the probabilistic model used to infer types
is entirely static.
In contrast,
our approach retrieves types directly \textit{during} dynamic executions through
recording of attribute accesses.
Combining both approaches might be promising to find possible
input types more quickly.
Since \guesswhat targets JavaScript and no equivalent
type-inference-guided test generator exists for Python, a direct
experimental comparison is beyond the scope of this study; we
therefore did not include
it as a baseline in our evaluation.

While a main difference between most of these prior approaches and our
work is that our technique is dynamic, there is also prior related
work using dynamic analysis.
One approach,
\toolname{MonkeyType}\footnote{\url{https://monkeytype.readthedocs.io/en/latest/},
last accessed 2026-04-30.} observes an SUT during runtime and collects
information which parameter types are passed to functions and what
their return types are. The approach for recording return types works
similarly to ours, but for parameter types, it relies on the premise
that if a developer implements a routine then there must be a location
where this new routine is called.
Another approach, \toolname{RightTyper}~\cite{pizzorno_2025_righttyper}
observes program behaviour through instrumentation and
combines observed traces with static analysis and name resolution
to infer types.
While \pynguin uses instrumentation to measure code coverage,
\typetracing does not use instrumentation, but instead
uses proxy objects to record type information.
A fundamental limitation of both
\toolname{MonkeyType} and
\toolname{RightTyper}
is their reliance on external, pre-existing executions of the SUT---such as an
existing test suite---to observe these types.
Our approach, in contrast, does not require
these existing executions because we generate test cases ourselves.

Recently, Large Language Models (LLMs) have been integrated into
\pynguin to predict the types of the SUT's parameters before
generating test cases~\cite{yang_llm-enhanced_2025}. More advanced
approaches such as \codamosa~\cite{lemieux_codamosa_2023} use LLMs to
complement the search-based generation when the evolutionary algorithm
stagnates. These tools typically use the LLM to retrieve the types of
the parameters of a function or to generate entire test case
skeletons, thus reducing the search space for the foundational
test generator. However, these predictive approaches do not iteratively
refine the types of the parameters during the core search process and are
limited by the LLM's initial predictive capabilities. Furthermore,
LLMs may introduce confounding effects in evaluations, as they can
often guess types from semantic context (e.g., parameter names), which
disguises the performance of the underlying type-inference mechanisms.
A promising hybrid approach would be to combine these methods: the
LLM's predictions could provide a strong initial starting point, which
\typetracing could then dynamically refine during test generation. We
plan to explore this hybridization in future work.

Several type checkers exist for Python, such as \toolname{PyType} and
\toolname{MyPy}, which employ static analysis to perform type
inference. A core principle of these tools is soundness: they only
infer types they can definitively prove. Consequently, if the exact
type of an expression is unclear during static analysis, these tools
may not infer a type at all. While soundness is crucial, this
characteristic, combined with Python's dynamic nature and the
inherent over-approximation~\cite{Mad15} required by static
analysis, means inferred types can sometimes be imprecise or overly
general~\cite{Pav19}. Static type inference itself is an established
field with ongoing research (e.g.,~\cite{HUE+18}) aiming to address
these challenges. Our approach differs by focusing on inferring or
recording types that are pragmatically close to the ground truth,
even if this means occasionally sacrificing strict soundness for
broader coverage or more specific (though potentially unverified)
type information.
We compared the quality of type predictions from \typetracing with
several static approaches using \toolname{TypeEvalPy}~\cite{VSW+23}, a
recently proposed standard micro-benchmark for type-inference
approaches for the Python programming language. Our results (see
\cref{sec:rq4_results}) demonstrated that \typetracing achieves
comparable or superior results to several static type inference tools.

\section{Conclusions}\label{sec:conclusions}

In this paper, we introduced \typetracing, a novel approach to infer
missing parameter and return type information during unit-test
generation for Python programs. To infer parameter types, \typetracing
wraps arguments in proxies which track how they are used during test
execution and then aims to infer types that match these usages. For
return types, we simply observe the types of the values returned from
routines. We implemented \typetracing into the \pynguin
test-generation framework, and in our experiments observed type
predictions that are of similar quality compared to other
state-of-the-art type-inference approaches.  Our experiments also show
that observed (relative) coverage can increase by up to
\qty{\RelCoverageTypeHintsTypeTracing}{\percent} on modules from
real-world projects, with minor effects also on fault-finding
capability.

While our presentation and implementation focusses on the Python programming
language and the \pynguin test-generation framework, in principle the approach
itself can be applied to other programming languages, such as JavaScript, or
test generators, such as \toolname{Hypothesis}~\cite{MH19,MD20}.
However, doing so would require a dedicated automated test-generation framework
for the target language; since no such framework exists for most
dynamically-typed languages beyond Python, we leave this as future work.
Considering the observed accuracy of LLMs during type inference,
querying an LLM for initial type hints and then refining those using
\typetracing could also be a promising approach for future work.

However, knowing the type of a parameter may not be sufficient for
generating effective tests: There is also the challenge of
instantiating valid and interesting objects matching these types,
which is an aspect that automated unit test generation tools still struggle
with~\cite{LKF23,SJR15}. While this problem is orthogonal to the
question of type inference, the tracing functionality provided by
\typetracing may be useful for supporting approaches that aim to
identify how to construct complex objects (e.g.,~\cite{LOS21}).

\section{Data Availability}\label{sec:eval:data}

We provide a supplementary artefact consisting of the Docker images used, the
projects used for evaluation, the raw result data, and our data analysis from
Zenodo to allow for future usage~\cite{dataset}.

\begin{acks}
  This work was partially supported by the German Research Foundation (DFG)
  under grant FR 2955/5-1~(TYPES4STRINGS: Types For Strings).
\end{acks}

\bibliographystyle{ACM-Reference-Format}
\bibliography{related.bib}

@string{acm = {{ACM}}}

@string{acsc = {Australasian Computer Science Conference~(ACSC)}}

@string{ase = {International Conference on Automated Software
Engineering~(ASE)}}

@string{cav = {International Conference on Computer Aided Verification~(CAV)}}

@string{computer = {Computer}}

@string{dagstuhl = {Schloss Dagstuhl – Leibnitz-Zentrum für Informatik}}

@string{dls = {{ACM} {SIGPLAN} International Symposium on Dynamic
Languages~(DLS)}}

@string{ecoop = {European Conference on Object-Oriented Programming~(ECOOP)}}

@string{fse = {International Symposium on Foundations of Software
Engineering~(FSE)}}

@string{gecco = {Annual Conference on Genetic and Evolutionary
Computation~(GECCO)}}

@string{icpc = {International Conference on Program Comprehension~(ICPC)}}

@string{icse = {International Conference on Software Engineering~(ICSE)}}

@string{ieee = {{IEEE}}}

@string{issta = {International Symposium on Software Testing and
Analysis~(ISSTA)}}

@string{LIPIcs = {Leibnitz International Proceedings in Informatics~(LIPIcs)}}

@string{popl = {Symposium on Principles of Programming Languages~(POPL)}}

@string{saner = {International Conference on Software Analysis,
Evolution, and Reengineering~(SANER)}}

@string{sigplan = {{ACM} {SIGPLAN} Notices}}

@string{springer = {Springer}}

@string{ssbse = {International Symposium on Search Based Software
Engineering~(SSBSE)}}

@string{toplas = {{ACM} Transactions on Programming Languages and Systems}}

@string{tosem = {{ACM} Transactions on Software Engineering and Methodology}}

@string{zenodo = {Zenodo}}

@inproceedings{MGN17,
  author    = {Nevena Milojkovic and Mohammad Ghafari and Oscar Nierstrasz},
  booktitle = {International Conference on Program Comprehension~(ICPC)},
  doi       = {10.1109/ICPC.2017.10},
  pages     = {312--315},
  publisher = {{IEEE} Computer Society},
  title     = {It's Duck (Typing) Season!},
  year      = {2017}
}

@inproceedings{SJR15,
  author    = {Sina Shamshiri and René Just and José Miguel Rojas and
  Gordon Fraser and Phil McMinn and Andrea Arcuri},
  booktitle = {International Conference on Automated Software
  Engineering~(ASE)},
  doi       = {10.1109/ASE.2015.86},
  pages     = {201--211},
  publisher = {{IEEE} Computer Society},
  title     = {Do Automatically Generated Unit Tests Find Real Faults? An
  Empirical Study of Effectiveness and Challenges},
  year      = {2015}
}

@inproceedings{PLE+07,
  author    = {Carlos Pacheco and Shuvendu K. Lahiri and Michael D.
  Ernst and Thomas Ball},
  booktitle = {International Conference on Software Engineering~(ICSE)},
  doi       = {10.1109/ICSE.2007.37},
  pages     = {75--84},
  publisher = {{IEEE} Computer Society},
  title     = {Feedback-Directed Random Test Generation},
  year      = {2007}
}

@inproceedings{MPP19,
  author    = {Rabee Sohail Malik and Jibesh Patra and Michael Pradel},
  booktitle = {International Conference on Software Engineering~(ICSE)},
  doi       = {10.1109/ICSE.2019.00045},
  pages     = {304--315},
  publisher = {{IEEE}/{ACM}},
  title     = {NL2Type: Inferring JavaScript Function Types from Natural
  Language Information},
  year      = {2019}
}

@article{FA13,
  author    = {Gordon Fraser and Andrea Arcuri},
  doi       = {10.1109/TSE.2012.14},
  number    = {2},
  pages     = {276--291},
  publisher = {{IEEE}},
  title     = {Whole Test Suite Generation},
  volume    = {39},
  journal   = {{IEEE} Transactions on Software Engineering},
  year      = {2013}
}

@article{PKT18b,
  author    = {Annibale Panichella and Fitsum Meshesha Kifetew and
  Paolo Tonella},
  doi       = {10.1109/TSE.2017.2663435},
  number    = {2},
  pages     = {122--158},
  publisher = {{IEEE}},
  title     = {Automated Test Case Generation as a Many-Objective
  Optimisation Problem with Dynamic Selection of the Targets},
  volume    = {44},
  journal   = {{IEEE} Transactions on Software Engineering},
  year      = {2018}
}

@article{VD00,
  author    = {Andr{\'a}s Vargha and Harold D. Delaney},
  doi       = {10.3102/10769986025002101},
  number    = {2},
  pages     = {101--132},
  publisher = {Sage Publications},
  title     = {A critique and improvement of the CL common language
  effect size statistics of McGraw and Wong},
  volume    = {25},
  journal   = {journaltitle of Educational and Behavioral Statistics},
  year      = {2000}
}

@inproceedings{Ton04,
  author    = {Paolo Tonella},
  booktitle = {International Symposium on Software Testing and
  Analysis~(ISSTA)},
  doi       = {10.1145/1007512.1007528},
  pages     = {119--128},
  publisher = {{ACM}},
  title     = {Evolutionary testing of classes},
  year      = {2004}
}

@inproceedings{LKF20,
  author     = {Stephan Lukasczyk and Florian Kroi\ss{} and Gordon Fraser},
  booktitle  = {International Symposium on Search Based Software
  Engineering~(SSBSE)},
  doi        = {10.1007/978-3-030-59762-7_2},
  pages      = {9--24},
  publisher  = {Springer},
  readstatus = {read},
  series     = {Lecture Notes in Computer Science},
  title      = {Automated Unit Test Generation for Python},
  volume     = {12420},
  year       = {2020}
}

@inproceedings{GPS+15,
  author    = {Liang Gong and Michael Pradel and Manu Sridharan and
  Koushik Sen},
  booktitle = {International Symposium on Software Testing and
  Analysis~(ISSTA)},
  doi       = {10.1145/2771783.2771809},
  pages     = {94--105},
  publisher = {{ACM}},
  title     = {{DLint:} Dynamically Checking Bad Coding Practices in
  {JavaScript}},
  year      = {2015}
}

@inproceedings{KHR12,
  author    = {Sebastian Kleinschmager and Stefan Hanenberg and Romain
  Robbes and Éric Tanter and Andreas Stefik},
  booktitle = {International Conference on Program Comprehension~(ICPC)},
  doi       = {10.1109/ICPC.2012.6240483},
  pages     = {153--162},
  publisher = {{IEEE} Computer Society},
  title     = {Do Static Type Systems Improve the Maintainability of
  Software Systems? An Empirical Study},
  year      = {2012}
}

@article{MH19,
  author  = {David MacIver and Zac Hatfield{-}Dodds},
  doi     = {10.21105/joss.01891},
  number  = {43},
  pages   = {1891},
  title   = {Hypothesis: {A} new approach to property-based testing},
  volume  = {4},
  journal = {Journal of Open Source Software},
  year    = {2019}
}

@inproceedings{MD20,
  author    = {David MacIver and Alastair F. Donaldson},
  booktitle = {European Conference on Object-Oriented Programming~(ECOOP)},
  doi       = {10.4230/LIPIcs.ECOOP.2020.13},
  pages     = {13:1--13:27},
  publisher = {Schloss Dagstuhl – Leibnitz-Zentrum für Informatik},
  series    = {Leibnitz International Proceedings in Informatics~(LIPIcs)},
  title     = {Test-Case Reduction via Test-Case Generation: Insights
  from the Hypothesis Reducer (Tool Insights Paper)},
  volume    = {166},
  year      = {2020}
}

@article{JH11,
  author  = {Yue Jia and Mark Harman},
  doi     = {10.1109/TSE.2010.62},
  number  = {5},
  pages   = {649--678},
  title   = {An Analysis and Survey of the Development of Mutation Testing},
  volume  = {37},
  journal = {{IEEE} Transactions on Software Engineering},
  year    = {2011}
}

@article{MW47,
  author  = {Henry B. Mann and Donald R. Whitney},
  doi     = {10.1214/aoms/1177730491},
  number  = {1},
  pages   = {50--60},
  title   = {On a Test of Whether one of Two Random Variables is
  Stochastically Larger than the Other},
  volume  = {18},
  journal = {The Annals of Mathematical Statistics},
  year    = {1947}
}

@article{AF13,
  author  = {Andrea Arcuri and Gordon Fraser},
  doi     = {10.1007/s10664-013-9249-9},
  number  = {3},
  pages   = {594--623},
  title   = {Parameter tuning or default values? An empirical
  investigation in search-based software engineering},
  volume  = {18},
  journal = {Empirical Software Engineering},
  year    = {2013}
}

@inproceedings{Pea95,
  author    = {Karl Pearson},
  booktitle = {Proceedings of the Royal Society of London},
  pages     = {240--242},
  title     = {Note on Regression and Inheritance in the Case of Two Parents},
  volume    = {58},
  year      = {1895}
}

@inproceedings{TMG22,
  author    = {Daniel Trübenbach and Sebastian Müller and Lars Grunske},
  booktitle = {International Workshop on Search-Based Software
  Testing~(SBST@ICSE)},
  doi       = {10.1145/3526072.3527523},
  pages     = {6--13},
  publisher = {{IEEE}/{ACM}},
  title     = {A Comparative Evaluation on the Quality of Manual and
    Automatic Test Case Generation Techniques for Scientific Software—A
  Case Study of a Python Project for Material Science Workflows},
  year      = {2022}
}

@article{CGA18,
  author  = {José Campos and Yan Ge and Nasser Albunian and Gordon
  Fraser and Marcelo Eler and Andrea Arcuri},
  doi     = {10.1016/j.infsof.2018.08.010},
  pages   = {207--235},
  title   = {An empirical evaluation of evolutionary algorithms for
  unit test suite generation},
  volume  = {104},
  journal = {Information {\&} Software Technology},
  year    = {2018}
}

@inproceedings{LF22,
  author     = {Stephan Lukasczyk and Gordon Fraser},
  booktitle  = {International Conference on Software Engineering
  Companion~(ICSE Companion)},
  doi        = {10.1145/3510454.3516829},
  pages      = {168--172},
  publisher  = {{IEEE}/{ACM}},
  readstatus = {read},
  title      = {Pynguin: Automated Unit Test Generation for Python},
  year       = {2022}
}

@inproceedings{JJI14,
  author    = {René Just and Darioush Jalali and Laura Inozemtseva and
  Michael D. Ernst and Reid Holmes and Gordon Fraser},
  booktitle = {International Symposium on Foundations of Software
  Engineering~(FSE)},
  doi       = {10.1145/2635868.2635929},
  pages     = {654--665},
  publisher = {{ACM}},
  title     = {Are Mutants a Valid Substitute for Real Faults in
  Software Testing?},
  year      = {2014}
}

@inproceedings{RMM20,
  author    = {Ingkarat Rak{-}amnouykit and Daniel McCrevan and Ana L.
  Milanova and Martin Hirzel and Julian Dolby},
  booktitle = {{ACM} {SIGPLAN} International Symposium on Dynamic
  Languages~(DLS)},
  doi       = {10.1145/3426422.3426981},
  pages     = {57--70},
  publisher = {{ACM}},
  title     = {Python 3 Types in the Wild: A Tale of Two Type Systems},
  year      = {2020}
}

@inproceedings{HH09,
  author    = {Alex Holkner and James Harland},
  booktitle = {Australasian Computer Science Conference~(ACSC)},
  pages     = {17--25},
  publisher = {Australian Computer Society},
  series    = {{CRPIT}},
  title     = {Evaluating the dynamic behaviour of Python applications},
  url       =
  {http://crpit.scem.westernsydney.edu.au/abstracts/CRPITV91Holkner.html},
  volume    = {91},
  year      = {2009}
}

@inproceedings{GP22,
  author     = {Luca Di Grazia and Michael Pradel},
  booktitle  = {Joint Meeting of the European Software Engineering
    Conference and the Symposium on the Foundations of Software
  Engineering~(ESEC/FSE)},
  doi        = {10.1145/3540250.3549114},
  pages      = {209--220},
  publisher  = {{ACM}},
  readstatus = {read},
  title      = {The Evolution of Type Annotations in Python: An
  Empirical Study},
  year       = {2022}
}

@inproceedings{HBB+18,
  author    = {Vincent J. Hellendoorn and Christian Bird and Earl T.
  Barr and Miltiadis Allamanis},
  booktitle = {Joint Meeting of the European Software Engineering
    Conference and the Symposium on the Foundations of Software
  Engineering~(ESEC/FSE)},
  doi       = {10.1145/3236024.3236051},
  pages     = {152--162},
  publisher = {{ACM}},
  title     = {Deep Learning Type Inference},
  year      = {2018}
}

@inproceedings{MLP+22,
  author    = {Amir M. Mir and Evaldas Latoskinas and Sebastian Proksch
  and Georgios Gousios},
  booktitle = {International Conference on Software Engineering~(ICSE)},
  doi       = {10.1145/3510003.3510124},
  pages     = {2241--2252},
  publisher = {{ACM}},
  title     = {Type4Py: Practical Deep Similarity Learning-Based Type
  Inference for Python},
  year      = {2022}
}

@inproceedings{WL05,
  author    = {Stefan Wappler and Frank Lammermann},
  booktitle = {Annual Conference on Genetic and Evolutionary
  Computation~(GECCO)},
  doi       = {10.1145/1068009.1068187},
  pages     = {1053--1060},
  publisher = {{ACM}},
  title     = {Using Evolutionary Algorithms for the Unit Testing of
  Object-Oriented Software},
  year      = {2005}
}

@inproceedings{LOS21,
  author    = {Yun Lin and You Sheng Ong and Jun Sun and Gordon Fraser
  and Jin Song Dong},
  booktitle = {Joint Meeting of the European Software Engineering
    Conference and the Symposium on the Foundations of Software
  Engineering~(ESEC/FSE)},
  doi       = {10.1145/3468264.3468619},
  pages     = {1068--1080},
  publisher = {{ACM}},
  title     = {Graph-based Seed Object Synthesis for Seach-Based Unit Testing},
  year      = {2021}
}

@inproceedings{SMC13,
  author    = {Samir Sapra and Marius Minea and Sagar Chaki and Arie
  Gurfinkel and Edmund M. Clarke},
  booktitle = {{IFIP} International Conference on Testing Software and Systems},
  doi       = {10.1007/978-3-642-41707-8_20},
  pages     = {283--289},
  publisher = {Springer},
  series    = {Lecture Notes in Computer Science},
  title     = {Finding Errors in Python Programs Using Dynamic
  Symbolic Execution},
  volume    = {8254},
  year      = {2013}
}

@inproceedings{HUE+18,
  author    = {Mostafa Hassan and Caterina Urban and Marco Eilers and
  Peter Müller},
  booktitle = {International Conference on Computer Aided Verification~(CAV)},
  doi       = {10.1007/978-3-319-96142-2_2},
  pages     = {12--19},
  publisher = {Springer},
  series    = {Lecture Notes in Computer Science},
  title     = {MaxSMT-Based Type Inference for Python 3},
  volume    = {10982},
  year      = {2018}
}

@inproceedings{SOP22,
  author    = {Dimitri Michel Stallenberg and Mitchell Olsthoorn and
  Annibale Panichella},
  booktitle = {International Symposium on Search Based Software
  Engineering~(SSBSE)},
  doi       = {10.1007/978-3-031-21251-2_5},
  pages     = {67--82},
  publisher = {Springer},
  series    = {Lecture Notes in Computer Science},
  title     = {Guess What: Test Case Generation for Javascript with
  Unsupervised Probabilistic Type Inference},
  volume    = {13711},
  year      = {2022}
}

@inproceedings{ST07,
  author    = {Jeremy G. Siek and Walid Taha},
  booktitle = {European Conference on Object-Oriented Programming~(ECOOP)},
  doi       = {10.1007/978-3-540-73589-2_2},
  pages     = {2--27},
  publisher = {Springer},
  series    = {Lecture Notes in Computer Science},
  title     = {Gradual Typing for Objects},
  volume    = {4609},
  year      = {2007}
}

@inproceedings{PGL22,
  author    = {Yun Peng and Cuiyun Gao and Zongjie Li and Bowei Gao and
  David Lo and Qirun Zhang and Michael Lyu},
  booktitle = {International Conference on Software Engineering~(ICSE)},
  doi       = {10.1145/3510003.3510038},
  pages     = {2019--2030},
  publisher = {{ACM}},
  title     = {Static Inference Meets Deep Learning: A Hybrid Type
  Inference Approach for Python},
  year      = {2022}
}

@inproceedings{PGL+20,
  author    = {Michael Pradel and Georgios Gousios and Jason Liu and
  Satish Chendra},
  booktitle = {Joint Meeting of the European Software Engineering
    Conference and the Symposium on the Foundations of Software
  Engineering~(ESEC/FSE)},
  doi       = {10.1145/3368089.3409715},
  pages     = {209--220},
  publisher = {{ACM}},
  title     = {TypeWriter: Neural Type Prediction with Search-Based Validation},
  year      = {2020}
}

@inproceedings{DDS21,
  author    = {Swaroopa Dola and Matthew B. Dwyer and Mary Lou Soffa},
  booktitle = {International Conference on Software Engineering~(ICSE)},
  doi       = {10.1109/ICSE43902.2021.00032},
  pages     = {226--237},
  publisher = {{IEEE}},
  title     = {Distribution-Aware Testing of Neural Networks Using
  Generative Models},
  year      = {2021}
}

@phdthesis{Pav19,
  author      = {Zvonimir Pavlinovic},
  institution = {New York University, {USA}},
  title       = {Leveraging Program Analysis for Type Inference},
  type        = {phdthesis},
  year        = {2019}
}

@phdthesis{Mad15,
  author      = {Magnus Madsen},
  institution = {Aarhus University, Denmark},
  title       = {Static Analysis of Dynamic Languages},
  type        = {phdthesis},
  year        = {2015}
}

@inproceedings{RVK15,
  author    = {Veselin Raychev and Martin Vechev and Andreas Krause},
  booktitle = {Symposium on Principles of Programming Languages~(POPL)},
  doi       = {10.1145/2676726.2677009},
  pages     = {111--124},
  publisher = {{ACM}},
  title     = {Predicting Program Properties from “Big Code”},
  year      = {2015}
}

@inproceedings{VPK04,
  author    = {Willem Visser and Corina S. Păsăreanu and Sarfraz Khurshid},
  booktitle = {International Symposium on Software Testing and
  Analysis~(ISSTA)},
  doi       = {10.1145/1007512.1007526},
  pages     = {97--107},
  publisher = {{ACM}},
  title     = {Test Input Generation with Java PathFinder},
  year      = {2004}
}

@inproceedings{CGO15,
  author    = {Shauvik Roy Choudhary and Alessandra Gorla and Alessandro Orso},
  booktitle = {International Conference on Automated Software
  Engineering~(ASE)},
  doi       = {10.1109/ASE.2015.89},
  pages     = {429--440},
  publisher = {{IEEE} Computer Society},
  title     = {Automated Test Input Generation for Android: Are We There Yet?},
  year      = {2015}
}

@article{VSW+23,
  author     = {Ashwin Prasad Shivarpatna Venkatesh and Samkutty Sabu and
  Jiawei Wang and Amir M. Mir and Li Li and Eric Bodden},
  eprint     = {2312.16882},
  eprinttype = {arXiv},
  title      = {TypeEvalPy: {A} Micro-benchmarking Framework for Python
  Type Inference},
  volume     = {abs/2312.16882},
  journal    = {CoRR},
  year       = {2023}
}

@inproceedings{GRP+24,
  author    = {Martin Gruber and Muhammad Firhard Roslan and Owain Parry
  and Fabian Scharnb{\"{o}}ck and Phil McMinn and Gordon Fraser},
  booktitle = {International Conference on Software Engineering~(ICSE)},
  doi       = {10.1145/3597503.3608138},
  pages     = {47:1--47:12},
  publisher = ACM,
  title     = {Do Automatc Test Generation Tools Generate Flaky Tests},
  year      = {2024}
}

@article{EMB+24,
  author     = {Nicolas Erni and Al{-}Ameen Mohammed Ali Mohammed and
    Christian Birchler and Pouria Derakhshanfar and Stephan Lukasczyk
  and Sebastiano Panichella},
  eprint     = {2401.15189},
  eprinttype = {arXiv},
  title      = {{SBFT} Tool Competition 2024 - Python Test Case
  Generation Track},
  volume     = {abs/2401.15189},
  journal    = {CoRR},
  year       = {2024}
}

@article{OGK+23,
  author  = {Milos Ojdanic and Aayush Garg and Ahmed Khanfir and Renzo
  Degiovanni and Mike Papadakis and Yves Le Traon},
  doi     = {10.1109/TSE.2023.3277564},
  number  = {7},
  pages   = {3922--3938},
  title   = {Syntactic Versus Semantic Similarity of Artificial andReal
  Faults in Mutation Testing Studies},
  volume  = {49},
  journal = {{IEEE} Transactions on Software Engineering},
  year    = {2023}
}

@inproceedings{ABL05,
  author    = {James H. Andrews and Lionel C. Briand and Yvan Labiche},
  booktitle = {International Conference on Software Engineering~(ICSE)},
  doi       = {10.1145/1062455.1062530},
  pages     = {402--411},
  publisher = {{ACM}},
  title     = {Is Mutation an Appropriate Tool for Testing Experiments?},
  year      = {2005}
}

@article{LKF23,
  author     = {Stephan Lukasczyk and Florian Kroiß and Gordon Fraser},
  doi        = {10.1007/s10664-022-10248-w},
  number     = {2},
  pages      = {36:1--36:46},
  readstatus = {read},
  title      = {An empirical study of automated unit test generation
  for Python},
  volume     = {28},
  journal    = {Empirical Software Engineering},
  year       = {2023}
}

@inproceedings{PSY+18,
  author    = {Mike Papadakis and Donghwan Shin and Shin Yoo and
  Doo{-}Hwan Bae},
  booktitle = {International Conference on Software Engineering~(ICSE)},
  doi       = {10.1145/3180155.3180183},
  pages     = {537--548},
  publisher = {{ACM}},
  subtitle  = {A Large Scale Empirical study on the Relationship
  Between Mutants and Real Faults},
  title     = {Are Mutation Scores Correlated with Real Fault Detection?},
  year      = {2018}
}

@inproceedings{PWW+23,
  author    = {Yun Peng and Chaozhen Wang and Wenxuan Wang and Cuiyun
  Gao and Michael R. Lyu},
  booktitle = {International Conference on Automated Software
  Engineering~(ASE)},
  doi       = {10.1109/ASE56229.2023.00031},
  pages     = {988--999},
  publisher = {{IEEE}},
  title     = {Generative Type Inference for Python},
  year      = {2023}
}

@inproceedings{VWL+23,
  author    = {Ashwin Prasad Shivarpatna Venkatesh and Jiawei Wang and
  Li Li and Eric Bodden},
  booktitle = {International Conference on Software Analysis,
  Evolution, and Reengineering~(SANER)},
  doi       = {10.1109/SANER56733.2023.00044},
  pages     = {391--401},
  publisher = {{IEEE}},
  title     = {Enhancing Comprehension and Navigation in Jupyter
  Notebooks with Static Analysis},
  year      = {2023}
}

@article{LWQ22,
  author     = {Li Li and Jiawei Wang and Haowei Quan},
  eprint     = {2202.11840},
  eprinttype = {arXiv},
  title      = {Scalpel: The Python Static Analysis Framework},
  volume     = {abs/2202.11840},
  journal    = {CoRR},
  year       = {2022}
}

@dataset{dataset,
  author    = {Lukas Krodinger and Stephan Lukasczyk and Gordon Fraser},
  title     = {Artifact for the paper "Combining Type Inference
    and Automated Unit Test Generation for Python"
  initially submitted to TOSEM 2025},
  month     = jul,
  year      = 2026,
  publisher = {Zenodo},
  doi       = {10.5281/zenodo.15788695},
  url       = {https://doi.org/10.5281/zenodo.15788695}
}

@book{Pie02,
  author    = {Benjamin C. Pierce},
  title     = {Types and Programming Languages},
  publisher = {{MIT} Press},
  year      = {2002},
  isbn      = {978-0-262-16209-8}
}

@incollection{Car04,
  author     = {Luca Cardelli},
  booktitle  = {Computer Science Handbook},
  date       = {2004},
  title      = {Type Systems},
  edition    = {2},
  editor     = {Allen B. Tucker},
  isbn       = {978-1-584-88360-9},
  publisher  = {CRC Press},
  bookauthor = {Allen B. Tucker}
}

@inproceedings{yan_dlinfer_2023,
  address    = {Melbourne, Australia},
  title      = {{DLInfer}: {Deep} {Learning} with {Static} {Slicing} for
  {Python} {Type} {Inference}},
  copyright  = {https://doi.org/10.15223/policy-029},
  isbn       = {978-1-66545-701-9},
  shorttitle = {{DLInfer}},
  url        = {https://ieeexplore.ieee.org/document/10172544/},
  doi        = {10.1109/ICSE48619.2023.00170},
  language   = {en},
  urldate    = {2024-09-06},
  booktitle  = {2023 {IEEE}/{ACM} 45th {International} {Conference} on
  {Software} {Engineering} ({ICSE})},
  publisher  = {IEEE},
  author     = {Yan, Yanyan and Feng, Yang and Fan, Hongcheng and Xu, Baowen},
  month      = may,
  year       = {2023},
  note       = {3 citations (Crossref) [2024-09-06]},
  pages      = {2009--2021}
}

@article{wu_quac_2024,
  title      = {{QuAC}: {Quick} {Attribute}-{Centric} {Type} {Inference}
  for {Python}},
  volume     = {8},
  shorttitle = {{QuAC}},
  url        = {https://dl.acm.org/doi/10.1145/3689783},
  doi        = {10.1145/3689783},
  number     = {OOPSLA2},
  urldate    = {2025-03-21},
  journal    = {Reproduction Package for Article `QuAC: Quick
  Attribute-Centric Type Inference for Python`},
  author     = {Wu, Jifeng and Lemieux, Caroline},
  month      = oct,
  year       = {2024},
  pages      = {343:2040--343:2069}
}

@article{dakhel_effective_2024,
  title    = {Effective test generation using pre-trained {Large}
  {Language} {Models} and mutation testing},
  volume   = {171},
  issn     = {0950-5849},
  url      =
  {https://www.sciencedirect.com/science/article/pii/S0950584924000739},
  doi      = {10.1016/j.infsof.2024.107468},
  urldate  = {2024-11-14},
  journal  = {Information and Software Technology},
  author   = {Dakhel, Arghavan Moradi and Nikanjam, Amin and
  Majdinasab, Vahid and Khomh, Foutse and Desmarais, Michel C.},
  month    = jul,
  year     = {2024},
  pages    = {107468}
}

@misc{yang_enhancing_2024,
  title     = {Enhancing {LLM}-based {Test} {Generation} for
  {Hard}-to-{Cover} {Branches} via {Program} {Analysis}},
  url       = {http://arxiv.org/abs/2404.04966},
  doi       = {10.48550/arXiv.2404.04966},
  urldate   = {2024-11-14},
  publisher = {arXiv},
  author    = {Yang, Chen and Chen, Junjie and Lin, Bin and Zhou, Jianyi
  and Wang, Ziqi},
  month     = apr,
  year      = {2024},
  note      = {arXiv:2404.04966}
}

@misc{pizzorno_coverup_2024,
  title      = {{CoverUp}: {Coverage}-{Guided} {LLM}-{Based} {Test}
  {Generation}},
  shorttitle = {{CoverUp}},
  url        = {http://arxiv.org/abs/2403.16218},
  doi        = {10.48550/arXiv.2403.16218},
  urldate    = {2024-11-14},
  publisher  = {arXiv},
  author     = {Pizzorno, Juan Altmayer and Berger, Emery D.},
  month      = sep,
  year       = {2024},
  note       = {arXiv:2403.16218}
}

@inproceedings{lemieux_codamosa_2023,
  title      = {{CodaMosa}: {Escaping} {Coverage} {Plateaus} in {Test}
  {Generation} with {Pre}-trained {Large} {Language} {Models}},
  shorttitle = {{CodaMosa}},
  url        = {https://ieeexplore.ieee.org/document/10172800},
  doi        = {10.1109/ICSE48619.2023.00085},
  urldate    = {2024-11-19},
  booktitle  = {2023 {IEEE}/{ACM} 45th {International} {Conference} on
  {Software} {Engineering} ({ICSE})},
  author     = {Lemieux, Caroline and Inala, Jeevana Priya and Lahiri,
  Shuvendu K. and Sen, Siddhartha},
  month      = may,
  year       = {2023},
  note       = {ISSN: 1558-1225},
  pages      = {919--931}
}

@article{yang_llm-enhanced_2025,
  title    = {{LLM}-enhanced evolutionary test generation for untyped
  languages},
  volume   = {32},
  issn     = {1573-7535},
  url      = {https://doi.org/10.1007/s10515-025-00496-7},
  doi      = {10.1007/s10515-025-00496-7},
  language = {en},
  number   = {1},
  urldate  = {2025-02-27},
  journal  = {Automated Software Engineering},
  author   = {Yang, Ruofan and Xu, Xianghua and Wang, Ran},
  month    = feb,
  year     = {2025},
  pages    = {20}
}

@inproceedings{xiao_optimizing_2024,
  address    = {New York, NY, USA},
  series     = {Internetware '24},
  title      = {Optimizing {Search}-{Based} {Unit} {Test} {Generation}
  with {Large} {Language} {Models}: {An} {Empirical} {Study}},
  isbn       = {979-8-4007-0705-6},
  shorttitle = {Optimizing {Search}-{Based} {Unit} {Test}
  {Generation} with {Large} {Language} {Models}},
  url        = {https://dl.acm.org/doi/10.1145/3671016.3674813},
  doi        = {10.1145/3671016.3674813},
urldate    = {2025-03-23},
booktitle  = {Proceedings of the 15th {Asia}-{Pacific} {Symposium} on
{Internetware}},
publisher  = {Association for Computing Machinery},
author     = {Xiao, Danni and Guo, Yimeng and Li, Yanhui and Chen, Lin},
month      = jul,
year       = {2024},
pages      = {71--80}
}

@article{bucur_prototyping_2014,
title    = {Prototyping symbolic execution engines for interpreted languages},
volume   = {42},
issn     = {0163-5964},
url      = {https://dl.acm.org/doi/10.1145/2654822.2541977},
doi      = {10.1145/2654822.2541977},
number   = {1},
urldate  = {2025-05-29},
journal  = {SIGARCH Comput. Archit. News},
author   = {Bucur, Stefan and Kinder, Johannes and Candea, George},
month    = feb,
year     = {2014},
pages    = {239--254}
}

@inproceedings{ding_dynamic_2016,
title     = {Dynamic {Symbolic} {Execution} {Tool} for {Python} {Programs}},
url       = {https://ieeexplore.ieee.org/abstract/document/8047141},
doi       = {10.1109/ICITBS.2016.88},
urldate   = {2025-05-29},
booktitle = {2016 {International} {Conference} on {Intelligent}
{Transportation}, {Big} {Data} \& {Smart} {City} ({ICITBS})},
author    = {Ding, Xuefeng and Huang, Wanyu and Liu, Ying and Wantao,
Chen and Xuyang, Ding},
month     = dec,
year      = {2016},
pages     = {212--217}
}

@article{ryan_code_aware_2024,
title      = {Code-{Aware} {Prompting}: {A} {Study} of {Coverage}-{Guided}
{Test} {Generation} in {Regression} {Setting} using {LLM}},
volume     = {1},
issn       = {2994-970X},
shorttitle = {Code-{Aware} {Prompting}},
url        = {https://dl.acm.org/doi/10.1145/3643769},
doi        = {10.1145/3643769},
language   = {en},
number     = {FSE},
urldate    = {2024-11-26},
journal    = {Proceedings of the ACM on Software Engineering},
author     = {Ryan, Gabriel and Jain, Siddhartha and Shang, Mingyue and
Wang, Shiqi and Ma, Xiaofei and Ramanathan, Murali Krishna and Ray, Baishakhi},
month      = jul,
year       = {2024},
pages      = {951--971}
}

@book{PY07,
author    = {Mauro Pezz{\`{e}} and Michal Young},
title     = {Software testing and analysis - process, principles and
techniques},
publisher = {Wiley},
year      = {2007},
isbn      = {978-0-471-45593-6}
}

@inproceedings{boonstoppel_rwset_2008,
author    = {Boonstoppel, Peter
and Cadar, Cristian
and Engler, Dawson},
title     = {RWset: Attacking Path Explosion in Constraint-Based Test
Generation},
booktitle = {Tools and Algorithms for the Construction and Analysis of Systems},
year      = {2008},
publisher = springer,
pages     = {351--366}
}

@article{ramalingam1994undecidability,
title     = {The undecidability of aliasing},
author    = {Ramalingam, Ganesan},
journal   = toplas,
volume    = {16},
number    = {5},
pages     = {1467--1471},
year      = {1994},
publisher = acm
}

@inproceedings{waychal_practitioners_2021,
author    = {Waychal, Pradeep and Capretz, Luiz Fernando and Jia, Jingdong
and Varona, Daniel and Lizama, Yadira},
booktitle = {2021 IEEE International Conference on Software Analysis,
Evolution and Reengineering (SANER)},
title     = {Practitioners’ Testimonials about Software Testing},
year      = {2021},
pages     = {582-589}
}

@inproceedings{straubinger_engaging_2024,
author    = {Straubinger, Philipp and Fraser, Gordon},
title     = {Engaging Developers in Exploratory Unit Testing through
Gamification},
year      = {2024},
publisher = {Association for Computing Machinery},
booktitle = {Proceedings of the 3rd ACM International Workshop on
Gamification in Software Development, Verification, and Validation},
pages     = {2--9},
numpages  = {8}
}

@inproceedings{santos_would_2017,
author    = {Santos, Ronnie Edson de Souza and Magalhães, Cleyton Vanut
Cordeiro de and Correia-Neto, Jorge da Silva and Silva, Fabio Queda
Bueno da and Capretz, Luiz Fernando and Souza, Rodrigo},
booktitle = {2017 ACM/IEEE International Symposium on Empirical
Software Engineering and Measurement (ESEM)},
title     = {Would You Like to Motivate Software Testers? Ask Them How},
year      = {2017},
pages     = {95-104}
}

@article{deak_challenges_2016,
author  = {Deak, Anca and St\r{a}lhane, Tor and Sindre, Guttorm},
title   = {Challenges and strategies for motivating software testing personnel},
year    = {2016},
volume  = {73},
number  = {C},
pages   = {1--15},
journal = {Information and Software Technology}
}

@article{weyuker_clearing_2000,
author  = {Weyuker, Elaine J and Ostrand, Thomas J and Brophy, J and Prasad, B},
title   = {Clearing a career path for software testers},
year    = {2000},
volume  = {17},
pages   = {76--82},
journal = {IEEE Software}
}

@article{wei_typet5_2023,
title   = {Typet5: Seq2seq type inference using static analysis},
author  = {Wei, Jiayi and Durrett, Greg and Dillig, Isil},
journal = {arXiv preprint arXiv:2303.09564},
year    = {2023}
}

@article{dong_unmasking_2026,
author    = {Dong, Yiwen and Xu, Zhenyang and Tian, Yongqiang and
Sun, Chengnian},
title     = {Unmasking the Type Inference Capabilities of LLMs for
Java Code Snippets},
year      = {2026},
publisher = acm,
journal   = {ACM Trans. Softw. Eng. Methodol.},
month     = jan
}

@article{pizzorno_2025_righttyper,
title={RightTyper: Effective and Efficient Type Annotation for Python},
author={Pizzorno, Juan Altmayer and Berger, Emery D},
journal={arXiv preprint arXiv:2507.16051},
year={2025}
}

@inproceedings{LF25,
author = {Lukasczyk, Stephan and Fraser, Gordon},
booktitle = ssbse,
title = {Search-based Hyperparameter Tuning for Python Unit Test Generation},
year = {2025},
series = {Lecture Notes in Computer Science},
volume = {16228},
pages = {66--81},
publisher = springer,
doi = {10.1007/978-3-032-24839-8\_5},
}

\end{document}